%% file: ApJ.tex
\documentclass[twocolumn]{aastex631}

\usepackage{mathtools}
\usepackage{floatrow}
\usepackage{multirow}
\usepackage{comment}
\usepackage{xspace}
\usepackage{tipa}

\usepackage{algorithm}
\usepackage{algorithmic}

\newcommand\maigo{\mbox{{\sc MaiGO}}\xspace}
\newcommand\cMpch{\ensuremath{\,\mathrm{cMpc}/h}}
\newcommand\hcMpc{\ensuremath{\,h/\mathrm{cMpc}}}
\newcommand\mujica{\mbox{{\sc Mujic\footnotesize{\ensuremath{\Lambda}}}}\xspace}
\newcommand\lamall{\lambda_G}
\newcommand\lamobs{\tilde\lambda_G}
\newcommand\LCDM{\ensuremath{\Lambda\mathrm{CDM}}}

\begin{document}

\title{\mujica : Reconstructing the Initial Conditions \\
from Selection-Function Dominated Spectroscopic Survey Volumes}


\title{\mujica : Initial Condition Reconstruction \\on Realistic Redshift Surveys with Augmented Optimization}

\author[0000-0002-8505-9815]{Chenze Dong}
\affiliation{Kavli IPMU (WPI), UTIAS, The University of Tokyo, Kashiwa, Chiba 277-8583, Japan}
\affiliation{Center for Data-Driven Discovery, Kavli IPMU (WPI), UTIAS, The University of Tokyo, Kashiwa, Chiba 277-8583, Japan}

\correspondingauthor{Chenze Dong}
\email{chenze.dong@ipmu.jp}

\author[0000-0001-7832-5372]{Benjamin Horowitz}
\affiliation{Kavli IPMU (WPI), UTIAS, The University of Tokyo, Kashiwa, Chiba 277-8583, Japan}
\affiliation{Center for Data-Driven Discovery, Kavli IPMU (WPI), UTIAS, The University of Tokyo, Kashiwa, Chiba 277-8583, Japan}
\affiliation{Lawrence Berkeley National Lab, 1 Cyclotron Road, Berkeley, CA 94720, USA}

\author[0000-0002-3568-3900]{Adrian E. Bayer}
\affiliation{Center for Computational Astrophysics, Flatiron Institute, 162 5th Avenue, New York NY 10010, USA}
\affiliation{Department of Astrophysical Sciences, Peyton Hall, Princeton University, Princeton, NJ 08544, USA}

\author[0000-0001-9299-5719]{Khee-Gan Lee}
\affiliation{Kavli IPMU (WPI), UTIAS, The University of Tokyo, Kashiwa, Chiba 277-8583, Japan}
\affiliation{Center for Data-Driven Discovery, Kavli IPMU (WPI), UTIAS, The University of Tokyo, Kashiwa, Chiba 277-8583, Japan}



\begin{abstract}
\input{sections/abstract}
\end{abstract}


\keywords{Large-scale structure of the universe (902) — Cosmic web (330) — Cosmological simulations (321) — Astronomy data analysis (1858) — Bayesian statistics (1900) — Redshift surveys (1378)}

\input{sections/maintext}

\section*{Acknowledgement}
\input{sections/acknowledgement}

\bibliography{biblio}{}
\bibliographystyle{aasjournal}


\appendix
\input{sections/appendix}


\end{document}

%% file: sections/abstract.tex
In this paper, we introduce \mujica{} (\textbf{M}apping the \textbf{U}niverse with \textbf{J}ax-based \textbf{I}nitial \textbf{C}ondition Reconstr$\mathbf{\Lambda}$ction), an optimization-based framework for reconstructing initial conditions from realistic galaxy spectroscopic redshift surveys.
Unlike standard optimization-based approaches, \mujica{} augments the L-BFGS algorithm with a projection operator and rank-order matching to enforce Gaussianity of the initial conditions and substantially improve robustness to incomplete survey geometries.
We validate \mujica{} on a mock lightcone catalog derived from semi-analytic models applied to the Millennium simulation. 
We construct a differentiable forward model that incorporates a fast particle-mesh simulation at megaparsec resolution and a comprehensive treatment of observational effects and survey incompleteness.
\mujica{} reaches good agreement with the true density field down to the scale of the forward model, while maintaining consistency with the Gaussian prior through the projection step.
It also broadly recovers the cosmic web classification, underscoring its value for deciphering environmental information in galaxy evolution studies. 
Beyond its key role in next-generation constrained simulations, the methodology offers a practical way to generate initial guesses and speed up field-level inference, especially for upcoming large-scale galaxy surveys.

%% file: sections/maintext.tex
\section{Introduction}

Large-scale structure forms a ``cosmic web'' organized into knots, filaments, sheets, and voids. 
Since being observed \citep{Joeveer:1978, Gregory:1978, Lapparent:1986} in early galaxy redshift surveys and analyzed in numerical simulations \citep{Bond:1996}, the cosmic web has become a key interface between cosmology and galaxy evolution.
The cosmic web is widely studied from both perspectives of observation \citep{York:2000, Driver:2011, Laigle:2016, Malavasi:2017, Darvish:2017, Barsanti:2022} and simulation \citep{Cautun:2014, Libeskind:2018, Martizzi:2019, Singh:2020, Vurm:2023, Schimd:2024}, revealing its importance at the crossroads of cosmology \citep{Benson:2013, Bleem:2015, Detecting, Thiele:2024, Sunseri:2025jem, Verza:2025} and galaxy evolution \citep{Cen:2006, Martizzi:2019, Galarraga-Espinosa:2021, Lan:2021}.
Despite the great importance of the cosmic web for these studies, unambiguously identifying the cosmic web remains a challenging task. 

In practice, extracting the underlying cosmic web from  observables is non-trivial due to its requirement for full 3D information, in contrast to realistic observational data that might be limited by systematics. 
Many cosmic web finding algorithms (e.g., \cite{Forero-Romero:2009, Hoffman:2012, Cautun:2013}) rely on the density field from simulations, which greatly restricts their application to realistic astrophysical data. 
Even for those methods that allow galaxies as biased tracers \citep{Sousbie:2011, Tempel:2016}, the incompleteness and selection function of the galaxy survey remain challenging to incorporate into the algorithm. 
One must therefore take into account observational systematics to obtain a robust cosmic web.

Density reconstruction via forward-modeling partly addresses this issue since it explicitly connects the density field to observed tracers and incorporates survey selection functions.
Bayesian sampling frameworks, such as Elucid \citep{Wang:2014_1, Wang:2016_2, Wang:2016_3}, ARGO \citep{Ata:2015}, and BORG \citep{Jasche:2019}, have demonstrated significant success in reconstructing the density field. 
\cite{Seljak:2017} proposed an alternative to Bayesian inference, using MAP (maximum a posteriori) for density reconstruction and a Laplacian approximation at the MAP solution for cosmological parameter estimation (see also \cite{2019JCAP...10..035H}).
Recent progress in Lyman-$\alpha$ tomography \citep{Lee:2014, Lee:2016, Lee:2016a, Lee:2018, Newman:2020, Horowitz:2022} also enables the $z>2$ Lyman-$\alpha$ forest absorption to act as a tracer of the large-scale structure free from the galaxy selection function since foreground absorption is typically separated by hundreds of Megaparsecs from the background sources and thus has little correlation with the latters' position. 
The Lyman-$\alpha$ forest tracers have motivated the optimization-based density reconstruction code TARDIS \citep{Horowitz:2019, Horowitz:2021} following the MAP approach by \cite{Seljak:2017}, which recovers the cosmic web with good accuracy. Optimization-based density reconstruction has also been shown to robustly improve BAO reconstruction in galaxy surveys \citep{Bayer:2026zcr}. 
These developments reflect a broader trend of cosmological analysis toward the field-level, in which inference operates directly on the full data vector rather than on summary statistics.

However, it is difficult to quantify the reliability of the reconstructions and their inferred cosmic web. Structure formation models typically rely on strong simplifying assumptions when describing the tracer distribution to keep the computation efficient. 
At the same time, the incompleteness of input tracers renders the reconstruction intrinsically under-constrained.

In Bayesian sampling-based approaches, such as MCMC \citep{Jasche:2013, Jasche:2019, Bayer:2023, Nguyen:2024, Akitsu:2025boy, Simon-Onfroy:2025}, these difficulties are
mitigated by drawing many samples, providing built-in
uncertainty quantification at the expense of
incurring larger computational
costs, while relying on informative cosmological priors to regularize the inference.
However, sampling-based methods are not
immune to model misspecification: if the forward model is inaccurate,
the posterior itself will be biased, regardless of how thoroughly it
is sampled.
For optimization-based methods, the failure mode is more acute.
Because the optimizer seeks a single point estimate, model
misspecification and incompleteness can drive it towards
\textit{reward hacking}, exploiting deficiencies in the forward model to
reduce the loss function while deviating from the true underlying linear
field --- either in the recovered cosmological parameters or in the reconstructed field (in particular giving a non-Gaussian field).  Without continuous regularization from the
sampling process, an unconstrained optimizer can diverge catastrophically
from the prior in regions poorly constrained by data. This mismatch
leaves clear signatures in both the recovered power spectra and the
field-level features \citep[e.g.,][]{Horowitz:2019, Byrohl:2025}.

The objective of this study is to develop a novel optimization-based density reconstruction algorithm, denoted as \mujica{}, which is capable of explicitly maintaining the Gaussianity of the initial conditions. 
Instead of the commonly used Maximum a posteriori (MAP) approach, \mujica{} maximizes the likelihood under the constraints based on the prior.
This approach acts as a workaround for cases in which the prior overwhelms the likelihood, thereby dominating the posterior and leading MAP estimation to unreliable results.
The algorithm features a projection optimizer and a rank-order matching routine; both are crucial for correcting the deviations from the Gaussianity of the initial conditions. 
These features differentiate \mujica{} from previous approaches and significantly improve the stability and physical fidelity of the reconstruction.

A major motivation for developing \mujica{} is the ongoing Subaru Prime Focus Spectrograph (PFS) Subaru Strategic Program (SSP) survey \citep{Takada:2014}.
The PFS-CO (cosmology) component is focused on mapping [OII] emission line galaxies within the redshift range of $0.8<z<2.4$ across an area of $1400 \deg^2$, providing an ideal dataset for density reconstruction. 
In parallel, the PFS-GE (galaxy evolution; \cite{Greene:2022}) component is designed to observe a variety of galaxy types over $0.7\lesssim z\lesssim 6.7$, enabling joint reconstruction using both galaxies over the entire redshift range as well as Lyman-$\alpha$ forest tracers at $z\sim2-3$. 
\mujica{} is particularly suitable for these applications due to its Gaussianity-preserving formulation and robustness to complicated selection functions.

This manuscript is structured as follows. 
We start by introducing the data and methods used in Section \ref{sec:methods}, including the dataset that defines the reconstruction task (Section \ref{ssec:dataset}) and the forward model used to evaluate the similarity between the reconstruction and the observation (Section \ref{ssec:methods-fwd-n-lik}).
In Section \ref{ssec:mujica}, we explain the methodology behind \mujica. 
The results of the numerical test are detailed in Section \ref{sec:results-optimization}.
We discuss the features and potential applications of \mujica in Section \ref{sec:discussion}.

\section{Data and Methods} \label{sec:methods}

\subsection{Mock Dataset and the reconstructed volume} \label{ssec:dataset}
In this section, we introduce mock simulation datasets for validation. Following \cite{Lee:2022}, we utilized the semi-analytic lightcone catalog generated by \cite{Henriques:2015} (hereafter H15), which is based on the Millennium N-body simulation \citep{Springel:2005b}. 
Throughout the paper, we adopt the Planck13 cosmology \citep{Planck:2014} with $\sigma_8=0.829$, $h=0.673$, $(\Omega_\Lambda, \Omega_m, \Omega_b) =(0.685, 0.315, 0.0487)$ to be consistent with the main data set used in this study.

\subsubsection{Background N-body simulation}

The Millennium Simulation is a dark-matter-only simulation, run with GADGET-2 \citep{Springel:2005}, with a box size of $L=500\cMpch$ and a particle count of $2160^3$, using a WMAP Y1 cosmology \citep{Spergel:2003}. 
In H15, the simulation was rescaled to match the more recent Planck13 cosmology \citep{Planck:2014} with the methodology introduced in \cite{Angulo:2015}, which consequently modified the effective box size from $500\cMpch$ to $480.279\cMpch$ and adjusted the redshift of the Millennium snapshots.

\subsubsection{Semi-analytic galaxy catalogs} \label{sssec:semi-analytic-galaxy-catalogs}

The ``Munich'' semi-analytic model (SAM) of galaxy formation \citep{Guo:2011} was utilized in H15 to generate mock galaxy catalogs from the rescaled Millennium simulation. This SAM integrates star formation astrophysics, as well as supernova and AGN radio-mode feedback, and was calibrated to align with observational data up to $z\sim3$, with a focus on galaxy abundances and passive fractions for $8.0\le \log (M_\star /M_\odot) \le 12.0$. The mock galaxies' rest-frame spectral energy distributions (SED) were then computed on the basis of the stellar population synthesis (SPS) model described by \cite{Maraston:2005}.

Subsequently, the H15 lightcone galaxy catalogs were constructed employing the method by \cite{Blaizot:2005} that calculates the galaxy locations and the apparent multi-wavelength photometry using randomly selected positions and orientations within the simulation box. 
The approach of \cite{Kitzbichler:2007} is used to determine the line-of-sight direction in lightcone generation, which prevents the duplication of the same galaxy in the catalogs due to the periodic boundary conditions. 
H15 generated two ``all-sky'' lightcones (LC001, LC002) with a coverage of a solid angle of $4\pi$, but these are confined to galaxies with apparent magnitudes $i<21$. The observer on lightcone 001 was set at the origin position, $(x, y, z)=(0,0,0)$.

We populate the mock catalog for our reconstruction using the galaxy model defined by applying the following filters to the parent catalog, LC001:

\begin{itemize}
    \item Magnitude cut. We consider a magnitude limit of $r<21$, which is motivated by the HectoMAP survey \citep{Sohn:2023} and reaches a completeness of 70\% at $r=21.3$. 
    \item Angular cut. We adopted an angular cut with a half-opening angle of $20^\circ$. Such a survey would be considered as a ``wide-field" survey, but its sky coverage is still smaller than that of all-sky galaxy surveys like SDSS or DESI. 
    \item Redshift cut. The purpose of this cut is to avoid the edge effect due to the periodic simulation box: we set a redshift cut of $0.07<z<0.22$. 
\end{itemize}

\begin{figure}[htbp] 
    \centering
    \includegraphics[width=\textwidth]{figures/galaxy_catalog_stat.pdf}
    \caption{Radial and angular distribution of the galaxies in the mock catalog. 
    Upper panel: 1D histogram of the galaxy counts as a function of the comoving distance. The galaxies in the catalog spans the redshift range of $0.07<z<0.22$, corresponding to a comoving distance range between $d\approx200 \cMpch$ and $d\approx 625 \cMpch$. The distribution is used in the radial response function modeling. 
    Lower panel: the distribution of galaxies on the sky. The angular cut leads to a disk-shape mask centering at $(\alpha, \delta)=(0^\circ, 0^\circ)$ on the all-sky map.
    }\label{fig:gal_cat}
\end{figure}

The selection of the parent galaxy catalogs results in a catalog with $1.06 \times 10^6$ galaxies, covering about $1.24 \times 10^3$ square degrees in the sky and spanning the redshift range $0.07<z<0.22$.
Figure~\ref{fig:gal_cat} shows the distribution of galaxies in the radial direction (upper panel) and the sky coordinates. 
Following the methodology in \cite{Lee:2022}, we calculate the survey response function $R$ to account for the incompleteness of the mock survey. 
The angular response function in our model is, by design, a circular disk with a radius of $20^\circ$ in the sky; the radial response function is modeled via the radial distribution of the galaxies (see Figure 6 of \cite{Lee:2022} and the reference there).

\subsubsection{Reconstructed Volume} \label{sssec:reconstructed_volume}

For the reconstruction grid, we set up a cubic box with a size length $500\cMpch$ centered at the comoving distance $d=400\cMpch$. 
This choice ensures that all observed galaxies are included in the reconstructed volume and guarantees that all observed galaxies are at least $20 \cMpch$ away from the periodic boundary. 
At the same time, only $25.8\%$ of the reconstructed volume is populated with galaxy tracers, meaning that the corresponding MAP problem is highly underdetermined.

\begin{figure}[htbp]
    \centering
    \includegraphics[width=\textwidth]{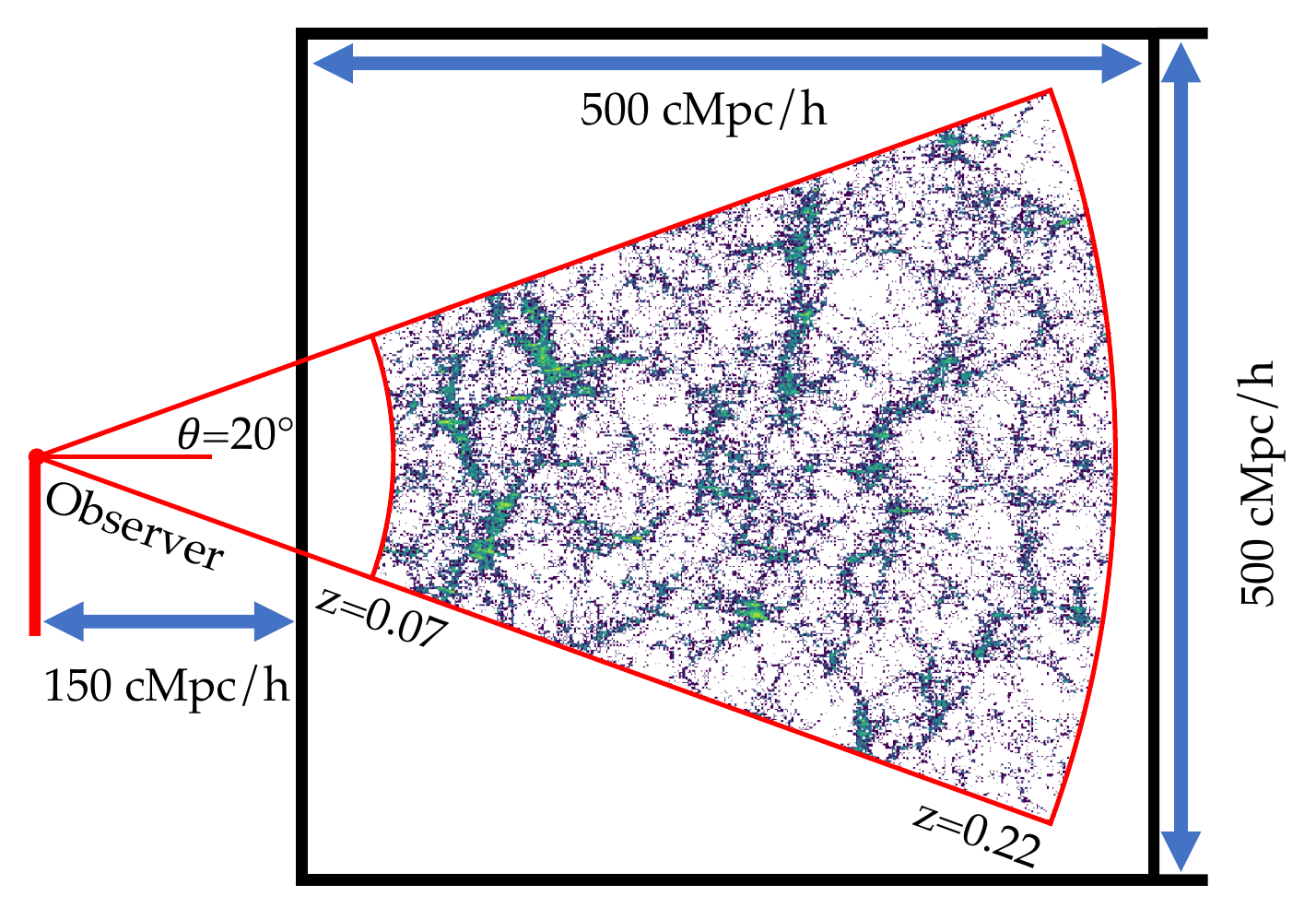}
    \caption{A schematic figure on the mock catalog and the reconstructed volume. 
    The fan-shaped area wrapped with red lines represents the radial and angular selection applied when populating the galaxy catalog; 
    a 20 cMpc/h thickness slice of galaxy distribution is shown in the area. 
    The square with black line shows the layout of the $(500 \cMpch)^3$ reconstructed volume. 
    }\label{fig:recon_volume}
\end{figure}

The relative position of the reconstructed volume and the mock tracer galaxies can be found in Figure~\ref{fig:recon_volume}, where the red lines define the redshift cut and the angular cut for the mock dataset, and the black box marks the layout of the reconstructed volume.

\subsection{Forward Model and Likelihood} \label{ssec:methods-fwd-n-lik}
In this section, we introduce the forward model $\mathcal{M_\theta}$ and the (log-)likelihood function to be optimized in the \mujica framework. 

\subsubsection{Forward model} \label{sssec:fwd-model}

The forward model translates initial input modes into the observed galaxy rate density $\lambda$. 
This process utilizes an N-body gravity solver and a galaxy bias model linking the density field to galaxy observations. 
Note that, \mujica is not limited to a certain type of N-body solver or galaxy bias model, provided that they adhere to time and space complexity constraints. 
The forward model is described below:

\begin{enumerate}
    \item Initial Matter Field $\delta_L(\mathbf{k})$. 
    In standard \LCDM{} cosmology, these modes follow a complex-Gaussian distribution $\delta_L(\mathbf{k})\sim \mathcal{CN}(0, P_k)$.
    In our forward model, we reparameterize $\delta_L(\mathbf{k})$ as 
    \begin{equation}
        \tilde\delta_{\mathrm{ic}} (\mathbf{k}) = \frac{\delta_L(\mathbf{k})}{\sqrt{P_k}} \sim \mathcal{CN}(0, 1)
    \end{equation}
    where the linear power spectra come from fiducial cosmology. In practice, the complex Hermitian $\tilde\delta_{\mathrm{ic}}$ is composed of real Gaussian variables $s\sim\mathcal{N}(0, 1)$.
    
    \item LSS formation. 
    We make use of JaxPM \footnote{\url{https://github.com/DifferentiableUniverseInitiative/JaxPM}}, a JAX implementation of the FastPM particle-mesh N-body solver \citep{Feng:2016, Bayer:2020tko}.
    Based on the reconstructed volume defined in Section~\ref{sssec:reconstructed_volume}, we choose a particle grid of $256^3$ for the JaxPM solver. This yields a spatial resolution of $\Delta L=1.95 \cMpch$.
    We utilize 2LPT to evolve the initial density field to $z=9$ ($a=0.1$), then perform an 18-step fastPM-style LSS evolution to the present-day $a=1$, adopting a leapfrog integrator with a constant step size $\Delta a=0.05$.
    
    \item Lightcone and RSD.
    JaxPM supports the production of the PM output at intermediate redshifts, which facilitates the modeling of lightcone effects via interpolation.
    For our task, we save the snapshot at $a=0.8, 0.9, 1.0$ (corresponding to $z = 0.25, 0.11, 0$). The lightcone effect, modeled via an iterative approach (detailed in the Appendix \ref{appendix:lightcone}), yields the scale factor of each particle in the lightcone $a_i$, and the corresponding position $\mathbf{x}=\mathbf{x}_i(a_i)$ and velocity $\mathbf{v}=\mathbf{v}_i(a_i)$.
    
    Then, redshift-space distortions (RSD) are applied to the lightcone particles.
    \begin{equation}
        \Delta \mathrm{x_i} = H^{-1}(a) (\mathbf{v}_i \cdot \hat{\mathbf{d}}_i) \hat{\mathbf{d}}_i,
    \end{equation}

    Where $\hat{\mathbf{d}}_i$ is the unit vector along the line-of-sight from the observer $\mathbf{x}_{\mathrm{obs}}$ to $\mathbf{x}_i$
    \begin{equation}
        \hat{\mathbf{d}}_i = \frac{\mathbf{x}_i-\mathbf{x}_{\mathrm{obs}}}{||\mathbf{x}_i-\mathbf{x}_{\mathrm{obs}}||}.
    \end{equation}

    \item Galaxy population model.
    We take a galaxy bias model that combines power-law bias and a (soft) cut-off to connect the density field and the galaxy rate density field $\lamall$ \citep{Ata:2015, Jasche:2019}: 
    \begin{equation} \label{eq:galaxy_bias}
    \lamall = \bar{N}_G (1+\delta_m)^{\alpha} \sigma(K(\delta_m-\delta_{th}))
    \end{equation}
    Where $\lamall$ is the Poisson parameter for the stochastic galaxy formation process (see Section \ref{sssec:lik-func}), $\bar{N}_G$ is a normalization factor to correct the total number count in the survey volume, and $\sigma(x)$ is the sigmoid function $\sigma(x) = \frac{1}{1 + e^{-x}}$. The model has three free parameters: the power-law index $\alpha$, the overdensity threshold $\delta_{th}$, and the steepness of the sigmoid $K$.
    We found that $(\alpha, \delta_{th}, K) = (0.7, 1.2, 3.0)$ yielded good agreement on the galaxy density power spectra $P_{gg}$ with the mock dataset, though it is mainly valid for large scales, hinting that a more complicated bias model is needed (e.g., \cite{McAlpine:2025}; later shown in Figure~\ref{fig:lt_powerspec}). 
    \item Survey response function. 
    In any realistic spectroscopic galaxy survey, the survey will be incomplete, which we model with a response function (or equivalently, the selection function; Section \ref{sssec:semi-analytic-galaxy-catalogs}), $R$, defined as the probability of a galaxy at a certain position being observed by the survey.
    \begin{equation}
        \lamobs = R\lamall
    \end{equation}
    Although simple to model in the forward evolution, the survey response typically prevents common optimization methods from producing reasonable solutions outside the observation mask. 
\end{enumerate}

\subsubsection{Likelihood function} \label{sssec:lik-func}

We assume a Poisson model for the galaxy number count in each mesh grid; thus, the (log-)likelihood is
\begin{align}
\log\mathcal{L}
  &= \sum \log \frac{e^{-\lamobs} \lamobs^{N_G}}{N_G!} \\
  &= \sum \Big( -\lamobs + N_G \log\lamobs - \log (N_G!) \Big)
\end{align}
Here $N_G$ is the CiC-binned galaxy count from the mock observations. 

\subsection{\mujica Features} \label{ssec:mujica}

Different from the MAP approach in \cite{Seljak:2017, Horowitz:2019, Horowitz:2021}, \mujica aims to optimize this likelihood on a feasible set defined by the Gaussian prior. 
This section provides a concise overview of \mujica's key features that are essential for optimization-based density reconstruction in a selection-function-dominated volume. 
\mujica{} alternates between likelihood-based optimization updates (Section~\ref{sssec:pLBFGS}) and rank-order matching procedures driven by the imposed statistical constraints (Section~\ref{sssec:rank_order_match}).

\subsubsection{Projected L-BFGS} \label{sssec:pLBFGS}
In \mujica{}, we optimize the input parameters using the limited-memory Broyden-Fletcher-Goldfarb-Shanno algorithm (L-BFGS; \cite{Liu:1989, Wright:1999}) combined with projected gradient descent (PGD) \citep{Madry:2017}. 

L-BFGS is a generalized optimization method designed for large-scale nonlinear optimization tasks. 
By incorporating curvature information, it typically converges faster than purely first-order algorithms that use only gradient evaluations.
Rather than explicitly calculating the full Hessian matrix --- prohibitive for high-dimensional problems --- L-BFGS stores a compact approximation of the Hessian.
This approximation is iteratively updated using gradient differences and parameter changes from recent iterations.
The resulting approximate Hessian is then used to compute a search direction for the next update step. 
Given this direction, a line search procedure is performed to update the parameters. 

We incorporate the idea of PGD into L-BFGS by introducing a projection function $\mathcal{P}(s)$ in the linear search to keep the parameters as a physically meaningful cosmological initial conditions. 
Given the position $s$ and the search direction $\hat{p}$, the line search finds an optimal step size $\epsilon$ along $s' = \mathcal{P}(s+\epsilon\hat{p})$ instead of $s' = s+\epsilon\hat{p}$. 
Assuming that (i) a physically meaningful set of $s$ consistent with \LCDM{} satisfies $\mathrm{Mean}(s)=0$ and $\mathrm{Std}(s)=1$, 
(ii) the extreme values of $s$ are not valid and cause a malformed density field, we define the feasible set of the prior and construct the projection function onto the set as a two-step procedure:

\begin{enumerate}
    \item Z-score normalization. 
    Normalization removes the nonzero mean value from $s$ and rescales $s$ to enforce a standard deviation of one. 
    \begin{equation}
        \hat{s} \gets (\hat{s} - \mathrm{Mean}[\hat{s}]) / {\mathrm{Std}[\hat{s}]}
    \end{equation}
    \item Clip and reset the extreme values.
    The extreme $s$ values may appear when the step size is too large, which is a clear sign of a deviation from Gaussianity. We reset modes when their absolute value goes beyond a threshold $s_{th}=10$
    \begin{equation}
        s' \leftarrow \left\{
                \begin{array}{ll}
                \hat{s} & |\hat{s}| \le s_\mathrm{th} \\
                \hat{s}' \sim \mathcal{N}(0,1) & |\hat{s}| > s_\mathrm{th} \\
                \end{array}
              \right.
    \end{equation}
\end{enumerate}

The projection has only a minor impact on the convergence behavior of L-BFGS. 
Indeed, any solution that satisfies the Gaussian assumption $s \sim \mathcal{N}(0,1)$ is almost a fixed point of the projection operator, in the sense that $\mathcal{P}(s)=s+O(1/d)$, where $d$ is the dimensionality of the parameter space.
In addition, L‑BFGS is robust to perturbations of position updates \citep{Dennis:1977}.
Consequently, when the step size is sufficiently small, the projected L-BFGS method effectively reduces to the standard L-BFGS algorithm. 

The combination prevents nonphysical features from developing in optimization and, at the same time, retains the memory and time efficiency of L-BFGS.

\subsubsection{Rank-order Matching} \label{sssec:rank_order_match}
Although the transform function introduced in Section~\ref{sssec:pLBFGS} can suppress non-Gaussianity during optimization, its constraint is not strong enough. 
To further ensure agreement between the initial modes and the probability distribution described by the prior, we apply rank-order matching (also dubbed ``abundance matching'' when applied to halo/galaxy catalogs) in both Fourier space and real space. 
Such a matching procedure can be interpreted as a generalized projection operation that enforces the desired statistical constraints on the initial conditions. 
It may reduce the (log-)likelihood $\mathcal{L}$, counteracting the optimization of the linear search steps. 
However, the impact of this operation will be minimal if the input modes show good Gaussianity.
The use of rank-order matching to enforce Gaussianity in cosmological
initial conditions was first proposed by \cite{Weinberg:1992} in the
context of Gaussianization of the density field. Here, we apply the
same principle within an iterative optimization framework.

Generally, for N variables $\{x_i\}, i=1,2,\cdots,N$ following a probability distribution $p(x)$, rank-order matching (hereafter, we will call this $\mathcal{R}\left[x_i\right]$) works as follows. 

\begin{enumerate} \label{algorithm:rom}
    \item Find the descending rank order of the array $\{x_i\}$, $\{\pi_i\}$, where $\pi_i=j$ means that $x_i$ is the $j$-th largest element in the array. 
    \item Draw a random realization of $N$ values from the desired distribution $p(x)$ and sort them in descending order. 
    We call this $\{x'_i\}, x'_1>x'_2>\cdots>x'_N$.
    \item The output of rank-order matching, $\{\tilde x_i\}$, is
    \begin{equation}
        \tilde x_i = x'_{\pi_i}
    \end{equation}
\end{enumerate}

This formulation preserves the rank order in the original array $\{x_i\}$ while ensuring that the output $\{\tilde x_i\}$ aligns with the target probability distribution $p(x)$. 
Moreover, if the input array $\{x_i\}$ closely matches $p(x)$, the resulting match will only cause an error that scales with the inverse square root of the length of the array: $|x_i-\tilde x_i| \sim O(1/\sqrt{N})$. 
Crucially, such an operation perturbs the solution, which is beneficial in avoiding the solver being trapped at local minima.
In \mujica, we apply rank-order matching in both the real-space initial conditions and the Fourier-space initial conditions.

In real space, initial condition (IC) rank-order matching is applied to the region with tracers, i.e., $R(q)>0$: $\delta_{\mathrm{ic, matched}}(q) = \mathcal{R}[\delta_{\mathrm{ic, matched}}(q)]$. 
In Fourier space, we match the modes within specific $k$-bins: the idea is an alternative to \cite{McAlpine:2025}, where extra terms for each k-bin are added to the prior in order to force Gaussianity. 
We segment the $k$ space modes into 100 bins of equal size based on amplitude $k$, applying $\mathcal{R}$ operators to each bin. 
We refrain from modifying large-scale modes with $k<0.1 \hcMpc$ as there are not enough large-scale modes for a reliable rank order matching, and also because they remain mostly linear.

\section{Results} \label{sec:results-optimization}
In this section, we show the numerical test of \mujica with the data and methods introduced in Section~\ref{sec:methods}. We show a direct comparison to vanilla L-BFGS, motivating the need for \mujica{}, in Appendix \ref{appendix:ablation}.

To accelerate the convergence of the algorithm, we adopt an annealing strategy for the reconstruction task. 
Specifically, we first apply \mujica{} to the same reconstruction problem but at a lower resolution (cell size of 4 cMpc/h) and subsequently use the resulting field as the initial condition for the full-resolution run. 
In practice, the low-resolution reconstruction is evolved for 250 pL-BFGS steps, with rank-order matching performed every 5 iterations; the full-resolution reconstruction is then executed for a total of 3600 iterations, with rank-order matching applied every 20 iterations. 
Overall, the procedure takes roughly $30$ hours to finish on one 32GB NVIDIA V100 GPU.

\subsection{Reconstructed initial conditions}
In this section, we compare the \mujica output with initial conditions generated randomly from Gaussian white noise to demonstrate the Gaussian nature of the \mujica output.

\begin{figure*}[htbp]
    \centering
    \includegraphics[width=\textwidth]{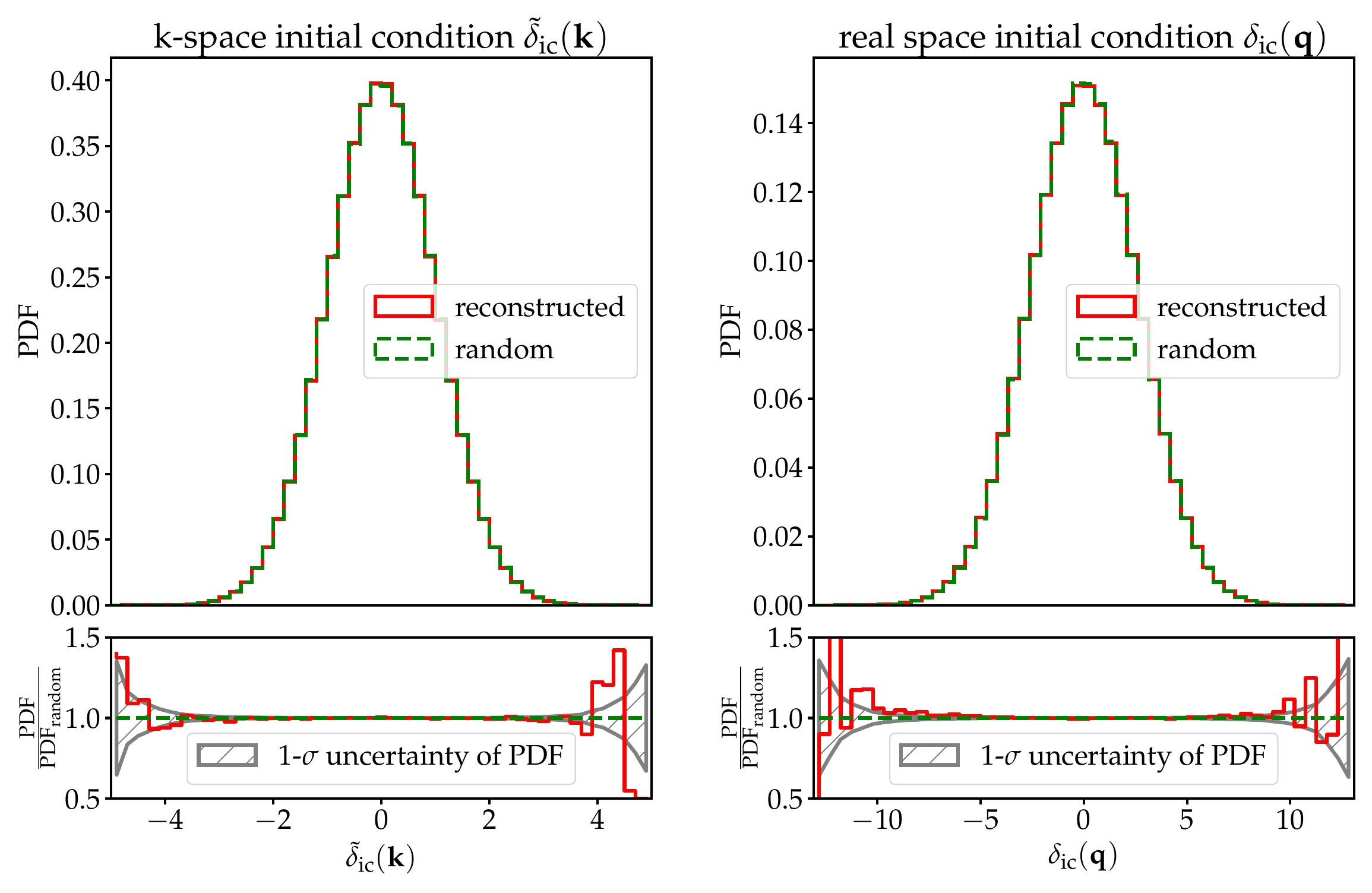}
    \caption{Comparison the distribution of the reconstructed initial condition (red solid line) and a random initial condition (green dashed line) in Fourier space (left) and the real space (right). Note that the two panels have different ranges in their abscissae.
    In both the Fourier space and the real space, the histogram of reconstructed IC has a good agreement with the randomly generated IC. 
    The 1-$\sigma$ uncertainty of PDFs is shown as gray hatched regions in the lower panels. } \label{fig:IC_hist}
\end{figure*}

Figure~\ref{fig:IC_hist} presents histograms that compare the reconstructed and random initial conditions in both Fourier and real spaces. The reconstructed initial conditions align well in both spaces. 
In Fourier space, the reconstructed mode distribution matches the Gaussian prediction for $|\tilde\delta_{\mathrm{ic}}(\mathbf{k})|<4$, with a deviation in the number of modes beyond this range. 
In real space, the agreement also extends to the 4$\sigma$ region with the random initial conditions.

\begin{figure*}[htbp]
    \centering
    \includegraphics[width=0.7\textwidth]{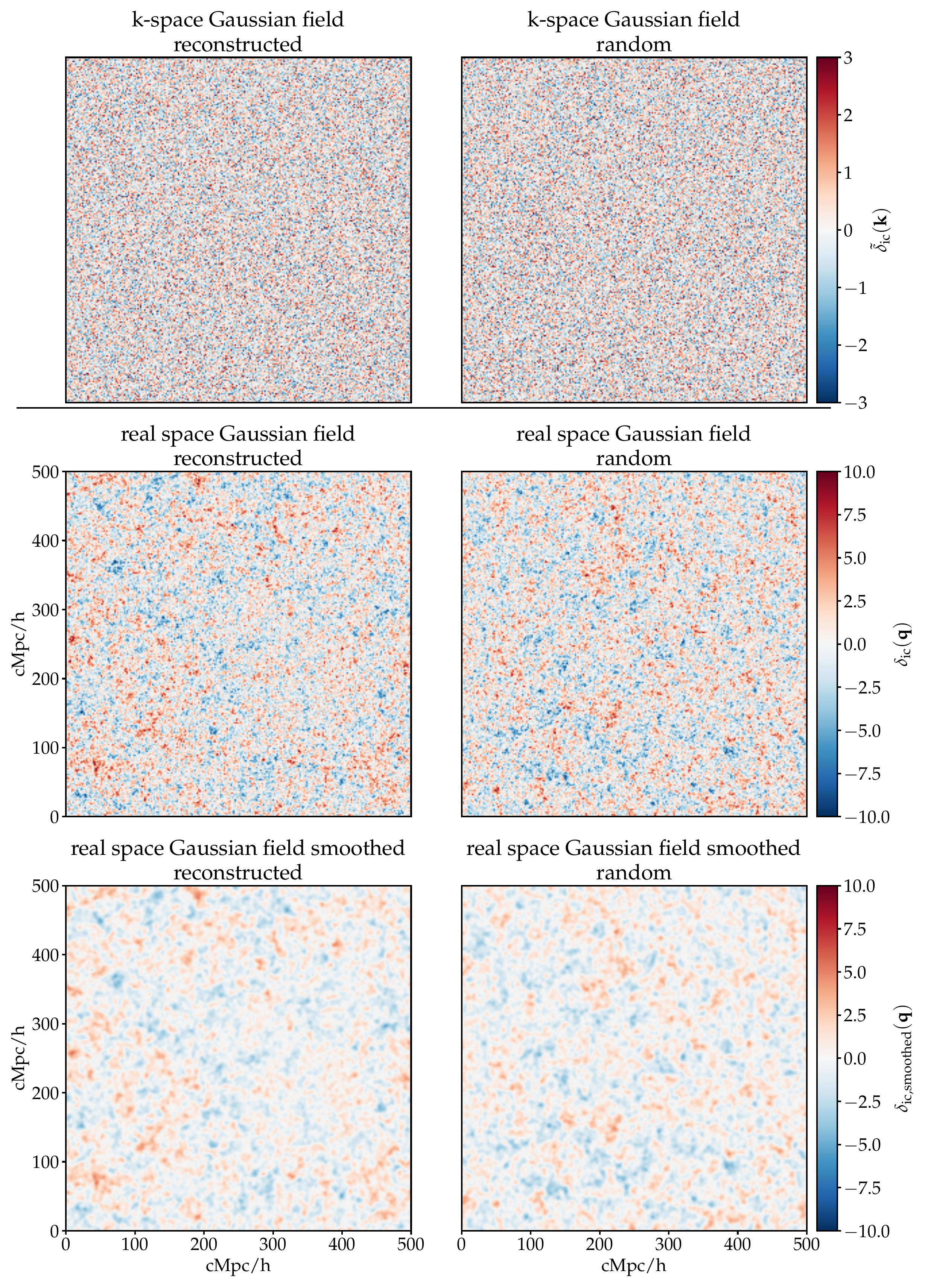}
    \caption{Visual inspection of the reconstructed initial condition. The upper two panels compares the reconstructed initial condition in Fourier space. 
    The middle panels and lower panels show the initial condition in real space, unsmoothed and smoothed with 2 cMpc/h Gaussian filter, respectively. 
    The reconstructed modes show good Gaussianity.}
    \label{fig:IC_visual}
\end{figure*}

Figure~\ref{fig:IC_visual} provides a visual illustration of the reconstructed initial conditions. 
We find no significant features suggesting a deviation from the Gaussian prior in the reconstructed maps. Visually, one can hardly tell the difference between the reconstructed IC and a randomly generated Gaussian field. 

\subsection{Late-time fields}
\label{subsec:ltf}
In this section, we examine the agreement of the late-time reconstructed field with the underlying truth from the mocks. 

\begin{figure*}[htbp]
    \centering
    \includegraphics[width=\textwidth]{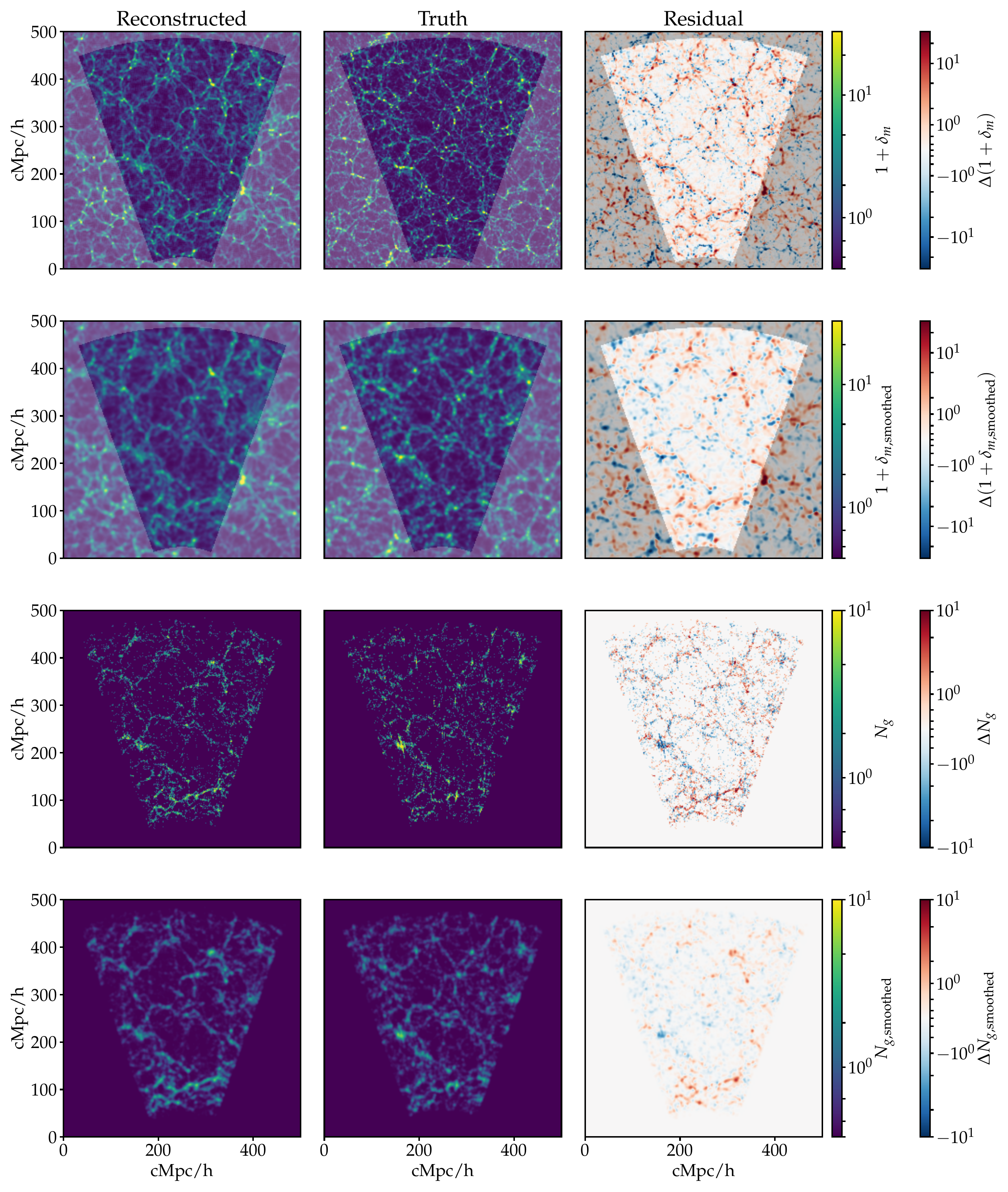}
    \caption{A visual inspection of the late-time density fields and the galaxy fields. From the top to the bottom, the panels in the first row shows the normalized density field $(1+\delta_m)$ of the reconstructed one, the truth and their residual. The second row is also about the density field, but a Gaussian filter with scale of $2 \cMpch$ is applied.
    The survey region is shown as a semitransparent mask.
    The third and the last row are the comparison for unsmoothed and Gaussian-smoothed galaxy field, respectively. } \label{fig:lt_vis_inspec}
\end{figure*}

Figure~\ref{fig:lt_vis_inspec} visualizes the late-time density field and the galaxy count field for both the reconstruction and the underlying truth. 

As shown in the upper two rows of Figure~\ref{fig:lt_vis_inspec}, the density field reconstruction by \mujica{} effectively recovers cosmic structures.
The position of a massive protocluster, located in the middle left of the panel, is well reproduced.
The filamentary structures are also visually comparable between the reconstructed field and the truth. 
Nevertheless, some residuals are evident along the filaments and knots, which we attribute to the misspecification of the galaxy bias model under nonlinear structural evolution. 
Additionally, the uncorrelated density field outside the survey window demonstrates the exclusive use of the tracers within the selection window for optimization. 
This feature points to the potential to produce multiple samples from a single target galaxy catalog, maintaining consistency within the selection window while allowing the unobserved regions to exhibit significant independence.

In the galaxy field (the two lower rows in Figure~\ref{fig:lt_vis_inspec}), the galaxy tracer distributions are also well recovered from the perspective of the cosmic web. 
However, two prominent features distinguish the true galaxy map from the reconstructed one:
(i) sharper, isolated galaxy spots, unlike the more continuous forward-modeled galaxy field resembling the cosmic web's morphology (as seen in density field plots). 
Such discreteness can be adjusted by fine-tuning the bias parameters, e.g., increasing the threshold $\delta_{th}$ to eliminate connecting filaments; however, it also reflects the stochastic nature of galaxy formation, which is typically complicated to model with a (differentiable) galaxy bias. 
(ii) The finger-of-god (FoG) effect around the massive structures pointing toward the observer, which is absent in the reconstructed galaxy map. 
Due to the absence of resolved satellite galaxy kinematics in the bias model, FoG is also challenging to incorporate into our forward model. 
Although FoG typically does not prevent the localization of massive halos, it alters the galaxy distribution along the line-of-sight, introducing systematics for potential applications on the constraints of cosmological parameters.

\begin{figure}[htbp]
    \centering
    \includegraphics[width=\textwidth]{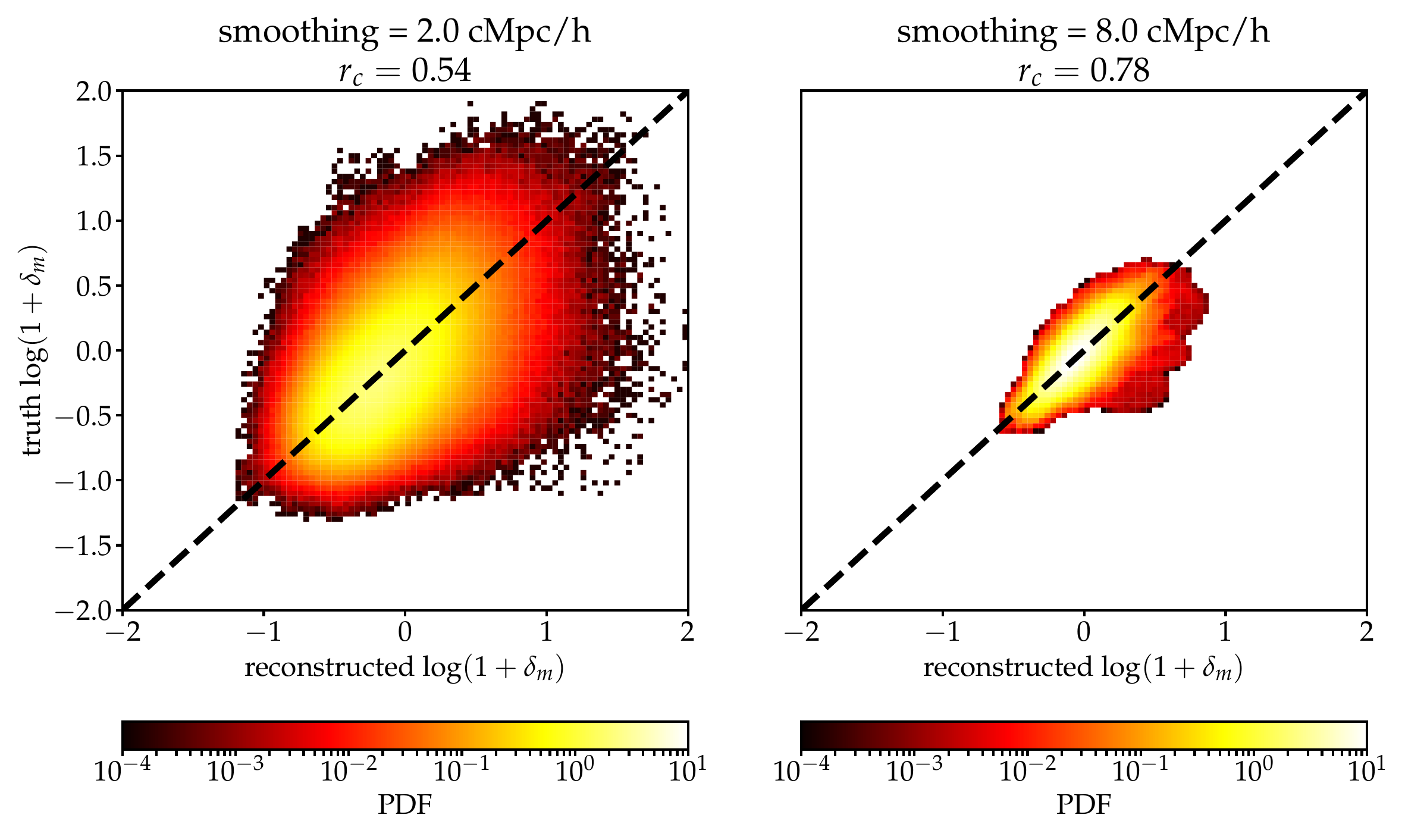}
    \caption{2D histogram of the density field inside the selection function between the reconstructed and the underlying truth. 
    Two smoothing scales are applied in the two panels: The left one with $R_s = 2 \cMpch$, while the right one is smoothed with $R_s=8 \cMpch$ scale. 
    The dashed line represents a perfect reconstruction $\delta_{m, \mathrm{rec}} = \delta_{m, \mathrm{truth}}$. } \label{fig:lt_2dhist}
\end{figure}

Aside from visual inspection, we also quantify the similarity between the reconstructed field and the truth using various methods. 
Figure~\ref{fig:lt_2dhist} is a scatter plot of the voxel-to-voxel comparison between the truth and the reconstructed matter density field under different smoothing scales. 
To avoid confusion from the density values outside the region, the smoothing is applied first, and then the histogram is plotted only for the voxels inside the survey mask. 
In the left panel, the 2D histogram of the minimally smoothed voxels (Gaussian filter with a scale of $2 \cMpch \approx \Delta L$) shows a scatter up to $\pm 1$ dex, but most of the voxels remain close to the one-to-one correlation line, yielding a correlation coefficient of $r_c=0.54$. 
In the right panel, we show the situation under more aggressive Gaussian smoothing at the $8 \cMpch$ scale -- the density fields are tightly correlated in this case, and the correlation coefficient rises to $r_c=0.78$.

\begin{figure}[htbp]
    \centering
    \includegraphics[width=8cm]{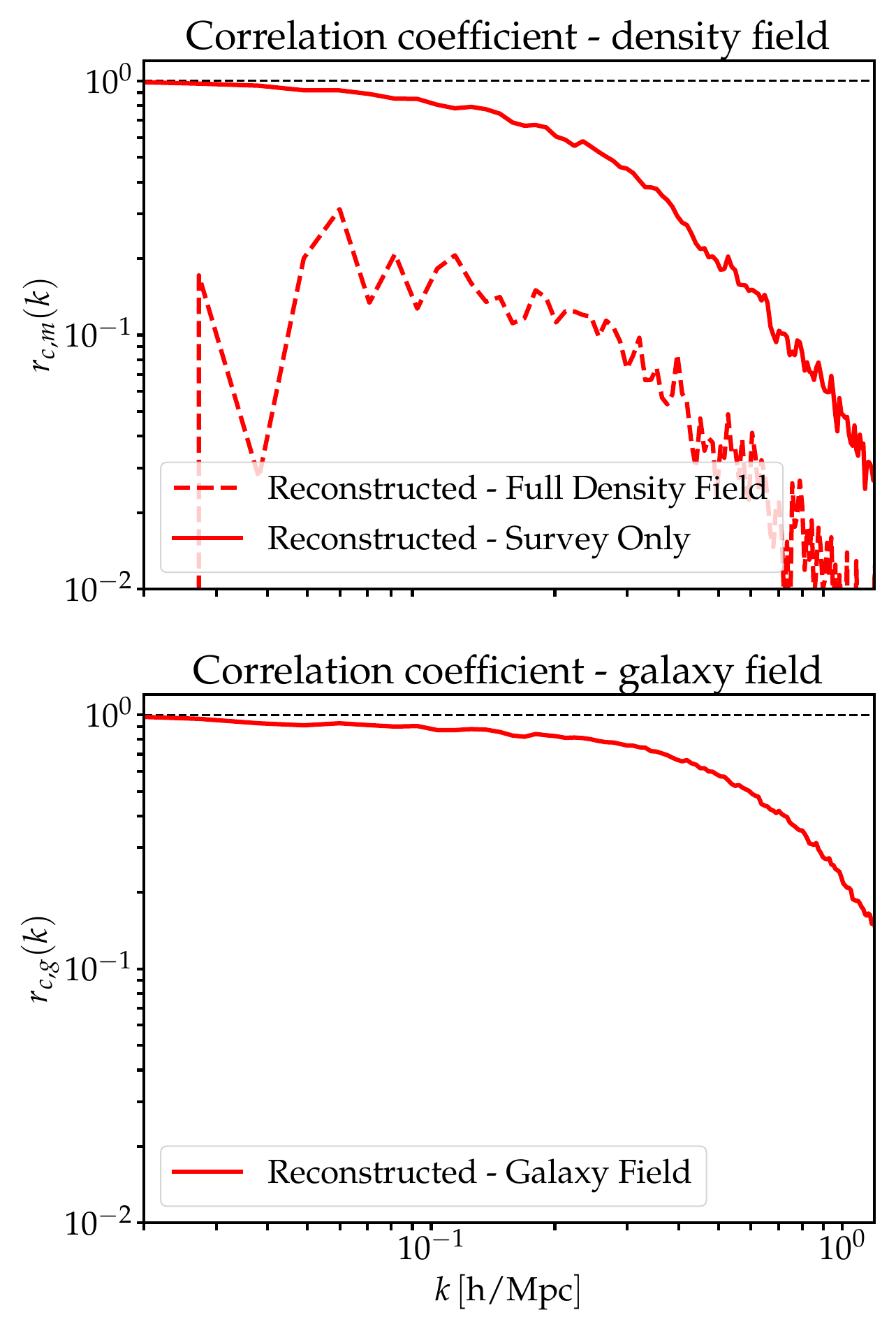}
    \caption{The correlation coefficient at different scales. On the left, the red solid line represents the coefficient calculated in the full reconstructed volume, while the red dash line represents the coefficient in the survey response window. On the right, we show the correlation of the galaxy field. For both the density field and the galaxy field, the coefficient $r>0.8$ at a scale of $k=0.1\,\hcMpc$.} \label{fig:lt_pearson_coeff}
\end{figure}

In Figure~\ref{fig:lt_pearson_coeff}, we further extend the comparison to two-point statistics. 
We define a correlation coefficient $r_c(k)$ based on the wavenumber $k$, with the auto-power spectra of the reconstructed field $P_{\mathrm{r-r}}(k)$, the auto-power spectra of the true field $P_{\mathrm{t-t}}(k)$, and the cross-power spectra between the actual and reconstructed fields $P_{\mathrm{r-t}}(k)$.

\begin{equation}
    r_c(k) = \frac{P_{\mathrm{r-t}}}{\sqrt{P_{\mathrm{r-r}}P_{\mathrm{t-t}}}}
\end{equation}

The correlation coefficient $r_c(k)$ is illustrated in Figure~\ref{fig:lt_pearson_coeff}. In the left panel, two methods were used to compute the coefficient $r_{c,m}(k)$ for the matter density field. 
The first method, represented by the red dashed line, calculated it using only the survey-covered volume. Here, the correlation remains close to 1 until approximately $k=0.03 \hcMpc$, then decreases to $r_{c,m} = 0.8$ at $k=0.1 \hcMpc$. 
The coefficient maintains a value of $r_{c,m} = 0.3$ at $k=0.5 \hcMpc$. 
The second method, depicted by the red solid line, involves the entire reconstructed volume. 
This approach shows that the correlation coefficient is up to $0.2$ at all scales, indicating that the reconstructed field is not correlated with the truth outside the survey region. 
The right panel of Figure~\ref{fig:lt_pearson_coeff} presents the galaxy field's correlation coefficient, which generally exhibits a correlation stronger than that of the masked density field, with $r_{c.g}=0.2$ at $k=1 \cMpch$. 
The stronger correlation is attributed to the likelihood being directly defined on observable galaxies, reducing the impact of the misspecification on the galaxy bias model.

\begin{figure*}[htbp]
    \centering
    \includegraphics[width=\textwidth]{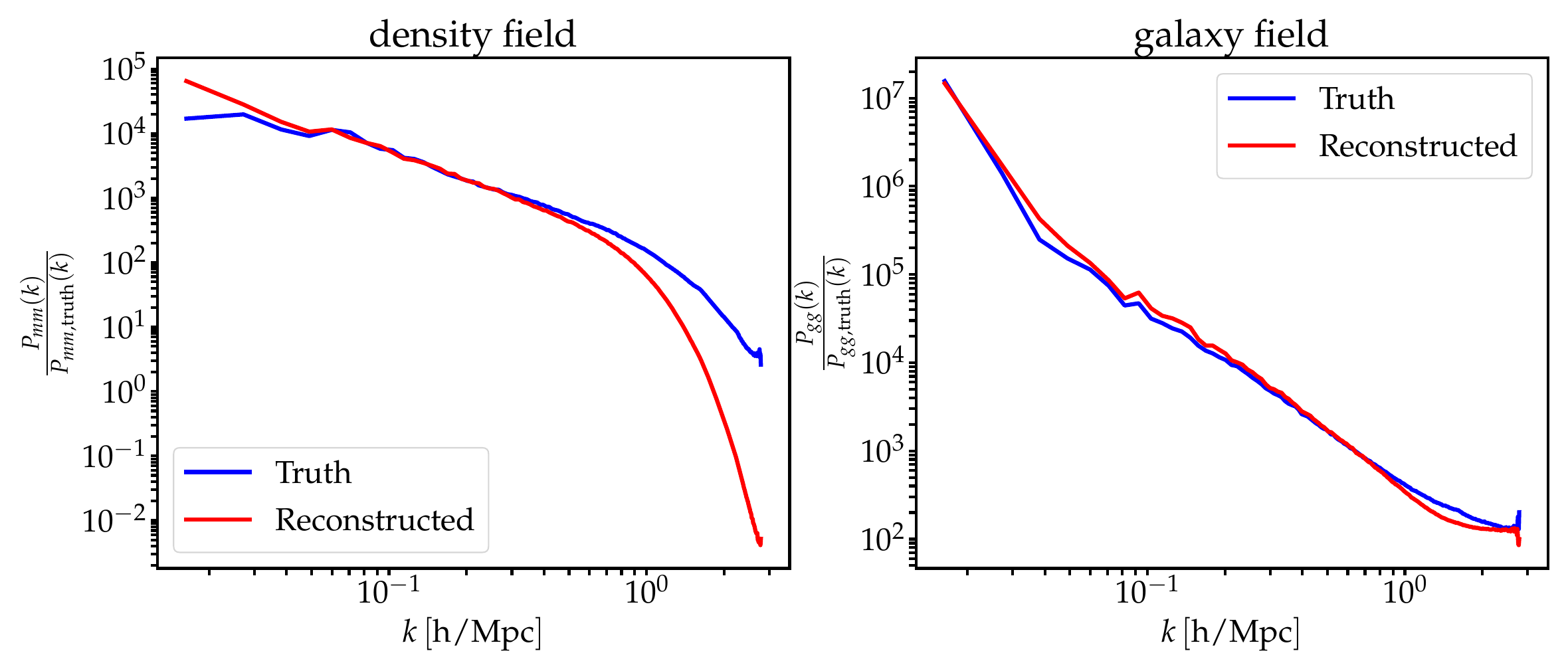}
    \caption{Comparison between the power spectra of density field (left) and galaxy field (right). 
    In the both panels, the power spectra of the underlying truth is marked with blue solid lines, and the power spectra of the reconstructed field is shown as red solid lines. 
    The reconstructed field shows a agreement within 0.3 dex with the underlying truth down to a scale of $k=1\,h/\mathrm{cMpc}$. } \label{fig:lt_powerspec}
\end{figure*}

Finally, we show the auto power spectra of the truth field and the reconstructed field in Figure~\ref{fig:lt_powerspec}. 
In the left panel of Figure~\ref{fig:lt_powerspec}, the power spectra of the  matter density fields $P_{mm}(k)$ are compared. 
The power spectrum of the reconstruction deviates from the truth at the largest two k-bins, likely due to the limit on $k=0.1 \hcMpc$ in the rank-order matching of the k-space mode (see Section~\ref{sssec:rank_order_match}). 
In the range $0.1 \hcMpc < k < 0.3 \hcMpc$, the reconstructed fields match closely with the true fields. 
For $k > 0.3 \hcMpc$, the power spectra of reconstructed fields decrease with scale, indicating simulation resolution effects.
The right panel of the same figure displays the comparison of the galaxy density power spectra $P_{gg}$. 
In large-scale modes, the reconstructed field exhibits slightly enhanced power relative to the truth field.
Nevertheless, within the wavenumber range $0.2\,h\,\mathrm{cMpc}^{-1} < k < 1\,h\,\mathrm{cMpc}^{-1}$, the reconstructed power spectra show better agreement with the true power spectrum.
On smaller scales ($k>1 \hcMpc$), the reconstructed field is lower than the truth due to the limited resolution of forward modeling. 

\subsection{Cosmic web recovery} \label{ssec:cosmic_web}

In this section, we extend the analysis of the reconstructed field to the cosmic web following the `T-web' formulation (e.g., \cite{Lee:2016a, Krolewski:2017, Horowitz:2019, Horowitz:2021}). We start by introducing the formulation and then presenting the results. 

The T-web relies on the eigenvalue and eigenvectors of the (pseudo)deformation tensor $\psi$, i.e., the Hessian of the gravitational potential $\phi = \nabla^{-2} \delta_m$,
\begin{equation}
    \psi_{ij} = \frac{\partial^2\phi}{\partial x_i \partial x_j}.
\end{equation}
The deformation tensor has strong physical implications in the Zel'dovich approximation \citep{Zeldovich:1970}. 
For eigenvalues $\lambda_1<\lambda_2<\lambda_3$ with respective eigenvectors $\mathbf{\hat{e}_1}, \mathbf{\hat{e}_2}, \mathbf{\hat{e}_3}$, a positive $\lambda_i$ indicates dynamic contraction along $\mathbf{\hat{e}_i}$. 
This concept is utilized in \cite{Hahn:2007} and \cite{Forero-Romero:2009}, leading to the T-web method for classifying the cosmic web by the eigenvalue count above zero. Structures are labeled as `voids' if all eigenvalues are non-positive, `sheets' for one positive eigenvalue, and `filaments' and `knots' for two and three positive eigenvalues, respectively.

In practice, the deformation tensor is derived in Fourier space as $\tilde\psi_{ij} = \frac{k_ik_j}{k^2}\delta_m(\mathbf{k})$. 
To mitigate resolution effects, an initial Gaussian smoothing $G(R_{sm})$ is applied to the density field; additionally, a threshold $\lambda_{th}$ is introduced to account for the nonlinear structural growth (the Zel'dovich approximation gives $\lambda_{th}=0$). 
As an initial test for cosmic web recovery, we adopt the parameter set $(\lambda_{th}, R_{sm}) = (0.16, 2\cMpch)$ as validated by \cite{Dong:2025}, without exploring alternative parameter choices.

\begin{figure*}[htbp]
    \centering
    \includegraphics[width=\textwidth]{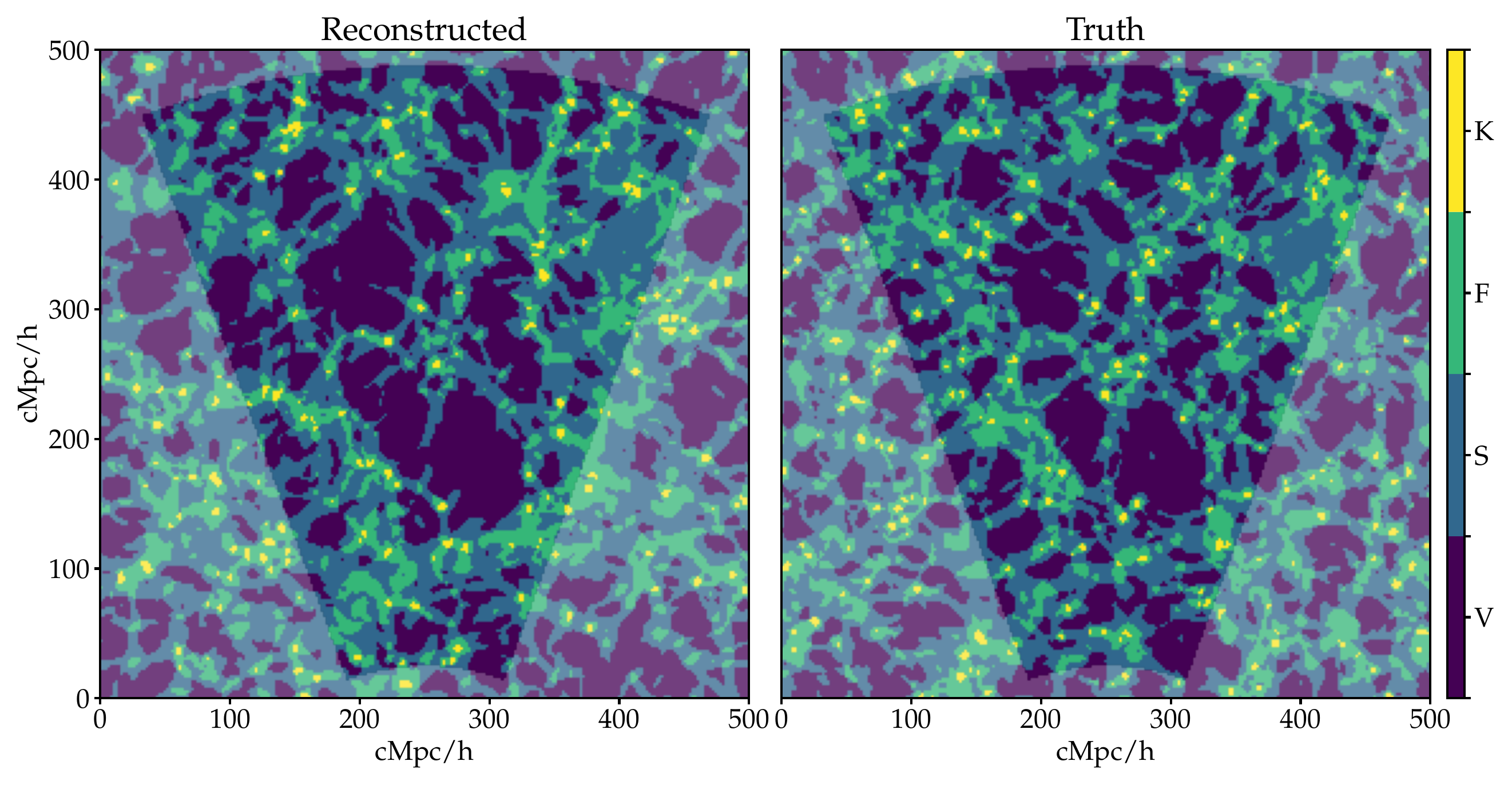}
    \caption{A visual inspection of one slice of the cosmic web classification. The yellow color represents knots ("K"), green color represents filaments ("F"), dark green color represents sheets ("S"), and purple color represents void ("V"). 
    The galaxy survey response is shown as a fan-shaped, semitransparent mask. 
    The reconstructed density field recovers the voids and sheets, but it fails to capture the position of knots. } \label{fig:cw_visual_inspec}
\end{figure*}

Figure~\ref{fig:cw_visual_inspec} provides a visual inspection of the cosmic web classification. 
Note that the classification is completely determined by the density field, so one would expect two similar density fields to yield similar cosmic web classifications. 
Actually, at this level of smoothing, the resulting cosmic web has a morphology quite similar to the density field (c.f. Figure~\ref{fig:lt_vis_inspec}). 
This can be confirmed in Figure~\ref{fig:cw_visual_inspec}. We find a good match for the size and position of the voids, and the large scale distribution of the sheets and filaments is also in good agreement. The small scale filaments and knots, however, show a clear mismatch between the truth and the reconstruction. Similar to the density field, the reconstructed field outside the selection window shows no correlation with the ground truth. 

\begin{figure}[htbp]
    \centering
    \includegraphics[width=8cm]{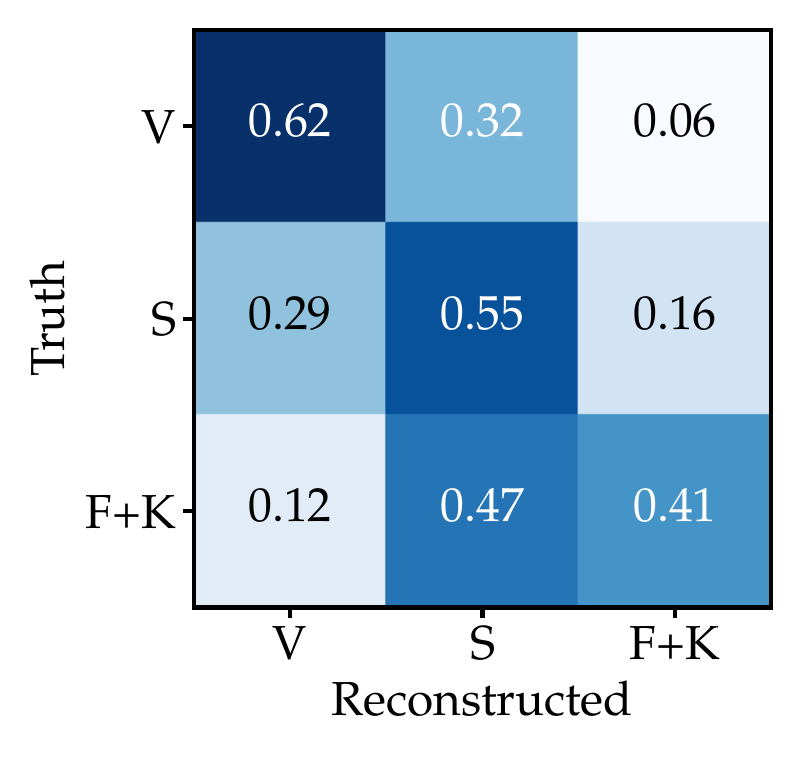}
    \caption{The confusion matrix of the reconstructed cosmic web versus the underlying truth, inside the survey response mask. 
    The notation of "V", "S", "F and "K" is the same as Figure~\ref{fig:cw_visual_inspec}. 
    The values are normalized to the truth density fields, meaning they indicate the volume fraction of a structure in the truth classified as that structure in the reconstructed volume. 
    The reconstructed density accurately recovers over half of the voids and sheets, yet it significantly confuses sheets and filaments+knots.
    } \label{fig:cw_confusion_matrix}
\end{figure}

Figure~\ref{fig:cw_confusion_matrix} presents the classification agreement of the cosmic web with the confusion matrix. It shows that $\sim$60\% of voids and sheets are accurately classified in the reconstructed cosmic web, which is decent in the context of a 3D volumetric comparison. In contrast, filaments and knots show less accuracy, as they are often misclassified as sheets. 
This result is in agreement with our findings on the density field, indicating imperfect reproduction of overdensities in the reconstructed density field.

At the end of this section, we compare the matching of the eigenvalues. 
This is achieved by calculating the `cosine similarity' of two vectors:
\begin{equation}
    \left|\cos \theta_{i}\right| = \left|\mathbf{\hat{e}_{i,rec}}\cdot\mathbf{\hat{e}_{i, truth}}\right|, i=1,2,3
\end{equation}
The absolute value is needed, as both $\mathbf{\hat{e}_{i}}$ and $-\mathbf{\hat{e}_{i}}$ are valid eigenvalues of the given eigenvalue $\lambda_i$. 

\begin{figure*}[htbp]
    \centering
    \includegraphics[width=\textwidth]{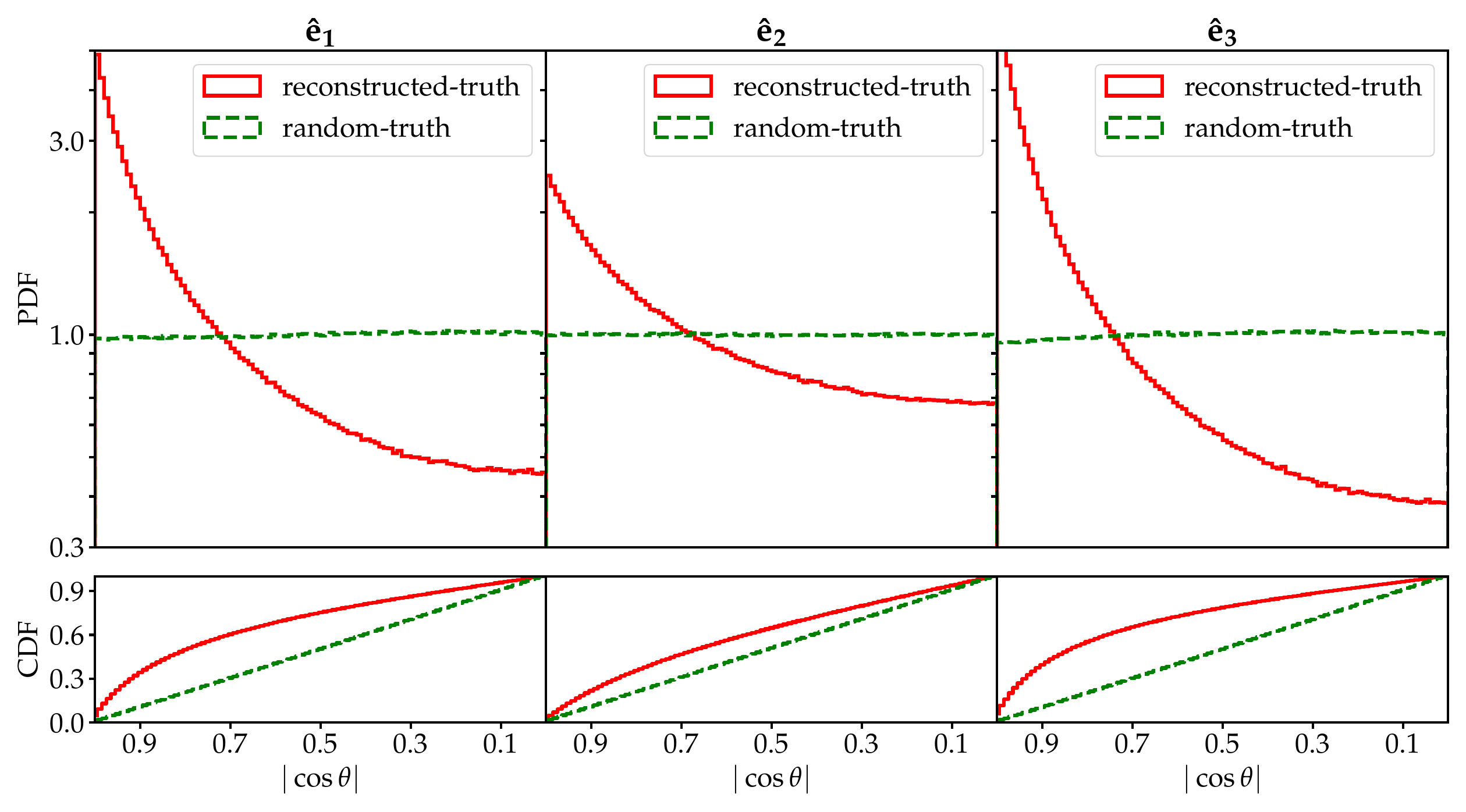}
    \caption{A test on the recovery of the deformation tensor eigenvectors. 
    The red solid curves show the distribution of the inner product of the eigenvectors of the truth and the reconstructed field (three eigenvectors $\mathbf{\hat{e}_1}, \mathbf{\hat{e}_2}, \mathbf{\hat{e}_3}$ in three panels); the green dash lines are the same distribution but for the inner product between the truth and a randomly generated density field.
    Only the eigenvectors inside the survey mask is taken into account. In the lower panels, we show the cumulative distribution function for the eigenvectors' inner product. Note the abscissa of the three panels are reversed, to highlight totally agreement implied by $\left|\cos\theta\right|=1$. 
    } \label{fig:cw_eigvec}
\end{figure*}

Figure~\ref{fig:cw_eigvec} presents the cosine similarity between the reconstructed field and the true field for the eigenvalues across all voxels within the selection window. 
The eigenvector of the largest eigenvalue, $\mathbf{\hat{e}_{3}}$, which is known to be of great importance for intrinsic alignment analyzes, demonstrates the highest recovery rate, with more than 70\% of voxels having $\left|\cos \theta_{3}\right| > 0.6$.
Meanwhile, for $\mathbf{\hat{e}_{1}}$, over 60\% of the voxels achieve $\left|\cos \theta_{3}\right| > 0.6$. 
Recovery of $\mathbf{\hat{e}_{2}}$ is the least effective, as evident from its cumulative distribution function, which resembles the distribution of a random initial condition.

\section{Discussion} \label{sec:discussion}
In this work, we introduce a novel optimization-based technique for density reconstruction, \mujica. By incorporating the transform function and rank-order matching within the optimizer, \mujica can generate initial conditions that evolve into structures traced by galaxies at late times while simultaneously suppressing deviations from the Gaussianity assumption of \LCDM{} cosmology due to the optimization steps.
Here, we discuss the potential extension of \mujica{}, show its application in galaxy formation and large-scale structure studies, and compare it with previous density reconstruction efforts.

\subsection{Extensions of \mujica model}
This section explores several directions for extending the data model in \mujica and their potential impact on \mujica's performance.

First of all, we emphasize that the optimization used by \mujica is independent of the detailed forward model and observables; therefore, there is potential for great flexibility with the implementation of more sophisticated or scalable models, as long as they take standard Gaussian variables as input and remain differentiable. 
In the context of structure formation models, alternative approaches, such as differentiable particle-mesh codes like pmwd \citep{Li:2024} and Bullfrog \citep{Rampf:2025}, are available. 
In terms of the galaxy model, \mujica{} also allows for different types of formulations. 
In addition to the Eulerian power-law bias commonly adopted by the density reconstruction community, the galaxy field can also come from a Lagrangian approach (e.g., \cite{Schmidt:2021, Horowitz:2021, Bayer:2023a, Simon-Onfroy:2025, Akitsu:2025boy, Bayer:2026zcr})
LPT-based methods, from 2LPT/nLPT (\texttt{LEFTfield} \cite{Schmidt:2021, Nguyen:2024, Stadler:2025}) to ALPT \citep{Kitaura:2013}, offer improved memory and computational efficiency, although at the expense of reduced accuracy at smaller scales. 
These alternative methods have the potential to significantly accelerate reconstruction processes, thereby enabling their application to larger volumes and finer resolutions.

Furthermore, it is important to note that the forward model does not necessarily need to be strictly differentiable with a modified version of the line search. 
As discussed in Section~\ref{sssec:pLBFGS}, the line search is based solely on the direction of the likelihood gradient $\hat{s} = \nabla\mathcal{L} / ||\nabla\mathcal{L}||$. 
Adopting approximate gradients of the likelihood, especially for the halo-based approaches, offers more flexibility in modeling the galaxy-density connection. 
For example, neural network techniques (like \cite{Doeser:2024}) can be used to estimate the gradient directions, while more nuanced forward models, like halo occupation distribution models (for example, the implementations \cite{Horowitz:2024, Horowitz:2025a}), can be used for the likelihood.
We will test the integration of those approximate gradient methods in follow-up work. 

In Section~\ref{ssec:dataset}, our validation run is limited to one survey of a single tracer (galaxies). However, extending the likelihood of integrating multiple surveys (combining several catalogs; see \cite{Ata:2015, Ata:2022}) or multiple tracers (such as the Ly$\alpha$ forest, line intensity mapping, or fast radio bursts; refer to \cite{2021JCAP...10..056M,Horowitz:2021} for example) is straightforward. 
The use of multiple survey methods optimizes the utilization of existing observational data (see \cite{Lee:2022} for a related practice).
As another example, combining galaxy redshifts with other tracers, such as the Ly$\alpha$ forest absorption, can enhance the efficiency of the algorithm by better tracing under-dense regions \citep{Horowitz:2025}.


\subsection{Uniqueness of \mujica} \label{ssec:comp-density-recon-code}

\mujica is one of the only optimization-based density reconstruction codes that pushes all the way to megaparsec scales, using galaxy tracers while fully modeling realistic observational effects.
Among optimization-based density reconstruction practices \citep{Seljak:2017, Horowitz:2019, Horowitz:2021, Bayer:2023a}, \mujica is one of the few works to deploy a full particle-mesh code, explicitly model observational effects, and robustly produce solutions consistent with the designated Gaussian \LCDM{} cosmology. 
Those advancements are attributed to the constrained backtrack line search (Section \ref{sssec:pLBFGS}) and the rank-order matching strategy (Section \ref{sssec:rank_order_match}). 
Although the projected L-BFGS is less efficient than the L-BFGS solver used in previous works, its robustness provides significant benefits in the context of density reconstruction. 

The primary advancement of \mujica{} lies in the introduction of additional correction steps that constrain the updates proposed by the optimizer so that the solution remains consistent with a Gaussian prior (Section~\ref{sssec:pLBFGS}). 
The idea for preserving Gaussianity can be traced back to Gaussianization \citep{Weinberg:1992}, wherein a similar rank-order matching procedure was employed to ensure the Gaussianity of the initial conditions.
As iteratively evaluating a forward model was computationally prohibitive at that time, the rank-order matching was performed directly between the late-time density field and the initial conditions. 
This limitation restricts the method to linear scales, leaving it inapplicable to realistic observational data. 

An alternative Gaussianization approach, for example, used in the
\texttt{Manticore} project \citep{McAlpine:2025}, is to impose
constraints on the power spectrum, skewness, and kurtosis of the initial
conditions by introducing additional terms to the loss function that
enforce agreement between the sampled and target power spectra. These
extra terms enhance consistency with the assumption of Gaussianity while
retaining sufficient flexibility for the sampler to explore the
parameter space. The procedure is analogous to our implementation of
rank-order matching in $k$-space, but with a key distinction:
the additional loss terms constrain specific moments of each $k$-bin,
whereas rank-order matching explicitly enforces agreement of the full
probability density function.
Both approaches involve tradeoffs. Penalty terms in the loss function
are smooth and differentiable, making them straightforward to
incorporate into a sampler or optimizer; however, the extent to which
these additional terms may bias the posterior remains an open question.
By contrast, rank-order matching enforces the target distribution
exactly, but the operation is discontinuous and can temporarily reduce
the likelihood, counteracting the optimizer's progress --- though this
effect diminishes as the solution approaches Gaussianity
(Section~\ref{sssec:rank_order_match}). A systematic comparison of
these strategies in the context of field-level inference is left to
future work.

Although \mujica{} itself is not a sampling method, the reconstructed density field still encodes substantial information at that level. 
In contrast to the optimization-based technique, codes such as ELUCID \citep{Wang:2014_1, Wang:2016_2}, ARGO \citep{Ata:2015}, COSMIC-BIRTH \citep{Kitaura:2021, Ata:2021}, and BORG \citep{Jasche:2013, Jasche:2019, McAlpine:2025} adopt a Bayesian, sampling-based approach. 
The \mujica{} reconstruction can serve as a high-quality initial guess for a subsequent Bayesian sampler, which, in turn, can significantly accelerate its convergence and improve the efficiency of the overall inference procedure.
In addition, \mujica is capable of producing a set of uncorrelated solutions (but not necessarily following the posterior) by setting different initial positions, which would be optimal for the sampling task to explore the parameter space with better efficiency. 
Currently, we are developing a novel field-level inference sampling framework, \maigo (Dong et al., in prep.), which will work in conjunction with \mujica. 

Lastly, it is worth noting that the use of neural networks is attracting growing attention within the density reconstruction community (e.g., see \cite{Modi:2021, List:2023, Jindal:2023, Legin:2024, cuesta2024joint, Bottema:2025, Doeser:2025, Parker:2025mtg, Bayer:2026zcr}).
These machine learning-based methodologies --- at the moment --- are designed for application on the late-time fields (halos, galaxies) in simulation volumes, without the incorporation of various observational effects such as incompleteness and selection functions. 
This represents a gap between the idealized tests in these studies and practical applications where the observational effects of the tracers are important, and also requires careful consideration of robustness and distributions-shifts between simulation and observation \citep{2022arXiv221114788H,Bayer:2025ija}. 
However, neural networks can help bridge the gap between density fields and tracers within optimization frameworks (e.g. \citep{2018JCAP...10..028M,2022ApJ...941...42H}).
Furthermore, they can enhance the performance of optimization, which will be explored in future work.

\subsection{Application of \mujica method}
The features of \mujica, as stated in Section~\ref{ssec:comp-density-recon-code}, make it one of the most applicable codes with realistic observational data. The density reconstruction has extensive applications in galaxy and large-scale structure analysis. 

The direct product of density reconstruction, initial conditions, and the density field is useful for investigating galaxy evolution from an environmental perspective. 
The density field at the time of observation helps define the cosmic web (Section~\ref{ssec:cosmic_web}) and provides context for the galaxy environment. 
At the same time, initial conditions offer insights beyond the late-time density field. 
By re-simulating the initial conditions (see \cite{Wang:2016_3, Ata:2022, Byrohl:2025}), it is feasible to trace the evolutionary history of galaxies within the reconstruction volume, thereby constraining galaxy physics at the field level. 

The density field is also essential for the study of the intergalactic medium (IGM). 
At intermediate to high redshifts ($2<z<4$), the synergy between galaxy and Ly$\alpha$ forest tracers can break the degeneracy between the IGM density and temperature \citep{Kooistra:2022, Dong:2023, Dong:2025}, allowing for the investigation of the IGM thermal evolution and its interaction with galaxies. 
At low redshifts ($z<1$), the reconstructed density field helps characterize the baryon distribution along fast radio burst (FRB) sightlines, enhancing the effectiveness of FRB dispersion measures as a probe of baryonic matter in the IGM (see, e.g., \cite{Lee:2022, Khrykin:2024}). 
Especially, the stability of \mujica{} could play a key role in the reconstruction tasks of the so-called `pencil-beam' surveys \footnote{While no longer a commonly-used term, `pencil-beam' surveys can be thought of as any survey that is not expected to provide a fair sampling of cosmic environments.}, where the survey footprint is typically on the order of $O(1 \deg^2)$ and thus provides only limited angular sampling of the reconstruction volume.

Furthermore, \mujica{} can also be used to extend field-level inference of cosmological parameters, such as with BAO \citep{Bayer:2026zcr}, to complex survey geometries. In addition, while we have focused on using bias models for galaxy modeling, more nuanced forward models could be used provided there is (possibly noisy) gradient information available (i.e. halo finding \citep{Horowitz:2025a} and galaxy populating \cite{Horowitz:2024,2026diffhod2}). More nuanced halo-based modeling could be useful for capturing the FoG features as discussed in Sec. \ref{subsec:ltf}.

\section{Conclusion}

In this study, we introduce a new projected-optimization based framework for density reconstruction, named \mujica. 
This framework is applied to a realistic galaxy catalog with incomplete sky coverage, maintaining consistency with the Gaussian initial conditions of \LCDM{} cosmology. 
Our findings indicate good agreement between the reconstructed density field and the actual density field, as well as the power spectra. 
Furthermore, subsequent evaluations related to cosmic web recovery suggest its extensive applicability in the study of galaxy evolution and large-scale structures.

With numerous ongoing and forthcoming galaxy observation programs, such as DESI \citep{DESI:2016}, PFS \citep{Takada:2014}, Roman \citep{Spergel:2015}, Euclid \citep{Laureijs:2011}, LSST \citep{Ivezic:2019}, and SPHEREx \citep{Dore:2014}, the astronomy community is moving closer to the goals of a Stage-IV survey, characterized by unprecedentedly rich galaxy datasets.
Looking ahead, the \mujica project will play an essential role in unraveling the formation and evolution of large-scale cosmic structures and imposing constraints on the understanding of galaxy physics and cosmology from the field level, thereby deepening our understanding of the universe.

%% file: sections/acknowledgement.tex
Kavli IPMU is supported by the World Premier International Research Center Initiative (WPI), MEXT, Japan. 
This work was performed in part at the Center for Data-Driven Discovery, Kavli IPMU (WPI).
We acknowledge support from the University of Tokyo-Princeton University Strategic Partnership Grant.
This study was supported by the Forefront Physics and Mathematics Program to Drive Transformation (FoPM), a World-leading Innovative Graduate Study (WINGS) Program, the University of Tokyo.
CD was supported by JSPS KAKENHI Grant Number 26KJ0695.
KGL acknowledges support from JSPS Kakenhi grant Nos. JP18H05868, JP19K14755 and JP24H00241.
CD thanks Dr. Metin Ata for his help in initializing this project and building up the forward model.
CD thanks Prof. Leander Thiele for a useful discussion on the connection between the optimizer and the sampler. 
CD thanks Prof. Guilhem Lavaux and Prof. Francisco-Shu Kitaura for their constructive suggestions in tuning the galaxy formation model. 
CD thanks Dr. Beatriz Tucci for her insightful comments on framing this work in a cosmology context. 

%% file: sections/appendix.tex
\section{Iterative Modeling of the Lightcone Effect} \label{appendix:lightcone}

For a particle $p_i$ in the JaxPM simulation, their trajectory and velocity can be interpolated from the snapshots $s_j|_{a_j}$
\begin{align}
    \mathbf{x}_i(a) &= \mathrm{Interp}(x_i; s_j|_{a_j})\\
    \mathbf{v}_i(a) &= \mathrm{Interp}(v_i; s_j|_{a_j})
\end{align}

Then one can calculate the distance from the particle to the observer located at $\mathbf{x}_{\mathrm{obs}}$ as
\begin{equation}
    d_i(a) = ||\mathbf{x}_i(a) - \mathbf{x}_{\mathrm{obs}}||
\end{equation}

Under a fixed $\theta_\mathrm{cosmo}$ cosmology, the relation between the comoving distance $d$ and the scale factor $a$ is known
\begin{equation}
    a_i = \mathrm{DistToScaleFactor}(d_i; \theta_\mathrm{cosmo})
\end{equation}

Combining those functions, one can construct an iterative formulation to derive $a_i$
\begin{equation}
    a_i^{(n)} = \mathrm{DistToScaleFactor}(d_i(a^{(n-1)}_i); \theta_\mathrm{cosmo})
\end{equation}

In this work, we start with $a^{(0)}_i=1.0$ and adopt $a^{(3)}_i$ (i.e.,  after $3$ iterations) as the scale factor of the particle in the lightcone.
The position and velocity of the particle in the lightcone are then given by the interpolations $\mathbf{x}_i(a_i)$ and $\mathbf{v}_i(a_i)$, respectively. 

\section{Ablation test} \label{appendix:ablation}

In this appendix, we compare the performance of \mujica{} and vanilla L-BFGS to highlight the robustness of our method against survey completeness. 

As a baseline for the experiments, we run \mujica{} for a total of $10^3$ projected L-BFGS iterations on the reconstruction task defined in Section \ref{ssec:dataset}; this configuration is referred to as \texttt{Mujica} in the remainder of this section. 
During this baseline run, we store a checkpoint after every $20$ projected L-BFGS iterations, immediately following a rank-order matching operation, yielding a total of $50$ checkpoints.

We additionally perform a series of ablation studies using a standard (unconstrained) L-BFGS method under different configurations. 
First, starting from the same initial guess as the \texttt{Mujica} baseline, we apply L-BFGS with various upper limits on linear search step sizes (\texttt{L-BFGS}, \texttt{L-BFGS-S-1}, \texttt{L-BFGS-S-2}, \texttt{L-BFGS-S-3}) designed to emulate the strategy of reducing the maximum allowed step size of L-BFGS to improve its robustness. 
Second, we initialize L-BFGS from selected early checkpoints of the \texttt{Mujica} baseline run (\texttt{M05L30}, \texttt{M10L30}, \texttt{M15L30}). 
A description of those runs can be found in Table \ref{tab:ablation}.
All these L-BFGS ablation configurations are executed for 1000 L-BFGS iterations, with 50 checkpoints recorded per run. 
Note that, in contrast to the \texttt{Mujica} baseline, rank-order matching is not applied in any of these L-BFGS runs. 

\begin{figure*}[htbp]
    \centering
    \includegraphics[width=0.78\linewidth]{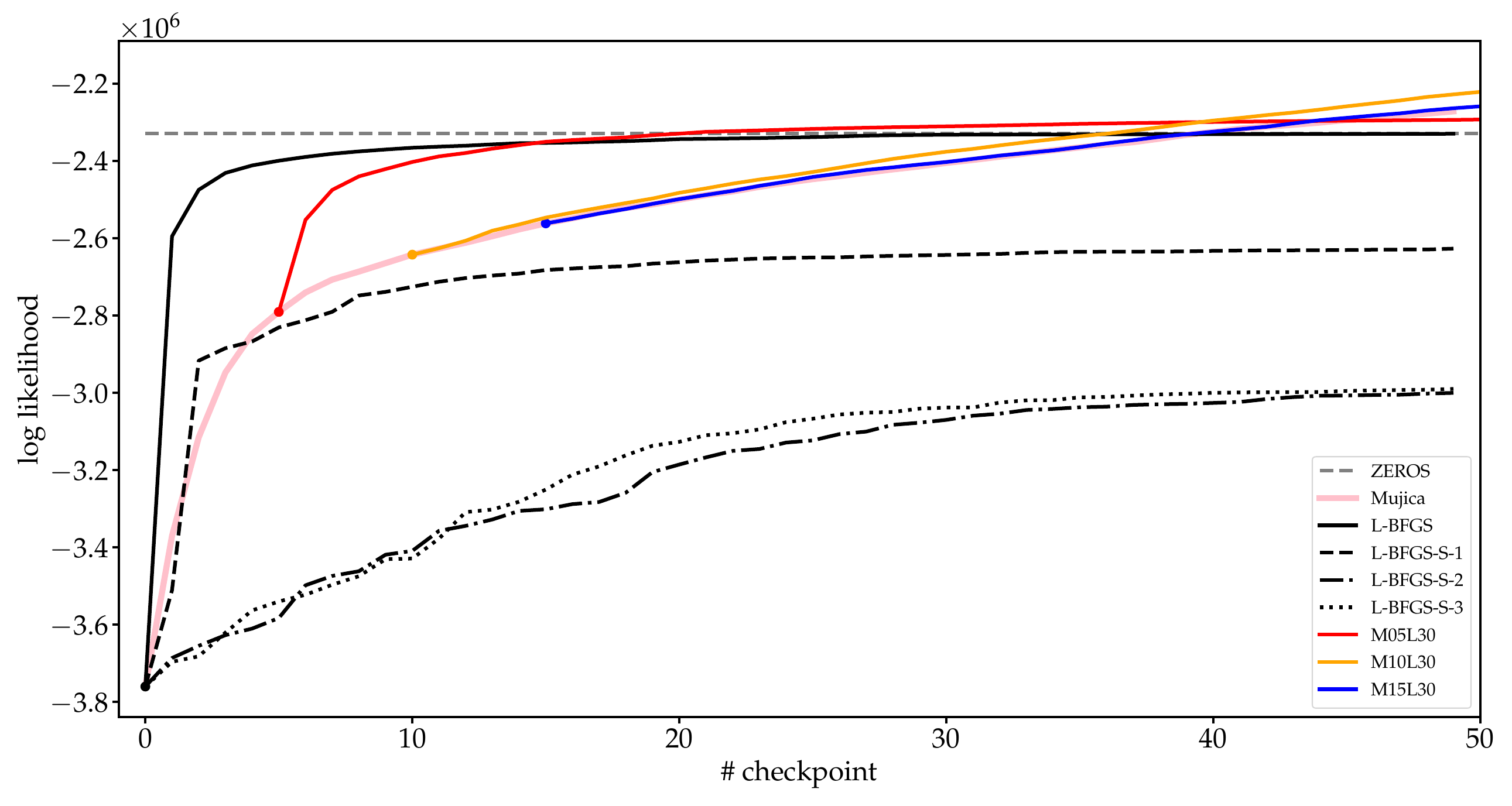}
    \caption{Log-likelihood as a function of optimization steps for the ablation experiments. 
    The gray dashed line denotes the likelihood given by the forward model when the initial conditions are prescribed to be strictly homogeneous.
    The pink solid curve corresponds to the fiducial \texttt{Mujica} run.
    The black curves show the results of a standard L-BFGS optimizer with different choices of line-search step size.
    The red, orange, and blue solid curves represent L-BFGS results starting from early checkpoints of the \texttt{Mujica} baseline. }
    \label{fig:lik_curve}
\end{figure*}

\begin{figure}[htbp]
    \centering
    \includegraphics[width=0.8\linewidth]{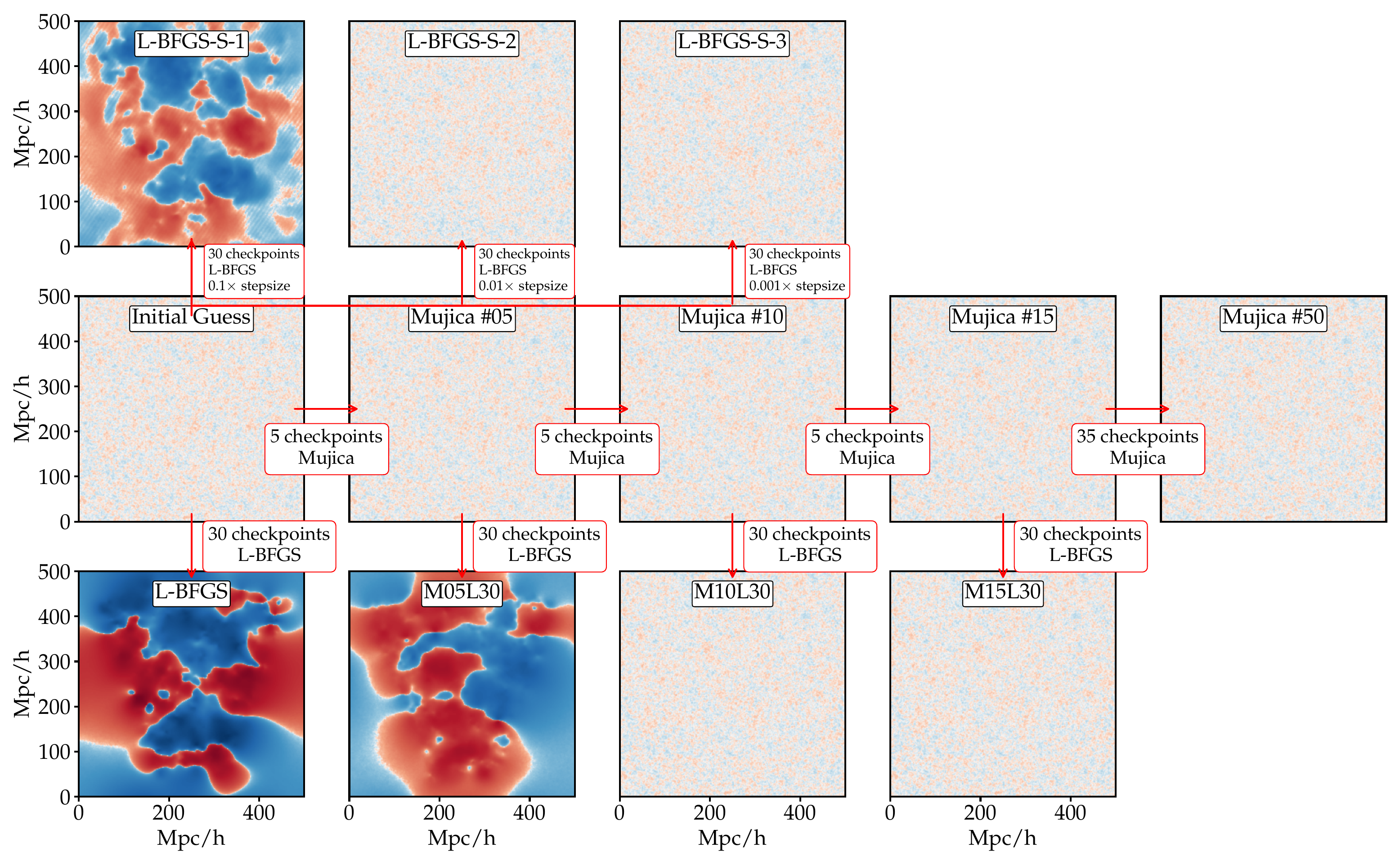}
    \caption{
    Visual inspection of the real-space initial conditions for the ablation runs after 30 checkpoints is presented. The middle row displays the output from the baseline \texttt{Mujica} run. From left to right, the panels correspond to: the initial realization drawn from a random Gaussian field, the reconstruction after 5 checkpoints, after 10 checkpoints, after 15 checkpoints, and the final reconstruction after 50 checkpoints. The same random Gaussian realization is also adopted as the initial condition for the runs including \texttt{L-BFGS} (bottom leftmost), \texttt{L-BFGS-S-1} (top left), \texttt{L-BFGS-S-2} (top middle), and \texttt{L-BFGS-S-3} (top right). Furthermore, the intermediate checkpoint outputs of the baseline \texttt{Mujica} run are used as initial conditions in the \texttt{M05L30}, \texttt{M10L30}, and \texttt{M15L30} runs. Unphysical structures emerge in the \texttt{L-BFGS}, \texttt{L-BFGS-S-1}, and \texttt{M05L30} runs, whereas the remaining initial conditions remain consistent with a Gaussian prior. In the \texttt{L-BFGS-S-2} and \texttt{L-BFGS-S-3} runs, the initial conditions appear physically plausible; however, the reconstruction fidelity is not as good as in the \texttt{Mujica} baseline, as shown by the likelihood in Figure~\ref{fig:lik_curve}.}
    \label{fig:abl_IC}
\end{figure}

\begin{figure}[htbp]
    \centering
    \includegraphics[width=0.8\linewidth]{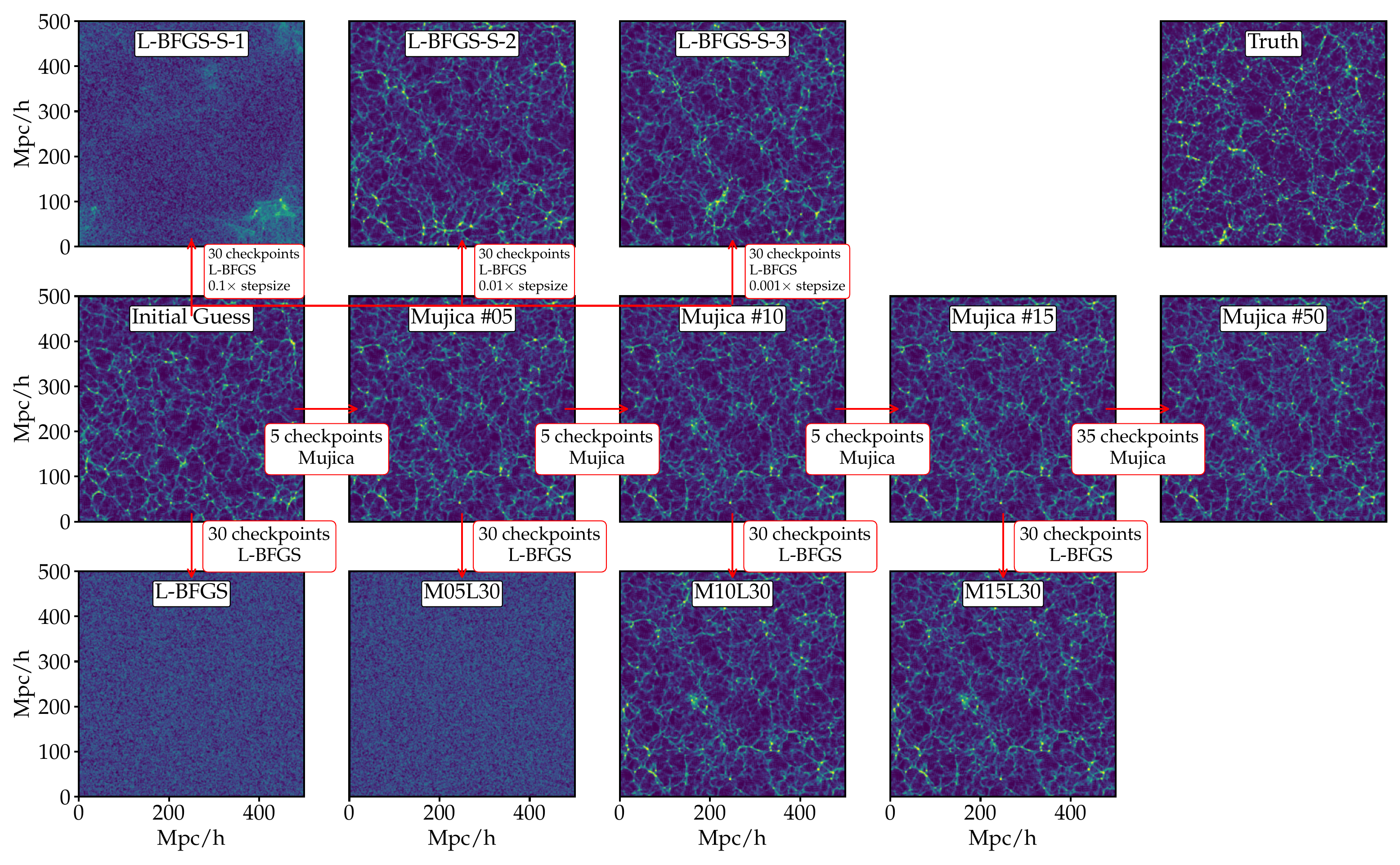}
    \caption{Similar to Figure~\ref{fig:abl_IC}, but for the late-time density field. The reference (true) density field is shown in the upper-right panel of the figure. 
    As expected, non-physical features develop in the \texttt{L-BFGS}, \texttt{L-BFGS-S-1}, and \texttt{M05L30} runs. In the \texttt{L-BFGS-S-2} and \texttt{L-BFGS-S-3} runs, the resulting structures appear physically plausible; however, the overall reconstruction fidelity is not as good as in the \texttt{Mujica} baseline, as also indicated by the likelihood in Figure~\ref{fig:lik_curve}. }
    \label{fig:abl_LT}
\end{figure}

\begin{figure}[htbp]
    \centering
    \includegraphics[width=0.8\linewidth]{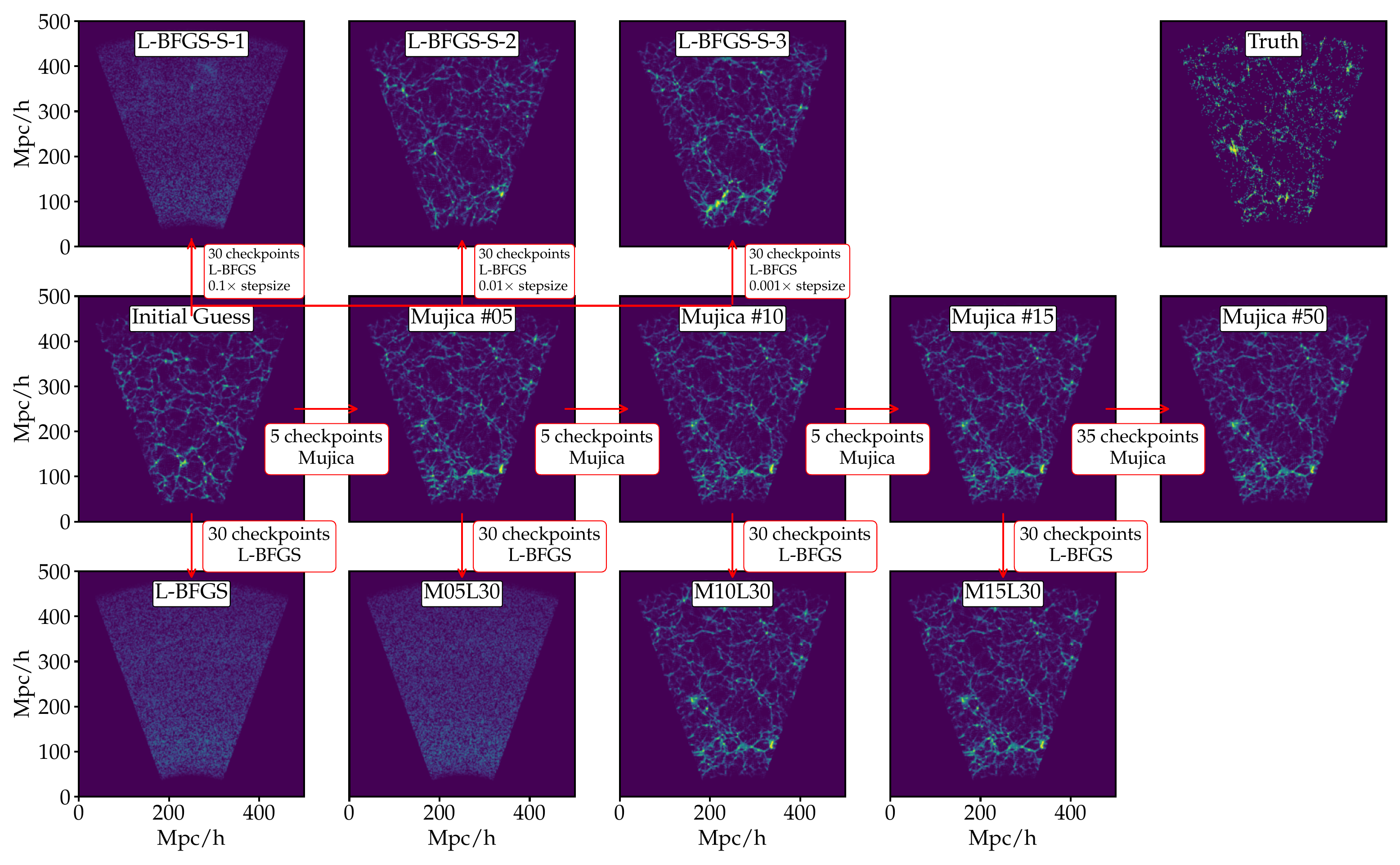}
    \caption{Similar to Figure~\ref{fig:abl_LT}, but for the galaxy distribution.}
    \label{fig:abl_GA}
\end{figure}

The evolution of the log-likelihood (negative loss) for the ablation runs is shown in Figure \ref{fig:lik_curve}. 
For runs initialized from a random Gaussian realization, the log-likelihood ascends quickly at the beginning of the optimization for all configurations with $\epsilon \ge 0.1$. After up to $10$ checkpoints, the rate of improvement in the likelihood diminishes, leading to pronounced plateaus in the curves. 
This behavior indicates that the projection and rank-order matching operations introduce a modest slowdown in early-stage convergence, as anticipated. 
However, as will be demonstrated later in Figures~\ref{fig:abl_IC} to \ref{fig:abl_GA}, the rapid increase in likelihood observed in the L-BFGS run corresponds to updates with an inappropriately large step size, which results in the catastrophic failure of the optimization procedure. 
This behavior is also evident from the \texttt{L-BFGS} likelihood: the likelihood ultimately converges to the likelihood calculated from the all-zero solution, in contrast to the \texttt{Mujica} run, which is capable of surpassing the likelihood corresponding to the all-zero solution.
It suggests that an initialization phase using L-BFGS followed by \mujica{} is not effective, despite the rapid improvement observed in \texttt{L-BFGS} run, because the updates proposed by L-BFGS do not eventually facilitate convergence.
Concerning the late-stage optimization efficiency of \mujica{}, it is comparable to that of L-BFGS with a reduced maximum allowed step size, suggesting that the step size limiter no longer governs the convergence behavior; in other words, the optimizer must take very small steps to further improve the likelihood. 
Comparing the baseline with the L-BFGS runs starting with early checkpoints of \texttt{Mujica}, one observes that \texttt{M05L30} exhibits a similarly rapid initial increase in log-likelihood as the standard \texttt{L-BFGS}. In contrast, the other two runs, \texttt{M10L30} and \texttt{M15L30}, display convergence efficiency comparable to the baseline. This implies that the performance of \mujica{} approaches that of vanilla L-BFGS in the regime of small step sizes, and given informed initial guess (here, informed by \mujica{}) consistent with the described Section \ref{ssec:mujica}.

In Figures \ref{fig:abl_IC}, \ref{fig:abl_LT}, and \ref{fig:abl_GA}, we present the real-space initial conditions, the late-time matter density field, and the corresponding galaxy count field, respectively.
It is notable that the malformed features emerge from the \texttt{L-BFGS}, \texttt{L-BFGS-S-1}, and \texttt{M05L30} runs. 
The initial conditions in these cases attain extreme amplitudes, which in turn yield unphysical density and galaxy count fields. 
Simultaneously, these runs exhibit a rapidly increasing likelihood in Figure \ref{fig:lik_curve}, indicating that the stability and robustness of L-BFGS are compromised in this regime.
In contrast, the ablation runs with reduced step sizes, $\epsilon = 0.01$ and $0.001$ (denoted \texttt{L-BFGS-S-2} and \texttt{L-BFGS-S-3}), remain broadly consistent with the Gaussian prior and do not show such pathological behavior. 
Nevertheless, their reconstruction quality is overall worse than that of \mujica{}: density peaks are not accurately recovered, and filamentary structures appear mismatched, as is also reflected in the corresponding likelihood (Figure~\ref{fig:lik_curve}). 
Finally, the two runs initialized from intermediate \mujica{} checkpoints yield results that are nearly indistinguishable from the final \mujica{} output, further suggesting that, given an informed guess for L-BFGS (here, informed by \mujica{}), \mujica{} and L-BFGS behave almost equivalently when adopting small step sizes.

Overall, this ablation study demonstrates that the \mujica{} algorithm is capable of preserving a physically meaningful solution while achieving rapid convergence during the initial stages of optimization, and subsequently exhibits convergence behavior comparable to that of standard L-BFGS in the later phases of the optimization process. Crucially, it also shows that L-BFGS alone is unable to converge to a physical solution without an informed starting point provided by \mujica{}.

\begin{table}[]
    \centering
\begin{tabular}{cccc}
Name                & Initial Guess                   & Optimizer & Max Step Size $\epsilon$ \\ \hline
\texttt{Mujica}     & Random Gaussian Realization     & \mujica{} & 1.0                     \\
\texttt{L-BFGS}     & Same as \texttt{Mujica} run     & L-BFGS    & 1.0                     \\
\texttt{L-BFGS-S-1} & Same as \texttt{Mujica} run     & L-BFGS    & 0.1                     \\
\texttt{L-BFGS-S-2} & Same as \texttt{Mujica} run     & L-BFGS    & 0.01                    \\
\texttt{L-BFGS-S-3} & Same as \texttt{Mujica} run     & L-BFGS    & 0.001                   \\
\texttt{M05L30}     & \texttt{Mujica} Checkpoint \#05 & L-BFGS    & 1.0                     \\
\texttt{M10L30}     & \texttt{Mujica} Checkpoint \#10 & L-BFGS    & 1.0                     \\
\texttt{M15L30}     & \texttt{Mujica} Checkpoint \#15 & L-BFGS    & 1.0                    
\end{tabular}
    \caption{Specification of the ablation test. 
    We run \mujica{} and a vanilla L-BFGS with the default setup, and change the initial guess and the step size of linear search for the rest runs. }
    \label{tab:ablation}
\end{table}

%% file: biblio.bib
@article{Dong:2023,
	adsurl = {https://ui.adsabs.harvard.edu/abs/2023ApJ...945L..28D},
	archiveprefix = {arXiv},
	author = {{Dong}, Chenze and {Lee}, Khee-Gan and {Ata}, Metin and {Horowitz}, Benjamin and {Momose}, Rieko},
	doi = {10.3847/2041-8213/acba89},
	eid = {L28},
	eprint = {2303.07619},
	journal = {\apjl},
	month = mar,
	number = {2},
	pages = {L28},
	primaryclass = {astro-ph.GA},
	title = {{Observational Evidence for Large-scale Gas Heating in a Galaxy Protocluster at z = 2.30}},
	volume = {945},
	year = 2023}

@article{Ata:2022,
	adsurl = {https://ui.adsabs.harvard.edu/abs/2022NatAs...6..857A},
	archiveprefix = {arXiv},
	author = {{Ata}, Metin and {Lee}, Khee-Gan and {Vecchia}, Claudio Dalla and {Kitaura}, Francisco-Shu and {Cucciati}, Olga and {Lemaux}, Brian C. and {Kashino}, Daichi and {M{\"u}ller}, Thomas},
	doi = {10.1038/s41550-022-01693-0},
	eprint = {2206.01115},
	journal = {Nature Astronomy},
	month = jun,
	pages = {857-865},
	primaryclass = {astro-ph.CO},
	title = {{Predicted future fate of COSMOS galaxy protoclusters over 11 Gyr with constrained simulations}},
	volume = {6},
	year = 2022}

@article{Horowitz:2022,
	adsurl = {https://ui.adsabs.harvard.edu/abs/2022ApJS..263...27H},
	archiveprefix = {arXiv},
	author = {{Horowitz}, Benjamin and {Lee}, Khee-Gan and {Ata}, Metin and {M{\"u}ller}, Thomas and {Krolewski}, Alex and {Prochaska}, J. Xavier and {Hennawi}, Joseph F. and {White}, Martin and {Schlegel}, David and {Rich}, R. Michael and {Nugent}, Peter E. and {Suzuki}, Nao and {Kashino}, Daichi and {Koekemoer}, Anton M. and {Lemaux}, Brian C.},
	doi = {10.3847/1538-4365/ac982d},
	eid = {27},
	eprint = {2109.09660},
	journal = {\apjs},
	month = dec,
	number = {2},
	pages = {27},
	primaryclass = {astro-ph.CO},
	title = {{Second Data Release of the COSMOS Ly{\ensuremath{\alpha}} Mapping and Tomography Observations: The First 3D Maps of the Detailed Cosmic Web at 2.05 < z < 2.55}},
	volume = {263},
	year = 2022}

@article{Newman:2020,
	adsurl = {https://ui.adsabs.harvard.edu/abs/2020ApJ...891..147N},
	archiveprefix = {arXiv},
	author = {{Newman}, Andrew B. and {Rudie}, Gwen C. and {Blanc}, Guillermo A. and {Kelson}, Daniel D. and {Rhoades}, Sunny and {Hare}, Tyson and {P{\'e}rez}, Victoria and {Benson}, Andrew J. and {Dressler}, Alan and {Gonzalez}, Valentino and {Kollmeier}, Juna A. and {Konidaris}, Nicholas P. and {Mulchaey}, John S. and {Rauch}, Michael and {Le F{\`e}vre}, Olivier and {Lemaux}, Brian C. and {Cucciati}, Olga and {Lilly}, Simon J.},
	doi = {10.3847/1538-4357/ab75ee},
	eid = {147},
	eprint = {2002.10676},
	journal = {\apj},
	month = mar,
	number = {2},
	pages = {147},
	primaryclass = {astro-ph.CO},
	title = {{LATIS: The Ly{\ensuremath{\alpha}} Tomography IMACS Survey}},
	volume = {891},
	year = 2020}

@article{Lee:2014,
	adsurl = {https://ui.adsabs.harvard.edu/abs/2014ApJ...795L..12L},
	archiveprefix = {arXiv},
	author = {{Lee}, Khee-Gan and {Hennawi}, Joseph F. and {Stark}, Casey and {Prochaska}, J. Xavier and {White}, Martin and {Schlegel}, David J. and {Eilers}, Anna-Christina and {Arinyo-i-Prats}, Andreu and {Suzuki}, Nao and {Croft}, Rupert A.~C. and {Caputi}, Karina I. and {Cassata}, Paolo and {Ilbert}, Olivier and {Garilli}, Bianca and {Koekemoer}, Anton M. and {Le Brun}, Vincent and {Le F{\`e}vre}, Olivier and {Maccagni}, Dario and {Nugent}, Peter and {Taniguchi}, Yoshiaki and {Tasca}, Lidia A.~M. and {Tresse}, Laurence and {Zamorani}, Gianni and {Zucca}, Elena},
	doi = {10.1088/2041-8205/795/1/L12},
	eid = {L12},
	eprint = {1409.5632},
	journal = {\apjl},
	month = nov,
	number = {1},
	pages = {L12},
	primaryclass = {astro-ph.CO},
	title = {{Ly{\ensuremath{\alpha}} Forest Tomography from Background Galaxies: The First Megaparsec-resolution Large-scale Structure Map at z > 2}},
	volume = {795},
	year = 2014}

@article{Lee:2016,
	adsurl = {https://ui.adsabs.harvard.edu/abs/2016ApJ...817..160L},
	archiveprefix = {arXiv},
	author = {{Lee}, Khee-Gan and {Hennawi}, Joseph F. and {White}, Martin and {Prochaska}, J. Xavier and {Font-Ribera}, Andreu and {Schlegel}, David J. and {Rich}, R. Michael and {Suzuki}, Nao and {Stark}, Casey W. and {Le F{\`e}vre}, Olivier and {Nugent}, Peter E. and {Salvato}, Mara and {Zamorani}, Gianni},
	doi = {10.3847/0004-637X/817/2/160},
	eid = {160},
	eprint = {1509.02833},
	journal = {\apj},
	month = feb,
	number = {2},
	pages = {160},
	primaryclass = {astro-ph.CO},
	title = {{Shadow of a Colossus: A z = 2.44 Galaxy Protocluster Detected in 3D Ly{\ensuremath{\alpha}} Forest Tomographic Mapping of the COSMOS Field}},
	volume = {817},
	year = 2016}

@article{Cen:2006,
	adsurl = {https://ui.adsabs.harvard.edu/abs/2006ApJ...650..560C},
	archiveprefix = {arXiv},
	author = {{Cen}, Renyue and {Ostriker}, Jeremiah P.},
	doi = {10.1086/506505},
	eprint = {astro-ph/0601008},
	journal = {\apj},
	month = oct,
	number = {2},
	pages = {560-572},
	primaryclass = {astro-ph},
	title = {{Where Are the Baryons? II. Feedback Effects}},
	volume = {650},
	year = 2006}

@article{Martizzi:2019,
	adsurl = {https://ui.adsabs.harvard.edu/abs/2019MNRAS.486.3766M},
	archiveprefix = {arXiv},
	author = {{Martizzi}, Davide and {Vogelsberger}, Mark and {Artale}, Maria Celeste and {Haider}, Markus and {Torrey}, Paul and {Marinacci}, Federico and {Nelson}, Dylan and {Pillepich}, Annalisa and {Weinberger}, Rainer and {Hernquist}, Lars and {Naiman}, Jill and {Springel}, Volker},
	doi = {10.1093/mnras/stz1106},
	eprint = {1810.01883},
	journal = {\mnras},
	month = jul,
	number = {3},
	pages = {3766-3787},
	primaryclass = {astro-ph.CO},
	title = {{Baryons in the Cosmic Web of IllustrisTNG - I: gas in knots, filaments, sheets, and voids}},
	volume = {486},
	year = 2019}

@article{Kooistra:2022,
	adsurl = {https://ui.adsabs.harvard.edu/abs/2022ApJ...938..123K},
	archiveprefix = {arXiv},
	author = {{Kooistra}, Robin and {Lee}, Khee-Gan and {Horowitz}, Benjamin},
	doi = {10.3847/1538-4357/ac92e8},
	eid = {123},
	eprint = {2201.10169},
	journal = {\apj},
	month = oct,
	number = {2},
	pages = {123},
	primaryclass = {astro-ph.CO},
	title = {{Constraining the Fluctuating Gunn-Peterson Approximation using Ly{\ensuremath{\alpha}} Forest Tomography at z = 2}},
	volume = {938},
	year = 2022}

@article{Takada:2014,
	adsurl = {https://ui.adsabs.harvard.edu/abs/2014PASJ...66R...1T},
	archiveprefix = {arXiv},
	author = {{Takada}, Masahiro and {Ellis}, Richard S. and {Chiba}, Masashi and {Greene}, Jenny E. and {Aihara}, Hiroaki and {Arimoto}, Nobuo and {Bundy}, Kevin and {Cohen}, Judith and {Dor{\'e}}, Olivier and {Graves}, Genevieve and {Gunn}, James E. and {Heckman}, Timothy and {Hirata}, Christopher M. and {Ho}, Paul and {Kneib}, Jean-Paul and {Le F{\`e}vre}, Olivier and {Lin}, Lihwai and {More}, Surhud and {Murayama}, Hitoshi and {Nagao}, Tohru and {Ouchi}, Masami and {Seiffert}, Michael and {Silverman}, John D. and {Sodr{\'e}}, Laerte and {Spergel}, David N. and {Strauss}, Michael A. and {Sugai}, Hajime and {Suto}, Yasushi and {Takami}, Hideki and {Wyse}, Rosemary},
	doi = {10.1093/pasj/pst019},
	eid = {R1},
	eprint = {1206.0737},
	journal = {\pasj},
	month = feb,
	number = {1},
	pages = {R1},
	primaryclass = {astro-ph.CO},
	title = {{Extragalactic science, cosmology, and Galactic archaeology with the Subaru Prime Focus Spectrograph}},
	volume = {66},
	year = 2014}

@article{Greene:2022,
	adsurl = {https://ui.adsabs.harvard.edu/abs/2022arXiv220614908G},
	archiveprefix = {arXiv},
	author = {{Greene}, Jenny and {Bezanson}, Rachel and {Ouchi}, Masami and {Silverman}, John and {the PFS Galaxy Evolution Working Group}},
	doi = {10.48550/arXiv.2206.14908},
	eid = {arXiv:2206.14908},
	eprint = {2206.14908},
	journal = {arXiv e-prints},
	month = jun,
	pages = {arXiv:2206.14908},
	primaryclass = {astro-ph.GA},
	title = {{The Prime Focus Spectrograph Galaxy Evolution Survey}},
	year = 2022}

@article{Ata:2021,
	adsurl = {https://ui.adsabs.harvard.edu/abs/2021MNRAS.500.3194A},
	archiveprefix = {arXiv},
	author = {{Ata}, Metin and {Kitaura}, Francisco-Shu and {Lee}, Khee-Gan and {Lemaux}, Brian C. and {Kashino}, Daichi and {Cucciati}, Olga and {Hern{\'a}ndez-S{\'a}nchez}, M{\'o}nica and {Le F{\`e}vre}, Oliver},
	doi = {10.1093/mnras/staa3318},
	eprint = {2004.11027},
	journal = {\mnras},
	month = jan,
	number = {3},
	pages = {3194-3212},
	primaryclass = {astro-ph.CO},
	title = {{BIRTH of the COSMOS field: primordial and evolved density reconstructions during cosmic high noon}},
	volume = {500},
	year = 2021}

@article{Krolewski:2017,
	adsurl = {https://ui.adsabs.harvard.edu/abs/2017ApJ...837...31K},
	archiveprefix = {arXiv},
	author = {{Krolewski}, Alex and {Lee}, Khee-Gan and {Luki{\'c}}, Zarija and {White}, Martin},
	doi = {10.3847/1538-4357/837/1/31},
	eid = {31},
	eprint = {1612.00067},
	journal = {\apj},
	month = mar,
	number = {1},
	pages = {31},
	primaryclass = {astro-ph.CO},
	title = {{Measuring Alignments between Galaxies and the Cosmic Web at z {\ensuremath{\sim}} 2-3 Using IGM Tomography}},
	volume = {837},
	year = 2017}

@article{Springel:2005,
	adsurl = {https://ui.adsabs.harvard.edu/abs/2005MNRAS.364.1105S},
	archiveprefix = {arXiv},
	author = {{Springel}, Volker},
	doi = {10.1111/j.1365-2966.2005.09655.x},
	eprint = {astro-ph/0505010},
	journal = {\mnras},
	month = dec,
	number = {4},
	pages = {1105-1134},
	primaryclass = {astro-ph},
	title = {{The cosmological simulation code GADGET-2}},
	volume = {364},
	year = 2005}

@article{Feng:2016,
	adsurl = {https://ui.adsabs.harvard.edu/abs/2016MNRAS.463.2273F},
	archiveprefix = {arXiv},
	author = {{Feng}, Yu and {Chu}, Man-Yat and {Seljak}, Uro{\v{s}} and {McDonald}, Patrick},
	doi = {10.1093/mnras/stw2123},
	eprint = {1603.00476},
	journal = {\mnras},
	month = dec,
	number = {3},
	pages = {2273-2286},
	primaryclass = {astro-ph.CO},
	title = {{FASTPM: a new scheme for fast simulations of dark matter and haloes}},
	volume = {463},
	year = 2016}

@article{Li:2024,
	adsurl = {https://ui.adsabs.harvard.edu/abs/2024ApJS..270...36L},
	archiveprefix = {arXiv},
	author = {{Li}, Yin and {Modi}, Chirag and {Jamieson}, Drew and {Zhang}, Yucheng and {Lu}, Libin and {Feng}, Yu and {Lanusse}, Fran{\c{c}}ois and {Greengard}, Leslie},
	doi = {10.3847/1538-4365/ad0ce7},
	eid = {36},
	eprint = {2211.09815},
	journal = {\apjs},
	month = feb,
	number = {2},
	pages = {36},
	primaryclass = {astro-ph.IM},
	title = {{Differentiable Cosmological Simulation with the Adjoint Method}},
	volume = {270},
	year = 2024}

@article{Kitaura:2013,
	adsurl = {https://ui.adsabs.harvard.edu/abs/2013MNRAS.435L..78K},
	archiveprefix = {arXiv},
	author = {{Kitaura}, F. -S. and {Hess}, S.},
	doi = {10.1093/mnrasl/slt101},
	eprint = {1212.3514},
	journal = {\mnras},
	month = aug,
	pages = {L78-L82},
	primaryclass = {astro-ph.CO},
	title = {{Cosmological structure formation with augmented lagrangian perturbation theory.}},
	volume = {435},
	year = 2013}

@article{Jasche:2013,
	adsurl = {https://ui.adsabs.harvard.edu/abs/2013MNRAS.432..894J},
	archiveprefix = {arXiv},
	author = {{Jasche}, Jens and {Wandelt}, Benjamin D.},
	doi = {10.1093/mnras/stt449},
	eprint = {1203.3639},
	journal = {\mnras},
	month = jun,
	number = {2},
	pages = {894-913},
	primaryclass = {astro-ph.CO},
	title = {{Bayesian physical reconstruction of initial conditions from large-scale structure surveys}},
	volume = {432},
	year = 2013}

@article{Kitaura:2021,
	adsurl = {https://ui.adsabs.harvard.edu/abs/2021MNRAS.502.3456K},
	archiveprefix = {arXiv},
	author = {{Kitaura}, Francisco-Shu and {Ata}, Metin and {Rodr{\'\i}guez-Torres}, Sergio A. and {Hern{\'a}ndez-S{\'a}nchez}, M{\'o}nica and {Balaguera-Antol{\'\i}nez}, A. and {Yepes}, Gustavo},
	doi = {10.1093/mnras/staa3774},
	eprint = {1911.00284},
	journal = {\mnras},
	month = apr,
	number = {3},
	pages = {3456-3475},
	primaryclass = {astro-ph.CO},
	title = {{COSMIC BIRTH: efficient Bayesian inference of the evolving cosmic web from galaxy surveys}},
	volume = {502},
	year = 2021}

@article{Ata:2015,
	adsurl = {https://ui.adsabs.harvard.edu/abs/2015MNRAS.446.4250A},
	archiveprefix = {arXiv},
	author = {{Ata}, Metin and {Kitaura}, Francisco-Shu and {M{\"u}ller}, Volker},
	doi = {10.1093/mnras/stu2347},
	eprint = {1408.2566},
	journal = {\mnras},
	month = feb,
	number = {4},
	pages = {4250-4259},
	primaryclass = {astro-ph.CO},
	title = {{Bayesian inference of cosmic density fields from non-linear, scale-dependent, and stochastic biased tracers}},
	volume = {446},
	year = 2015}

@article{Horowitz:2021,
	adsurl = {https://ui.adsabs.harvard.edu/abs/2021ApJ...906..110H},
	archiveprefix = {arXiv},
	author = {{Horowitz}, Benjamin and {Zhang}, Benjamin and {Lee}, Khee-Gan and {Kooistra}, Robin},
	doi = {10.3847/1538-4357/abca35},
	eid = {110},
	eprint = {2007.15994},
	journal = {\apj},
	month = jan,
	number = {2},
	pages = {110},
	primaryclass = {astro-ph.CO},
	title = {{TARDIS. II. Synergistic Density Reconstruction from Ly{\ensuremath{\alpha}} Forest and Spectroscopic Galaxy Surveys with Applications to Protoclusters and the Cosmic Web}},
	volume = {906},
	year = 2021}

@article{Jasche:2019,
	adsurl = {https://ui.adsabs.harvard.edu/abs/2019A&A...625A..64J},
	archiveprefix = {arXiv},
	author = {{Jasche}, J. and {Lavaux}, G.},
	doi = {10.1051/0004-6361/201833710},
	eid = {A64},
	eprint = {1806.11117},
	journal = {\aap},
	month = may,
	pages = {A64},
	primaryclass = {astro-ph.CO},
	title = {{Physical Bayesian modelling of the non-linear matter distribution: New insights into the nearby universe}},
	volume = {625},
	year = 2019}

@ARTICLE{Bayer:2023,
       author = {{Bayer}, Adrian E. and {Seljak}, Uros and {Modi}, Chirag},
        title = "{Field-Level Inference with Microcanonical Langevin Monte Carlo}",
      journal = {arXiv e-prints},
         year = 2023,
        month = jul,
          eid = {arXiv:2307.09504},
        pages = {arXiv:2307.09504},
          doi = {10.48550/arXiv.2307.09504},
archivePrefix = {arXiv},
       eprint = {2307.09504},
 primaryClass = {astro-ph.CO},
       adsurl = {https://ui.adsabs.harvard.edu/abs/2023arXiv230709504B}
}

@ARTICLE{Wang:2016_2,
       author = {{Wang}, Huiyuan and {Mo}, H.~J. and {Yang}, Xiaohu and {Zhang}, Youcai and {Shi}, JingJing and {Jing}, Y.~P. and {Liu}, Chengze and {Li}, Shijie and {Kang}, Xi and {Gao}, Yang},
        title = "{ELUCID - Exploring the Local Universe with ReConstructed Initial Density Field III: Constrained Simulation in the SDSS Volume}",
      journal = {\apj},
         year = 2016,
        month = nov,
       volume = {831},
       number = {2},
          eid = {164},
        pages = {164},
          doi = {10.3847/0004-637X/831/2/164},
archivePrefix = {arXiv},
       eprint = {1608.01763},
 primaryClass = {astro-ph.CO},
       adsurl = {https://ui.adsabs.harvard.edu/abs/2016ApJ...831..164W}
}

@ARTICLE{Wang:2014_1,
       author = {{Wang}, Huiyuan and {Mo}, H.~J. and {Yang}, Xiaohu and {Jing}, Y.~P. and {Lin}, W.~P.},
        title = "{ELUCID{\textemdash}Exploring the Local Universe with the Reconstructed Initial Density Field. I. Hamiltonian Markov Chain Monte Carlo Method with Particle Mesh Dynamics}",
      journal = {\apj},
         year = 2014,
        month = oct,
       volume = {794},
       number = {1},
          eid = {94},
        pages = {94},
          doi = {10.1088/0004-637X/794/1/94},
archivePrefix = {arXiv},
       eprint = {1407.3451},
 primaryClass = {astro-ph.CO},
       adsurl = {https://ui.adsabs.harvard.edu/abs/2014ApJ...794...94W}
}

@ARTICLE{Wang:2016_3,
       author = {{Wang}, Huiyuan and {Mo}, H.~J. and {Yang}, Xiaohu and {Zhang}, Youcai and {Shi}, JingJing and {Jing}, Y.~P. and {Liu}, Chengze and {Li}, Shijie and {Kang}, Xi and {Gao}, Yang},
        title = "{ELUCID - Exploring the Local Universe with ReConstructed Initial Density Field III: Constrained Simulation in the SDSS Volume}",
      journal = {\apj},
         year = 2016,
        month = nov,
       volume = {831},
       number = {2},
          eid = {164},
        pages = {164},
          doi = {10.3847/0004-637X/831/2/164},
archivePrefix = {arXiv},
       eprint = {1608.01763},
 primaryClass = {astro-ph.CO},
       adsurl = {https://ui.adsabs.harvard.edu/abs/2016ApJ...831..164W}
}

@ARTICLE{Lee:2022,
       author = {{Lee}, Khee-Gan and {Ata}, Metin and {Khrykin}, Ilya S. and {Huang}, Yuxin and {Prochaska}, J. Xavier and {Cooke}, Jeff and {Zhang}, Jielai and {Batten}, Adam},
        title = "{Constraining the Cosmic Baryon Distribution with Fast Radio Burst Foreground Mapping}",
      journal = {\apj},
         year = 2022,
        month = mar,
       volume = {928},
       number = {1},
          eid = {9},
        pages = {9},
          doi = {10.3847/1538-4357/ac4f62},
archivePrefix = {arXiv},
       eprint = {2109.00386},
 primaryClass = {astro-ph.CO},
       adsurl = {https://ui.adsabs.harvard.edu/abs/2022ApJ...928....9L}
}

@ARTICLE{Henriques:2015,
       author = {{Henriques}, Bruno M.~B. and {White}, Simon D.~M. and {Thomas}, Peter A. and {Angulo}, Raul and {Guo}, Qi and {Lemson}, Gerard and {Springel}, Volker and {Overzier}, Roderik},
        title = "{Galaxy formation in the Planck cosmology - I. Matching the observed evolution of star formation rates, colours and stellar masses}",
      journal = {\mnras},
         year = 2015,
        month = aug,
       volume = {451},
       number = {3},
        pages = {2663-2680},
          doi = {10.1093/mnras/stv705},
archivePrefix = {arXiv},
       eprint = {1410.0365},
 primaryClass = {astro-ph.GA},
       adsurl = {https://ui.adsabs.harvard.edu/abs/2015MNRAS.451.2663H}
}

@ARTICLE{Springel:2005b,
       author = {{Springel}, Volker and {White}, Simon D.~M. and {Jenkins}, Adrian and {Frenk}, Carlos S. and {Yoshida}, Naoki and {Gao}, Liang and {Navarro}, Julio and {Thacker}, Robert and {Croton}, Darren and {Helly}, John and {Peacock}, John A. and {Cole}, Shaun and {Thomas}, Peter and {Couchman}, Hugh and {Evrard}, August and {Colberg}, J{\"o}rg and {Pearce}, Frazer},
        title = "{Simulations of the formation, evolution and clustering of galaxies and quasars}",
      journal = {\nat},
         year = 2005,
        month = jun,
       volume = {435},
       number = {7042},
        pages = {629-636},
          doi = {10.1038/nature03597},
archivePrefix = {arXiv},
       eprint = {astro-ph/0504097},
 primaryClass = {astro-ph},
       adsurl = {https://ui.adsabs.harvard.edu/abs/2005Natur.435..629S}
}

@ARTICLE{Spergel:2003,
       author = {{Spergel}, D.~N. and {Verde}, L. and {Peiris}, H.~V. and {Komatsu}, E. and {Nolta}, M.~R. and {Bennett}, C.~L. and {Halpern}, M. and {Hinshaw}, G. and {Jarosik}, N. and {Kogut}, A. and {Limon}, M. and {Meyer}, S.~S. and {Page}, L. and {Tucker}, G.~S. and {Weiland}, J.~L. and {Wollack}, E. and {Wright}, E.~L.},
        title = "{First-Year Wilkinson Microwave Anisotropy Probe (WMAP) Observations: Determination of Cosmological Parameters}",
      journal = {\apjs},
         year = 2003,
        month = sep,
       volume = {148},
       number = {1},
        pages = {175-194},
          doi = {10.1086/377226},
archivePrefix = {arXiv},
       eprint = {astro-ph/0302209},
 primaryClass = {astro-ph},
       adsurl = {https://ui.adsabs.harvard.edu/abs/2003ApJS..148..175S}
}

@ARTICLE{Planck:2014,
       author = {{Planck Collaboration} and {Ade}, P.~A.~R. and {Aghanim}, N. and {Armitage-Caplan}, C. and {Arnaud}, M. and {Ashdown}, M. and {Atrio-Barandela}, F. and {Aumont}, J. and {Baccigalupi}, C. and {Banday}, A.~J. and {Barreiro}, R.~B. and {Bartlett}, J.~G. and {Battaner}, E. and {Benabed}, K. and {Beno{\^\i}t}, A. and {Benoit-L{\'e}vy}, A. and {Bernard}, J. -P. and {Bersanelli}, M. and {Bielewicz}, P. and {Bobin}, J. and {Bock}, J.~J. and {Bonaldi}, A. and {Bond}, J.~R. and {Borrill}, J. and {Bouchet}, F.~R. and {Bridges}, M. and {Bucher}, M. and {Burigana}, C. and {Butler}, R.~C. and {Calabrese}, E. and {Cappellini}, B. and {Cardoso}, J. -F. and {Catalano}, A. and {Challinor}, A. and {Chamballu}, A. and {Chary}, R. -R. and {Chen}, X. and {Chiang}, H.~C. and {Chiang}, L. -Y. and {Christensen}, P.~R. and {Church}, S. and {Clements}, D.~L. and {Colombi}, S. and {Colombo}, L.~P.~L. and {Couchot}, F. and {Coulais}, A. and {Crill}, B.~P. and {Curto}, A. and {Cuttaia}, F. and {Danese}, L. and {Davies}, R.~D. and {Davis}, R.~J. and {de Bernardis}, P. and {de Rosa}, A. and {de Zotti}, G. and {Delabrouille}, J. and {Delouis}, J. -M. and {D{\'e}sert}, F. -X. and {Dickinson}, C. and {Diego}, J.~M. and {Dolag}, K. and {Dole}, H. and {Donzelli}, S. and {Dor{\'e}}, O. and {Douspis}, M. and {Dunkley}, J. and {Dupac}, X. and {Efstathiou}, G. and {Elsner}, F. and {En{\ss}lin}, T.~A. and {Eriksen}, H.~K. and {Finelli}, F. and {Forni}, O. and {Frailis}, M. and {Fraisse}, A.~A. and {Franceschi}, E. and {Gaier}, T.~C. and {Galeotta}, S. and {Galli}, S. and {Ganga}, K. and {Giard}, M. and {Giardino}, G. and {Giraud-H{\'e}raud}, Y. and {Gjerl{\o}w}, E. and {Gonz{\'a}lez-Nuevo}, J. and {G{\'o}rski}, K.~M. and {Gratton}, S. and {Gregorio}, A. and {Gruppuso}, A. and {Gudmundsson}, J.~E. and {Haissinski}, J. and {Hamann}, J. and {Hansen}, F.~K. and {Hanson}, D. and {Harrison}, D. and {Henrot-Versill{\'e}}, S. and {Hern{\'a}ndez-Monteagudo}, C. and {Herranz}, D. and {Hildebrandt}, S.~R. and {Hivon}, E. and {Hobson}, M. and {Holmes}, W.~A. and {Hornstrup}, A. and {Hou}, Z. and {Hovest}, W. and {Huffenberger}, K.~M. and {Jaffe}, A.~H. and {Jaffe}, T.~R. and {Jewell}, J. and {Jones}, W.~C. and {Juvela}, M. and {Keih{\"a}nen}, E. and {Keskitalo}, R. and {Kisner}, T.~S. and {Kneissl}, R. and {Knoche}, J. and {Knox}, L. and {Kunz}, M. and {Kurki-Suonio}, H. and {Lagache}, G. and {L{\"a}hteenm{\"a}ki}, A. and {Lamarre}, J. -M. and {Lasenby}, A. and {Lattanzi}, M. and {Laureijs}, R.~J. and {Lawrence}, C.~R. and {Leach}, S. and {Leahy}, J.~P. and {Leonardi}, R. and {Le{\'o}n-Tavares}, J. and {Lesgourgues}, J. and {Lewis}, A. and {Liguori}, M. and {Lilje}, P.~B. and {Linden-V{\o}rnle}, M. and {L{\'o}pez-Caniego}, M. and {Lubin}, P.~M. and {Mac{\'\i}as-P{\'e}rez}, J.~F. and {Maffei}, B. and {Maino}, D. and {Mandolesi}, N. and {Maris}, M. and {Marshall}, D.~J. and {Martin}, P.~G. and {Mart{\'\i}nez-Gonz{\'a}lez}, E. and {Masi}, S. and {Massardi}, M. and {Matarrese}, S. and {Matthai}, F. and {Mazzotta}, P. and {Meinhold}, P.~R. and {Melchiorri}, A. and {Melin}, J. -B. and {Mendes}, L. and {Menegoni}, E. and {Mennella}, A. and {Migliaccio}, M. and {Millea}, M. and {Mitra}, S. and {Miville-Desch{\^e}nes}, M. -A. and {Moneti}, A. and {Montier}, L. and {Morgante}, G. and {Mortlock}, D. and {Moss}, A. and {Munshi}, D. and {Murphy}, J.~A. and {Naselsky}, P. and {Nati}, F. and {Natoli}, P. and {Netterfield}, C.~B. and {N{\o}rgaard-Nielsen}, H.~U. and {Noviello}, F. and {Novikov}, D. and {Novikov}, I. and {O'Dwyer}, I.~J. and {Osborne}, S. and {Oxborrow}, C.~A. and {Paci}, F. and {Pagano}, L. and {Pajot}, F. and {Paladini}, R. and {Paoletti}, D. and {Partridge}, B. and {Pasian}, F. and {Patanchon}, G. and {Pearson}, D. and {Pearson}, T.~J. and {Peiris}, H.~V. and {Perdereau}, O. and {Perotto}, L. and {Perrotta}, F. and {Pettorino}, V. and {Piacentini}, F. and {Piat}, M. and {Pierpaoli}, E. and {Pietrobon}, D. and {Plaszczynski}, S. and {Platania}, P. and {Pointecouteau}, E.},
        title = "{Planck 2013 results. XVI. Cosmological parameters}",
      journal = {\aap},
         year = 2014,
        month = nov,
       volume = {571},
          eid = {A16},
        pages = {A16},
          doi = {10.1051/0004-6361/201321591},
archivePrefix = {arXiv},
       eprint = {1303.5076},
 primaryClass = {astro-ph.CO},
       adsurl = {https://ui.adsabs.harvard.edu/abs/2014A&A...571A..16P}
}

@ARTICLE{Angulo:2015,
       author = {{Angulo}, Raul E. and {Hilbert}, Stefan},
        title = "{Cosmological constraints from the CFHTLenS shear measurements using a new, accurate, and flexible way of predicting non-linear mass clustering}",
      journal = {\mnras},
         year = 2015,
        month = mar,
       volume = {448},
       number = {1},
        pages = {364-375},
          doi = {10.1093/mnras/stv050},
archivePrefix = {arXiv},
       eprint = {1405.5888},
 primaryClass = {astro-ph.CO},
       adsurl = {https://ui.adsabs.harvard.edu/abs/2015MNRAS.448..364A}
}

@ARTICLE{Guo:2011,
       author = {{Guo}, Qi and {White}, Simon and {Boylan-Kolchin}, Michael and {De Lucia}, Gabriella and {Kauffmann}, Guinevere and {Lemson}, Gerard and {Li}, Cheng and {Springel}, Volker and {Weinmann}, Simone},
        title = "{From dwarf spheroidals to cD galaxies: simulating the galaxy population in a {\ensuremath{\Lambda}}CDM cosmology}",
      journal = {\mnras},
         year = 2011,
        month = may,
       volume = {413},
       number = {1},
        pages = {101-131},
          doi = {10.1111/j.1365-2966.2010.18114.x},
archivePrefix = {arXiv},
       eprint = {1006.0106},
 primaryClass = {astro-ph.CO},
       adsurl = {https://ui.adsabs.harvard.edu/abs/2011MNRAS.413..101G}
}

@ARTICLE{Maraston:2005,
       author = {{Maraston}, Claudia},
        title = "{Evolutionary population synthesis: models, analysis of the ingredients and application to high-z galaxies}",
      journal = {\mnras},
         year = 2005,
        month = sep,
       volume = {362},
       number = {3},
        pages = {799-825},
          doi = {10.1111/j.1365-2966.2005.09270.x},
archivePrefix = {arXiv},
       eprint = {astro-ph/0410207},
 primaryClass = {astro-ph},
       adsurl = {https://ui.adsabs.harvard.edu/abs/2005MNRAS.362..799M}
}

@ARTICLE{Blaizot:2005,
       author = {{Blaizot}, J{\'e}r{\'e}my and {Wadadekar}, Yogesh and {Guiderdoni}, Bruno and {Colombi}, St{\'e}phane T. and {Bertin}, Emmanuel and {Bouchet}, Fran{\c{c}}ois R. and {Devriendt}, Julien E.~G. and {Hatton}, Steve},
        title = "{MoMaF: the Mock Map Facility}",
      journal = {\mnras},
         year = 2005,
        month = jun,
       volume = {360},
       number = {1},
        pages = {159-175},
          doi = {10.1111/j.1365-2966.2005.09019.x},
archivePrefix = {arXiv},
       eprint = {astro-ph/0309305},
 primaryClass = {astro-ph},
       adsurl = {https://ui.adsabs.harvard.edu/abs/2005MNRAS.360..159B}
}

@ARTICLE{Kitzbichler:2007,
       author = {{Kitzbichler}, M.~G. and {White}, S.~D.~M.},
        title = "{The high-redshift galaxy population in hierarchical galaxy formation models}",
      journal = {\mnras},
         year = 2007,
        month = mar,
       volume = {376},
       number = {1},
        pages = {2-12},
          doi = {10.1111/j.1365-2966.2007.11458.x},
archivePrefix = {arXiv},
       eprint = {astro-ph/0609636},
 primaryClass = {astro-ph},
       adsurl = {https://ui.adsabs.harvard.edu/abs/2007MNRAS.376....2K}
}

@ARTICLE{McAlpine:2025,
       author = {{McAlpine}, Stuart and {Jasche}, Jens and {Ata}, Metin and {Lavaux}, Guilhem and {Stiskalek}, Richard and {Frenk}, Carlos S. and {Jenkins}, Adrian},
        title = "{The Manticore Project I: a digital twin of our cosmic neighbourhood from Bayesian field-level analysis}",
      journal = {\mnras},
         year = 2025,
        month = jun,
       volume = {540},
       number = {1},
        pages = {716-745},
          doi = {10.1093/mnras/staf767},
archivePrefix = {arXiv},
       eprint = {2505.10682},
 primaryClass = {astro-ph.CO},
       adsurl = {https://ui.adsabs.harvard.edu/abs/2025MNRAS.540..716M}
}

@ARTICLE{Sohn:2023,
       author = {{Sohn}, Jubee and {Geller}, Margaret J. and {Hwang}, Ho Seong and {Fabricant}, Daniel G. and {Utsumi}, Yousuke and {Damjanov}, Ivana},
        title = "{HectoMAP: The Complete Redshift Survey (Data Release 2)}",
      journal = {\apj},
         year = 2023,
        month = mar,
       volume = {945},
       number = {2},
          eid = {94},
        pages = {94},
          doi = {10.3847/1538-4357/acb925},
archivePrefix = {arXiv},
       eprint = {2210.16499},
 primaryClass = {astro-ph.CO},
       adsurl = {https://ui.adsabs.harvard.edu/abs/2023ApJ...945...94S}
}

@ARTICLE{Lee:2016a,
       author = {{Lee}, Khee-Gan and {White}, Martin},
        title = "{Revealing the z \raisebox{-0.5ex}\textasciitilde 2.5 Cosmic Web with 3D Ly{\ensuremath{\alpha}} Forest Tomography: a Deformation Tensor Approach}",
      journal = {\apj},
         year = 2016,
        month = nov,
       volume = {831},
       number = {2},
          eid = {181},
        pages = {181},
          doi = {10.3847/0004-637X/831/2/181},
archivePrefix = {arXiv},
       eprint = {1603.04441},
 primaryClass = {astro-ph.CO},
       adsurl = {https://ui.adsabs.harvard.edu/abs/2016ApJ...831..181L}
}

@ARTICLE{Zeldovich:1970,
       author = {{Zel'dovich}, Ya. B.},
        title = "{Gravitational instability: An approximate theory for large density perturbations.}",
      journal = {\aap},
         year = 1970,
        month = mar,
       volume = {5},
        pages = {84-89},
       adsurl = {https://ui.adsabs.harvard.edu/abs/1970A&A.....5...84Z}
}

@ARTICLE{Hahn:2007,
       author = {{Hahn}, Oliver and {Carollo}, C. Marcella and {Porciani}, Cristiano and {Dekel}, Avishai},
        title = "{The evolution of dark matter halo properties in clusters, filaments, sheets and voids}",
      journal = {\mnras},
         year = 2007,
        month = oct,
       volume = {381},
       number = {1},
        pages = {41-51},
          doi = {10.1111/j.1365-2966.2007.12249.x},
archivePrefix = {arXiv},
       eprint = {0704.2595},
 primaryClass = {astro-ph},
       adsurl = {https://ui.adsabs.harvard.edu/abs/2007MNRAS.381...41H}
}

@ARTICLE{Forero-Romero:2009,
       author = {{Forero-Romero}, J.~E. and {Hoffman}, Y. and {Gottl{\"o}ber}, S. and {Klypin}, A. and {Yepes}, G.},
        title = "{A dynamical classification of the cosmic web}",
      journal = {\mnras},
         year = 2009,
        month = jul,
       volume = {396},
       number = {3},
        pages = {1815-1824},
          doi = {10.1111/j.1365-2966.2009.14885.x},
archivePrefix = {arXiv},
       eprint = {0809.4135},
 primaryClass = {astro-ph},
       adsurl = {https://ui.adsabs.harvard.edu/abs/2009MNRAS.396.1815F}
}

@ARTICLE{Dong:2025,
       author = {{Dong}, Chenze and {Dedieu}, Florian and {Gal{\'a}rraga-Espinosa}, Daniela and {Lee}, Khee-Gan and {Sorini}, Daniele and {Dav{\'e}}, Romeel},
        title = "{Simba Simulation: The Effect of Feedback Physics on Matter Distribution in the Cosmic Web}",
      journal = {arXiv e-prints},
         year = 2025,
        month = jul,
          eid = {arXiv:2507.16115},
        pages = {arXiv:2507.16115},
          doi = {10.48550/arXiv.2507.16115},
archivePrefix = {arXiv},
       eprint = {2507.16115},
 primaryClass = {astro-ph.CO},
       adsurl = {https://ui.adsabs.harvard.edu/abs/2025arXiv250716115D}
}

@ARTICLE{Rampf:2025,
       author = {{Rampf}, Cornelius and {List}, Florian and {Hahn}, Oliver},
        title = "{BULLFROG: multi-step perturbation theory as a time integrator for cosmological simulations}",
      journal = {\jcap},
         year = 2025,
        month = feb,
       volume = {2025},
       number = {2},
          eid = {020},
        pages = {020},
          doi = {10.1088/1475-7516/2025/02/020},
archivePrefix = {arXiv},
       eprint = {2409.19049},
 primaryClass = {astro-ph.CO},
       adsurl = {https://ui.adsabs.harvard.edu/abs/2025JCAP...02..020R}
}

@ARTICLE{Simon-Onfroy:2025,
       author = {{Simon-Onfroy}, Hugo and {Lanusse}, Fran{\c{c}}ois and {de Mattia}, Arnaud},
        title = "{Benchmarking field-level cosmological inference from galaxy redshift surveys}",
      journal = {arXiv e-prints},
         year = 2025,
        month = apr,
          eid = {arXiv:2504.20130},
        pages = {arXiv:2504.20130},
          doi = {10.48550/arXiv.2504.20130},
archivePrefix = {arXiv},
       eprint = {2504.20130},
 primaryClass = {astro-ph.CO},
       adsurl = {https://ui.adsabs.harvard.edu/abs/2025arXiv250420130S}
}

@ARTICLE{Doeser:2024,
       author = {{Doeser}, Ludvig and {Jamieson}, Drew and {Stopyra}, Stephen and {Lavaux}, Guilhem and {Leclercq}, Florent and {Jasche}, Jens},
        title = "{Bayesian inference of initial conditions from non-linear cosmic structures using field-level emulators}",
      journal = {\mnras},
         year = 2024,
        month = dec,
       volume = {535},
       number = {2},
        pages = {1258-1277},
          doi = {10.1093/mnras/stae2429},
archivePrefix = {arXiv},
       eprint = {2312.09271},
 primaryClass = {astro-ph.CO},
       adsurl = {https://ui.adsabs.harvard.edu/abs/2024MNRAS.535.1258D}
}

@ARTICLE{Horowitz:2025,
       author = {{Horowitz}, B. and {Melchior}, P.},
        title = "{Joint cosmic density reconstruction from photometric and spectroscopic samples}",
      journal = {\mnras},
         year = 2025,
        month = apr,
       volume = {538},
       number = {3},
        pages = {2050-2057},
          doi = {10.1093/mnras/staf390},
archivePrefix = {arXiv},
       eprint = {2311.18738},
 primaryClass = {astro-ph.CO},
       adsurl = {https://ui.adsabs.harvard.edu/abs/2025MNRAS.538.2050H}
}

@ARTICLE{2018JCAP...10..028M,
       author = {{Modi}, Chirag and {Feng}, Yu and {Seljak}, Uro{\v{s}}},
        title = "{Cosmological reconstruction from galaxy light: neural network based light-matter connection}",
      journal = {\jcap},
         year = 2018,
        month = oct,
       volume = {2018},
       number = {10},
          eid = {028},
        pages = {028},
          doi = {10.1088/1475-7516/2018/10/028},
archivePrefix = {arXiv},
       eprint = {1805.02247},
 primaryClass = {astro-ph.CO},
       adsurl = {https://ui.adsabs.harvard.edu/abs/2018JCAP...10..028M}
}

@ARTICLE{2022ApJ...941...42H,
       author = {{Horowitz}, Benjamin and {Dornfest}, Max and {Luki{\'c}}, Zarija and {Harrington}, Peter},
        title = "{HYPHY: Deep Generative Conditional Posterior Mapping of Hydrodynamical Physics}",
      journal = {\apj},
         year = 2022,
        month = dec,
       volume = {941},
       number = {1},
          eid = {42},
        pages = {42},
          doi = {10.3847/1538-4357/ac9ea7},
archivePrefix = {arXiv},
       eprint = {2106.12675},
 primaryClass = {astro-ph.CO},
       adsurl = {https://ui.adsabs.harvard.edu/abs/2022ApJ...941...42H}
}

@ARTICLE{2022arXiv221114788H,
       author = {{Horowitz}, Benjamin and {Melchior}, Peter},
        title = "{Plausible Adversarial Attacks on Direct Parameter Inference Models in Astrophysics}",
      journal = {arXiv e-prints},
         year = 2022,
        month = nov,
          eid = {arXiv:2211.14788},
        pages = {arXiv:2211.14788},
          doi = {10.48550/arXiv.2211.14788},
archivePrefix = {arXiv},
       eprint = {2211.14788},
 primaryClass = {astro-ph.CO},
       adsurl = {https://ui.adsabs.harvard.edu/abs/2022arXiv221114788H}
}

@ARTICLE{2026diffhod2,
       author = {{Pandya}, Sneh and {Blazek}, Jonathan},
        title = "{Differentiable Stochastic Halo Occupation Distribution with Galaxy Intrinsic Alignments}",
      journal = {The Open Journal of Astrophysics},
         year = 2026,
        month = mar,
       volume = {9},
        pages = {59889},
          doi = {10.33232/001c.159889},
archivePrefix = {arXiv},
       eprint = {2602.04977},
 primaryClass = {astro-ph.CO},
       adsurl = {https://ui.adsabs.harvard.edu/abs/2026OJAp....959889P}
}

@ARTICLE{Horowitz:2019,
       author = {{Horowitz}, Benjamin and {Lee}, Khee-Gan and {White}, Martin and {Krolewski}, Alex and {Ata}, Metin},
        title = "{TARDIS. I. A Constrained Reconstruction Approach to Modeling the z {\ensuremath{\sim}} 2.5 Cosmic Web Probed by Ly{\ensuremath{\alpha}} Forest Tomography}",
      journal = {\apj},
         year = 2019,
        month = dec,
       volume = {887},
       number = {1},
          eid = {61},
        pages = {61},
          doi = {10.3847/1538-4357/ab4d4c},
archivePrefix = {arXiv},
       eprint = {1903.09049},
 primaryClass = {astro-ph.CO},
       adsurl = {https://ui.adsabs.harvard.edu/abs/2019ApJ...887...61H}
}

@ARTICLE{Byrohl:2025,
       author = {{Byrohl}, Chris and {Nelson}, Dylan and {Horowitz}, Benjamin and {Lee}, Khee-Gan and {Pillepich}, Annalisa},
        title = "{Introducing cosmosTNG: Simulating galaxy formation with constrained realizations of the COSMOS field}",
      journal = {\aap},
         year = 2025,
        month = jun,
       volume = {698},
          eid = {A103},
        pages = {A103},
          doi = {10.1051/0004-6361/202451970},
archivePrefix = {arXiv},
       eprint = {2409.19047},
 primaryClass = {astro-ph.GA},
       adsurl = {https://ui.adsabs.harvard.edu/abs/2025A&A...698A.103B}
}

@ARTICLE{Khrykin:2024,
       author = {{Khrykin}, Ilya S. and {Ata}, Metin and {Lee}, Khee-Gan and {Simha}, Sunil and {Huang}, Yuxin and {Prochaska}, J. Xavier and {Tejos}, Nicolas and {Bannister}, Keith W. and {Cooke}, Jeff and {Day}, Cherie K. and {Deller}, Adam and {Glowacki}, Marcin and {Gordon}, Alexa C. and {James}, Clancy W. and {Marnoch}, Lachlan and {Shannon}, Ryan. M. and {Zhang}, Jielai and {Bernales-Cortes}, Lucas},
        title = "{FLIMFLAM DR1: The First Constraints on the Cosmic Baryon Distribution from Eight Fast Radio Burst Sight Lines}",
      journal = {\apj},
         year = 2024,
        month = oct,
       volume = {973},
       number = {2},
          eid = {151},
        pages = {151},
          doi = {10.3847/1538-4357/ad6567},
archivePrefix = {arXiv},
       eprint = {2402.00505},
 primaryClass = {astro-ph.GA},
       adsurl = {https://ui.adsabs.harvard.edu/abs/2024ApJ...973..151K}
}

@ARTICLE{Seljak:2017,
       author = {{Seljak}, Uro{\v{s}} and {Aslanyan}, Grigor and {Feng}, Yu and {Modi}, Chirag},
        title = "{Towards optimal extraction of cosmological information from nonlinear data}",
      journal = {\jcap},
         year = 2017,
        month = dec,
       volume = {2017},
       number = {12},
          eid = {009},
        pages = {009},
          doi = {10.1088/1475-7516/2017/12/009},
archivePrefix = {arXiv},
       eprint = {1706.06645},
 primaryClass = {astro-ph.CO},
       adsurl = {https://ui.adsabs.harvard.edu/abs/2017JCAP...12..009S}
}

@ARTICLE{Bayer:2023a,
       author = {{Bayer}, Adrian E. and {Modi}, Chirag and {Ferraro}, Simone},
        title = "{Joint velocity and density reconstruction of the Universe with nonlinear differentiable forward modeling}",
      journal = {\jcap},
         year = 2023,
        month = jun,
       volume = {2023},
       number = {6},
          eid = {046},
        pages = {046},
          doi = {10.1088/1475-7516/2023/06/046},
archivePrefix = {arXiv},
       eprint = {2210.15649},
 primaryClass = {astro-ph.CO},
       adsurl = {https://ui.adsabs.harvard.edu/abs/2023JCAP...06..046B}
}

@ARTICLE{Modi:2021,
       author = {{Modi}, Chirag and {Lanusse}, Fran{\c{c}}ois and {Seljak}, Uro{\v{s}} and {Spergel}, David N. and {Perreault-Levasseur}, Laurence},
        title = "{CosmicRIM : Reconstructing Early Universe by Combining Differentiable Simulations with Recurrent Inference Machines}",
      journal = {arXiv e-prints},
         year = 2021,
        month = apr,
          eid = {arXiv:2104.12864},
        pages = {arXiv:2104.12864},
          doi = {10.48550/arXiv.2104.12864},
archivePrefix = {arXiv},
       eprint = {2104.12864},
 primaryClass = {astro-ph.CO},
       adsurl = {https://ui.adsabs.harvard.edu/abs/2021arXiv210412864M}
}

@ARTICLE{List:2023,
       author = {{List}, Florian and {Anau Montel}, Noemi and {Weniger}, Christoph},
        title = "{Bayesian Simulation-based Inference for Cosmological Initial Conditions}",
      journal = {arXiv e-prints},
         year = 2023,
        month = oct,
          eid = {arXiv:2310.19910},
        pages = {arXiv:2310.19910},
          doi = {10.48550/arXiv.2310.19910},
archivePrefix = {arXiv},
       eprint = {2310.19910},
 primaryClass = {astro-ph.CO},
       adsurl = {https://ui.adsabs.harvard.edu/abs/2023arXiv231019910L}
}

@ARTICLE{Legin:2024,
       author = {{Legin}, Ronan and {Ho}, Matthew and {Lemos}, Pablo and {Perreault-Levasseur}, Laurence and {Ho}, Shirley and {Hezaveh}, Yashar and {Wandelt}, Benjamin},
        title = "{Posterior sampling of the initial conditions of the universe from non-linear large scale structures using score-based generative models}",
      journal = {\mnras},
         year = 2024,
        month = jan,
       volume = {527},
       number = {1},
        pages = {L173-L178},
          doi = {10.1093/mnrasl/slad152},
archivePrefix = {arXiv},
       eprint = {2304.03788},
 primaryClass = {astro-ph.CO},
       adsurl = {https://ui.adsabs.harvard.edu/abs/2024MNRAS.527L.173L}
}

@ARTICLE{Jindal:2023,
       author = {{Jindal}, Vaibhav and {Liang}, Albert and {Singh}, Aarti and {Ho}, Shirley and {Jamieson}, Drew},
        title = "{Predicting the Initial Conditions of the Universe using a Deterministic Neural Network}",
      journal = {arXiv e-prints},
         year = 2023,
        month = mar,
          eid = {arXiv:2303.13056},
        pages = {arXiv:2303.13056},
          doi = {10.48550/arXiv.2303.13056},
archivePrefix = {arXiv},
       eprint = {2303.13056},
 primaryClass = {astro-ph.CO},
       adsurl = {https://ui.adsabs.harvard.edu/abs/2023arXiv230313056J}
}

@ARTICLE{Bottema:2025,
       author = {{Bottema}, Jelte and {Fl{\"o}ss}, Thomas and {Daniel Meerburg}, P.},
        title = "{Neural network reconstruction of non-Gaussian initial conditions from dark matter halos}",
      journal = {\jcap},
         year = 2025,
        month = aug,
       volume = {2025},
       number = {8},
          eid = {030},
        pages = {030},
          doi = {10.1088/1475-7516/2025/08/030},
archivePrefix = {arXiv},
       eprint = {2502.11846},
 primaryClass = {astro-ph.CO},
       adsurl = {https://ui.adsabs.harvard.edu/abs/2025JCAP...08..030B}
}

@ARTICLE{Doeser:2025,
       author = {{Doeser}, Ludvig and {Ata}, Metin and {Jasche}, Jens},
        title = "{Learning the Universe: learning to optimize cosmic initial conditions with non-differentiable structure formation models}",
      journal = {\mnras},
         year = 2025,
        month = sep,
       volume = {542},
       number = {2},
        pages = {1403-1422},
          doi = {10.1093/mnras/staf1289},
archivePrefix = {arXiv},
       eprint = {2502.13243},
 primaryClass = {astro-ph.CO},
       adsurl = {https://ui.adsabs.harvard.edu/abs/2025MNRAS.542.1403D}
}

@ARTICLE{DESI:2016,
       author = {{DESI Collaboration} and {Aghamousa}, Amir and {Aguilar}, Jessica and {Ahlen}, Steve and {Alam}, Shadab and {Allen}, Lori E. and {Allende Prieto}, Carlos and {Annis}, James and {Bailey}, Stephen and {Balland}, Christophe and {Ballester}, Otger and {Baltay}, Charles and {Beaufore}, Lucas and {Bebek}, Chris and {Beers}, Timothy C. and {Bell}, Eric F. and {Bernal}, Jos{\'e} Luis and {Besuner}, Robert and {Beutler}, Florian and {Blake}, Chris and {Bleuler}, Hannes and {Blomqvist}, Michael and {Blum}, Robert and {Bolton}, Adam S. and {Briceno}, Cesar and {Brooks}, David and {Brownstein}, Joel R. and {Buckley-Geer}, Elizabeth and {Burden}, Angela and {Burtin}, Etienne and {Busca}, Nicolas G. and {Cahn}, Robert N. and {Cai}, Yan-Chuan and {Cardiel-Sas}, Laia and {Carlberg}, Raymond G. and {Carton}, Pierre-Henri and {Casas}, Ricard and {Castander}, Francisco J. and {Cervantes-Cota}, Jorge L. and {Claybaugh}, Todd M. and {Close}, Madeline and {Coker}, Carl T. and {Cole}, Shaun and {Comparat}, Johan and {Cooper}, Andrew P. and {Cousinou}, M. -C. and {Crocce}, Martin and {Cuby}, Jean-Gabriel and {Cunningham}, Daniel P. and {Davis}, Tamara M. and {Dawson}, Kyle S. and {de la Macorra}, Axel and {De Vicente}, Juan and {Delubac}, Timoth{\'e}e and {Derwent}, Mark and {Dey}, Arjun and {Dhungana}, Govinda and {Ding}, Zhejie and {Doel}, Peter and {Duan}, Yutong T. and {Ealet}, Anne and {Edelstein}, Jerry and {Eftekharzadeh}, Sarah and {Eisenstein}, Daniel J. and {Elliott}, Ann and {Escoffier}, St{\'e}phanie and {Evatt}, Matthew and {Fagrelius}, Parker and {Fan}, Xiaohui and {Fanning}, Kevin and {Farahi}, Arya and {Farihi}, Jay and {Favole}, Ginevra and {Feng}, Yu and {Fernandez}, Enrique and {Findlay}, Joseph R. and {Finkbeiner}, Douglas P. and {Fitzpatrick}, Michael J. and {Flaugher}, Brenna and {Flender}, Samuel and {Font-Ribera}, Andreu and {Forero-Romero}, Jaime E. and {Fosalba}, Pablo and {Frenk}, Carlos S. and {Fumagalli}, Michele and {Gaensicke}, Boris T. and {Gallo}, Giuseppe and {Garcia-Bellido}, Juan and {Gaztanaga}, Enrique and {Pietro Gentile Fusillo}, Nicola and {Gerard}, Terry and {Gershkovich}, Irena and {Giannantonio}, Tommaso and {Gillet}, Denis and {Gonzalez-de-Rivera}, Guillermo and {Gonzalez-Perez}, Violeta and {Gott}, Shelby and {Graur}, Or and {Gutierrez}, Gaston and {Guy}, Julien and {Habib}, Salman and {Heetderks}, Henry and {Heetderks}, Ian and {Heitmann}, Katrin and {Hellwing}, Wojciech A. and {Herrera}, David A. and {Ho}, Shirley and {Holland}, Stephen and {Honscheid}, Klaus and {Huff}, Eric and {Hutchinson}, Timothy A. and {Huterer}, Dragan and {Hwang}, Ho Seong and {Illa Laguna}, Joseph Maria and {Ishikawa}, Yuzo and {Jacobs}, Dianna and {Jeffrey}, Niall and {Jelinsky}, Patrick and {Jennings}, Elise and {Jiang}, Linhua and {Jimenez}, Jorge and {Johnson}, Jennifer and {Joyce}, Richard and {Jullo}, Eric and {Juneau}, St{\'e}phanie and {Kama}, Sami and {Karcher}, Armin and {Karkar}, Sonia and {Kehoe}, Robert and {Kennamer}, Noble and {Kent}, Stephen and {Kilbinger}, Martin and {Kim}, Alex G. and {Kirkby}, David and {Kisner}, Theodore and {Kitanidis}, Ellie and {Kneib}, Jean-Paul and {Koposov}, Sergey and {Kovacs}, Eve and {Koyama}, Kazuya and {Kremin}, Anthony and {Kron}, Richard and {Kronig}, Luzius and {Kueter-Young}, Andrea and {Lacey}, Cedric G. and {Lafever}, Robin and {Lahav}, Ofer and {Lambert}, Andrew and {Lampton}, Michael and {Landriau}, Martin and {Lang}, Dustin and {Lauer}, Tod R. and {Le Goff}, Jean-Marc and {Le Guillou}, Laurent and {Le Van Suu}, Auguste and {Lee}, Jae Hyeon and {Lee}, Su-Jeong and {Leitner}, Daniela and {Lesser}, Michael and {Levi}, Michael E. and {L'Huillier}, Benjamin and {Li}, Baojiu and {Liang}, Ming and {Lin}, Huan and {Linder}, Eric and {Loebman}, Sarah R. and {Luki{\'c}}, Zarija and {Ma}, Jun and {MacCrann}, Niall and {Magneville}, Christophe and {Makarem}, Laleh and {Manera}, Marc and {Manser}, Christopher J. and {Marshall}, Robert and {Martini}, Paul and {Massey}, Richard and {Matheson}, Thomas and {McCauley}, Jeremy and {McDonald}, Patrick and {McGreer}, Ian D. and {Meisner}, Aaron and {Metcalfe}, Nigel and {Miller}, Timothy N. and {Miquel}, Ramon and {Moustakas}, John and {Myers}, Adam and {Naik}, Milind and {Newman}, Jeffrey A. and {Nichol}, Robert C. and {Nicola}, Andrina and {Nicolati da Costa}, Luiz and {Nie}, Jundan and {Niz}, Gustavo and {Norberg}, Peder and {Nord}, Brian and {Norman}, Dara and {Nugent}, Peter and {O'Brien}, Thomas and {Oh}, Minji and {Olsen}, Knut A.~G.},
        title = "{The DESI Experiment Part I: Science,Targeting, and Survey Design}",
      journal = {arXiv e-prints},
         year = 2016,
        month = oct,
          eid = {arXiv:1611.00036},
        pages = {arXiv:1611.00036},
          doi = {10.48550/arXiv.1611.00036},
archivePrefix = {arXiv},
       eprint = {1611.00036},
 primaryClass = {astro-ph.IM},
       adsurl = {https://ui.adsabs.harvard.edu/abs/2016arXiv161100036D}
}

@ARTICLE{Laureijs:2011,
       author = {{Laureijs}, R. and {Amiaux}, J. and {Arduini}, S. and {Augu{\`e}res}, J. -L. and {Brinchmann}, J. and {Cole}, R. and {Cropper}, M. and {Dabin}, C. and {Duvet}, L. and {Ealet}, A. and {Garilli}, B. and {Gondoin}, P. and {Guzzo}, L. and {Hoar}, J. and {Hoekstra}, H. and {Holmes}, R. and {Kitching}, T. and {Maciaszek}, T. and {Mellier}, Y. and {Pasian}, F. and {Percival}, W. and {Rhodes}, J. and {Saavedra Criado}, G. and {Sauvage}, M. and {Scaramella}, R. and {Valenziano}, L. and {Warren}, S. and {Bender}, R. and {Castander}, F. and {Cimatti}, A. and {Le F{\`e}vre}, O. and {Kurki-Suonio}, H. and {Levi}, M. and {Lilje}, P. and {Meylan}, G. and {Nichol}, R. and {Pedersen}, K. and {Popa}, V. and {Rebolo Lopez}, R. and {Rix}, H. -W. and {Rottgering}, H. and {Zeilinger}, W. and {Grupp}, F. and {Hudelot}, P. and {Massey}, R. and {Meneghetti}, M. and {Miller}, L. and {Paltani}, S. and {Paulin-Henriksson}, S. and {Pires}, S. and {Saxton}, C. and {Schrabback}, T. and {Seidel}, G. and {Walsh}, J. and {Aghanim}, N. and {Amendola}, L. and {Bartlett}, J. and {Baccigalupi}, C. and {Beaulieu}, J. -P. and {Benabed}, K. and {Cuby}, J. -G. and {Elbaz}, D. and {Fosalba}, P. and {Gavazzi}, G. and {Helmi}, A. and {Hook}, I. and {Irwin}, M. and {Kneib}, J. -P. and {Kunz}, M. and {Mannucci}, F. and {Moscardini}, L. and {Tao}, C. and {Teyssier}, R. and {Weller}, J. and {Zamorani}, G. and {Zapatero Osorio}, M.~R. and {Boulade}, O. and {Foumond}, J.~J. and {Di Giorgio}, A. and {Guttridge}, P. and {James}, A. and {Kemp}, M. and {Martignac}, J. and {Spencer}, A. and {Walton}, D. and {Bl{\"u}mchen}, T. and {Bonoli}, C. and {Bortoletto}, F. and {Cerna}, C. and {Corcione}, L. and {Fabron}, C. and {Jahnke}, K. and {Ligori}, S. and {Madrid}, F. and {Martin}, L. and {Morgante}, G. and {Pamplona}, T. and {Prieto}, E. and {Riva}, M. and {Toledo}, R. and {Trifoglio}, M. and {Zerbi}, F. and {Abdalla}, F. and {Douspis}, M. and {Grenet}, C. and {Borgani}, S. and {Bouwens}, R. and {Courbin}, F. and {Delouis}, J. -M. and {Dubath}, P. and {Fontana}, A. and {Frailis}, M. and {Grazian}, A. and {Koppenh{\"o}fer}, J. and {Mansutti}, O. and {Melchior}, M. and {Mignoli}, M. and {Mohr}, J. and {Neissner}, C. and {Noddle}, K. and {Poncet}, M. and {Scodeggio}, M. and {Serrano}, S. and {Shane}, N. and {Starck}, J. -L. and {Surace}, C. and {Taylor}, A. and {Verdoes-Kleijn}, G. and {Vuerli}, C. and {Williams}, O.~R. and {Zacchei}, A. and {Altieri}, B. and {Escudero Sanz}, I. and {Kohley}, R. and {Oosterbroek}, T. and {Astier}, P. and {Bacon}, D. and {Bardelli}, S. and {Baugh}, C. and {Bellagamba}, F. and {Benoist}, C. and {Bianchi}, D. and {Biviano}, A. and {Branchini}, E. and {Carbone}, C. and {Cardone}, V. and {Clements}, D. and {Colombi}, S. and {Conselice}, C. and {Cresci}, G. and {Deacon}, N. and {Dunlop}, J. and {Fedeli}, C. and {Fontanot}, F. and {Franzetti}, P. and {Giocoli}, C. and {Garcia-Bellido}, J. and {Gow}, J. and {Heavens}, A. and {Hewett}, P. and {Heymans}, C. and {Holland}, A. and {Huang}, Z. and {Ilbert}, O. and {Joachimi}, B. and {Jennins}, E. and {Kerins}, E. and {Kiessling}, A. and {Kirk}, D. and {Kotak}, R. and {Krause}, O. and {Lahav}, O. and {van Leeuwen}, F. and {Lesgourgues}, J. and {Lombardi}, M. and {Magliocchetti}, M. and {Maguire}, K. and {Majerotto}, E. and {Maoli}, R. and {Marulli}, F. and {Maurogordato}, S. and {McCracken}, H. and {McLure}, R. and {Melchiorri}, A. and {Merson}, A. and {Moresco}, M. and {Nonino}, M. and {Norberg}, P. and {Peacock}, J. and {Pello}, R. and {Penny}, M. and {Pettorino}, V. and {Di Porto}, C. and {Pozzetti}, L. and {Quercellini}, C. and {Radovich}, M. and {Rassat}, A. and {Roche}, N. and {Ronayette}, S. and {Rossetti}, E.},
        title = "{Euclid Definition Study Report}",
      journal = {arXiv e-prints},
         year = 2011,
        month = oct,
          eid = {arXiv:1110.3193},
        pages = {arXiv:1110.3193},
          doi = {10.48550/arXiv.1110.3193},
archivePrefix = {arXiv},
       eprint = {1110.3193},
 primaryClass = {astro-ph.CO},
       adsurl = {https://ui.adsabs.harvard.edu/abs/2011arXiv1110.3193L}
}

@ARTICLE{Spergel:2015,
       author = {{Spergel}, D. and {Gehrels}, N. and {Baltay}, C. and {Bennett}, D. and {Breckinridge}, J. and {Donahue}, M. and {Dressler}, A. and {Gaudi}, B.~S. and {Greene}, T. and {Guyon}, O. and {Hirata}, C. and {Kalirai}, J. and {Kasdin}, N.~J. and {Macintosh}, B. and {Moos}, W. and {Perlmutter}, S. and {Postman}, M. and {Rauscher}, B. and {Rhodes}, J. and {Wang}, Y. and {Weinberg}, D. and {Benford}, D. and {Hudson}, M. and {Jeong}, W. -S. and {Mellier}, Y. and {Traub}, W. and {Yamada}, T. and {Capak}, P. and {Colbert}, J. and {Masters}, D. and {Penny}, M. and {Savransky}, D. and {Stern}, D. and {Zimmerman}, N. and {Barry}, R. and {Bartusek}, L. and {Carpenter}, K. and {Cheng}, E. and {Content}, D. and {Dekens}, F. and {Demers}, R. and {Grady}, K. and {Jackson}, C. and {Kuan}, G. and {Kruk}, J. and {Melton}, M. and {Nemati}, B. and {Parvin}, B. and {Poberezhskiy}, I. and {Peddie}, C. and {Ruffa}, J. and {Wallace}, J.~K. and {Whipple}, A. and {Wollack}, E. and {Zhao}, F.},
        title = "{Wide-Field InfrarRed Survey Telescope-Astrophysics Focused Telescope Assets WFIRST-AFTA 2015 Report}",
      journal = {arXiv e-prints},
         year = 2015,
        month = mar,
          eid = {arXiv:1503.03757},
        pages = {arXiv:1503.03757},
          doi = {10.48550/arXiv.1503.03757},
archivePrefix = {arXiv},
       eprint = {1503.03757},
 primaryClass = {astro-ph.IM},
       adsurl = {https://ui.adsabs.harvard.edu/abs/2015arXiv150303757S}
}

@ARTICLE{Ivezic:2019,
       author = {{Ivezi{\'c}}, {\v{Z}}eljko and {Kahn}, Steven M. and {Tyson}, J. Anthony and {Abel}, Bob and {Acosta}, Emily and {Allsman}, Robyn and {Alonso}, David and {AlSayyad}, Yusra and {Anderson}, Scott F. and {Andrew}, John and {Angel}, James Roger P. and {Angeli}, George Z. and {Ansari}, Reza and {Antilogus}, Pierre and {Araujo}, Constanza and {Armstrong}, Robert and {Arndt}, Kirk T. and {Astier}, Pierre and {Aubourg}, {\'E}ric and {Auza}, Nicole and {Axelrod}, Tim S. and {Bard}, Deborah J. and {Barr}, Jeff D. and {Barrau}, Aurelian and {Bartlett}, James G. and {Bauer}, Amanda E. and {Bauman}, Brian J. and {Baumont}, Sylvain and {Bechtol}, Ellen and {Bechtol}, Keith and {Becker}, Andrew C. and {Becla}, Jacek and {Beldica}, Cristina and {Bellavia}, Steve and {Bianco}, Federica B. and {Biswas}, Rahul and {Blanc}, Guillaume and {Blazek}, Jonathan and {Blandford}, Roger D. and {Bloom}, Josh S. and {Bogart}, Joanne and {Bond}, Tim W. and {Booth}, Michael T. and {Borgland}, Anders W. and {Borne}, Kirk and {Bosch}, James F. and {Boutigny}, Dominique and {Brackett}, Craig A. and {Bradshaw}, Andrew and {Brandt}, William Nielsen and {Brown}, Michael E. and {Bullock}, James S. and {Burchat}, Patricia and {Burke}, David L. and {Cagnoli}, Gianpietro and {Calabrese}, Daniel and {Callahan}, Shawn and {Callen}, Alice L. and {Carlin}, Jeffrey L. and {Carlson}, Erin L. and {Chandrasekharan}, Srinivasan and {Charles-Emerson}, Glenaver and {Chesley}, Steve and {Cheu}, Elliott C. and {Chiang}, Hsin-Fang and {Chiang}, James and {Chirino}, Carol and {Chow}, Derek and {Ciardi}, David R. and {Claver}, Charles F. and {Cohen-Tanugi}, Johann and {Cockrum}, Joseph J. and {Coles}, Rebecca and {Connolly}, Andrew J. and {Cook}, Kem H. and {Cooray}, Asantha and {Covey}, Kevin R. and {Cribbs}, Chris and {Cui}, Wei and {Cutri}, Roc and {Daly}, Philip N. and {Daniel}, Scott F. and {Daruich}, Felipe and {Daubard}, Guillaume and {Daues}, Greg and {Dawson}, William and {Delgado}, Francisco and {Dellapenna}, Alfred and {de Peyster}, Robert and {de Val-Borro}, Miguel and {Digel}, Seth W. and {Doherty}, Peter and {Dubois}, Richard and {Dubois-Felsmann}, Gregory P. and {Durech}, Josef and {Economou}, Frossie and {Eifler}, Tim and {Eracleous}, Michael and {Emmons}, Benjamin L. and {Fausti Neto}, Angelo and {Ferguson}, Henry and {Figueroa}, Enrique and {Fisher-Levine}, Merlin and {Focke}, Warren and {Foss}, Michael D. and {Frank}, James and {Freemon}, Michael D. and {Gangler}, Emmanuel and {Gawiser}, Eric and {Geary}, John C. and {Gee}, Perry and {Geha}, Marla and {Gessner}, Charles J.~B. and {Gibson}, Robert R. and {Gilmore}, D. Kirk and {Glanzman}, Thomas and {Glick}, William and {Goldina}, Tatiana and {Goldstein}, Daniel A. and {Goodenow}, Iain and {Graham}, Melissa L. and {Gressler}, William J. and {Gris}, Philippe and {Guy}, Leanne P. and {Guyonnet}, Augustin and {Haller}, Gunther and {Harris}, Ron and {Hascall}, Patrick A. and {Haupt}, Justine and {Hernandez}, Fabio and {Herrmann}, Sven and {Hileman}, Edward and {Hoblitt}, Joshua and {Hodgson}, John A. and {Hogan}, Craig and {Howard}, James D. and {Huang}, Dajun and {Huffer}, Michael E. and {Ingraham}, Patrick and {Innes}, Walter R. and {Jacoby}, Suzanne H. and {Jain}, Bhuvnesh and {Jammes}, Fabrice and {Jee}, M. James and {Jenness}, Tim and {Jernigan}, Garrett and {Jevremovi{\'c}}, Darko and {Johns}, Kenneth and {Johnson}, Anthony S. and {Johnson}, Margaret W.~G. and {Jones}, R. Lynne and {Juramy-Gilles}, Claire and {Juri{\'c}}, Mario and {Kalirai}, Jason S. and {Kallivayalil}, Nitya J. and {Kalmbach}, Bryce and {Kantor}, Jeffrey P. and {Karst}, Pierre and {Kasliwal}, Mansi M. and {Kelly}, Heather and {Kessler}, Richard and {Kinnison}, Veronica and {Kirkby}, David and {Knox}, Lloyd and {Kotov}, Ivan V. and {Krabbendam}, Victor L. and {Krughoff}, K. Simon and {Kub{\'a}nek}, Petr and {Kuczewski}, John and {Kulkarni}, Shri and {Ku}, John and {Kurita}, Nadine R. and {Lage}, Craig S. and {Lambert}, Ron and {Lange}, Travis and {Langton}, J. Brian and {Le Guillou}, Laurent and {Levine}, Deborah and {Liang}, Ming and {Lim}, Kian-Tat and {Lintott}, Chris J. and {Long}, Kevin E. and {Lopez}, Margaux and {Lotz}, Paul J. and {Lupton}, Robert H. and {Lust}, Nate B. and {MacArthur}, Lauren A. and {Mahabal}, Ashish and {Mandelbaum}, Rachel and {Markiewicz}, Thomas W. and {Marsh}, Darren S. and {Marshall}, Philip J. and {Marshall}, Stuart and {May}, Morgan and {McKercher}, Robert and {McQueen}, Michelle and {Meyers}, Joshua and {Migliore}, Myriam and {Miller}, Michelle and {Mills}, David J.},
        title = "{LSST: From Science Drivers to Reference Design and Anticipated Data Products}",
      journal = {\apj},
         year = 2019,
        month = mar,
       volume = {873},
       number = {2},
          eid = {111},
        pages = {111},
          doi = {10.3847/1538-4357/ab042c},
archivePrefix = {arXiv},
       eprint = {0805.2366},
 primaryClass = {astro-ph},
       adsurl = {https://ui.adsabs.harvard.edu/abs/2019ApJ...873..111I}
}

@ARTICLE{Dore:2014,
       author = {{Dor{\'e}}, Olivier and {Bock}, Jamie and {Ashby}, Matthew and {Capak}, Peter and {Cooray}, Asantha and {de Putter}, Roland and {Eifler}, Tim and {Flagey}, Nicolas and {Gong}, Yan and {Habib}, Salman and {Heitmann}, Katrin and {Hirata}, Chris and {Jeong}, Woong-Seob and {Katti}, Raj and {Korngut}, Phil and {Krause}, Elisabeth and {Lee}, Dae-Hee and {Masters}, Daniel and {Mauskopf}, Phil and {Melnick}, Gary and {Mennesson}, Bertrand and {Nguyen}, Hien and {{\"O}berg}, Karin and {Pullen}, Anthony and {Raccanelli}, Alvise and {Smith}, Roger and {Song}, Yong-Seon and {Tolls}, Volker and {Unwin}, Steve and {Venumadhav}, Tejaswi and {Viero}, Marco and {Werner}, Mike and {Zemcov}, Mike},
        title = "{Cosmology with the SPHEREX All-Sky Spectral Survey}",
      journal = {arXiv e-prints},
         year = 2014,
        month = dec,
          eid = {arXiv:1412.4872},
        pages = {arXiv:1412.4872},
          doi = {10.48550/arXiv.1412.4872},
archivePrefix = {arXiv},
       eprint = {1412.4872},
 primaryClass = {astro-ph.CO},
       adsurl = {https://ui.adsabs.harvard.edu/abs/2014arXiv1412.4872D}
}

@ARTICLE{Horowitz:2025a,
       author = {{Horowitz}, Benjamin and {Bayer}, Adrian E.},
        title = "{jFoF: GPU Cluster Finding with Gradient Propagation}",
      journal = {arXiv e-prints},
         year = 2025,
        month = oct,
          eid = {arXiv:2510.26851},
        pages = {arXiv:2510.26851},
          doi = {10.48550/arXiv.2510.26851},
archivePrefix = {arXiv},
       eprint = {2510.26851},
 primaryClass = {astro-ph.IM},
       adsurl = {https://ui.adsabs.harvard.edu/abs/2025arXiv251026851H}
}

@ARTICLE{2021JCAP...10..056M,
       author = {{Modi}, Chirag and {White}, Martin and {Castorina}, Emanuele and {Slosar}, An{\v{z}}e},
        title = "{Mind the gap: the power of combining photometric surveys with intensity mapping}",
      journal = {\jcap},
         year = 2021,
        month = oct,
       volume = {2021},
       number = {10},
          eid = {056},
        pages = {056},
          doi = {10.1088/1475-7516/2021/10/056},
archivePrefix = {arXiv},
       eprint = {2102.08116},
 primaryClass = {astro-ph.CO},
       adsurl = {https://ui.adsabs.harvard.edu/abs/2021JCAP...10..056M}
}

@ARTICLE{2019JCAP...10..035H,
       author = {{Horowitz}, B. and {Seljak}, U. and {Aslanyan}, G.},
        title = "{Efficient optimal reconstruction of linear fields and band-powers from cosmological data}",
      journal = {\jcap},
         year = 2019,
        month = oct,
       volume = {2019},
       number = {10},
          eid = {035},
        pages = {035},
          doi = {10.1088/1475-7516/2019/10/035},
archivePrefix = {arXiv},
       eprint = {1810.00503},
 primaryClass = {astro-ph.CO},
       adsurl = {https://ui.adsabs.harvard.edu/abs/2019JCAP...10..035H}
}

@ARTICLE{Horowitz:2024,
       author = {{Horowitz}, Benjamin and {Hahn}, ChangHoon and {Lanusse}, Francois and {Modi}, Chirag and {Ferraro}, Simone},
        title = "{Differentiable stochastic halo occupation distribution}",
      journal = {\mnras},
         year = 2024,
        month = apr,
       volume = {529},
       number = {3},
        pages = {2473-2482},
          doi = {10.1093/mnras/stae350},
archivePrefix = {arXiv},
       eprint = {2211.03852},
 primaryClass = {astro-ph.CO},
       adsurl = {https://ui.adsabs.harvard.edu/abs/2024MNRAS.529.2473H}
}

@ARTICLE{Weinberg:1992,
       author = {{Weinberg}, David H.},
        title = "{Reconstructing primordial density fluctuations. I - Method}",
      journal = {\mnras},
         year = 1992,
        month = jan,
       volume = {254},
        pages = {315-342},
          doi = {10.1093/mnras/254.2.315},
       adsurl = {https://ui.adsabs.harvard.edu/abs/1992MNRAS.254..315W}
}

@ARTICLE{Lee:2018,
       author = {{Lee}, Khee-Gan and {Krolewski}, Alex and {White}, Martin and {Schlegel}, David and {Nugent}, Peter E. and {Hennawi}, Joseph F. and {M{\"u}ller}, Thomas and {Pan}, Richard and {Prochaska}, J. Xavier and {Font-Ribera}, Andreu and {Suzuki}, Nao and {Glazebrook}, Karl and {Kacprzak}, Glenn G. and {Kartaltepe}, Jeyhan S. and {Koekemoer}, Anton M. and {Le F{\`e}vre}, Olivier and {Lemaux}, Brian C. and {Maier}, Christian and {Nanayakkara}, Themiya and {Rich}, R. Michael and {Sanders}, D.~B. and {Salvato}, Mara and {Tasca}, Lidia and {Tran}, Kim-Vy H.},
        title = "{First Data Release of the COSMOS Ly{\ensuremath{\alpha}} Mapping and Tomography Observations: 3D Ly{\ensuremath{\alpha}} Forest Tomography at 2.05 < z < 2.55}",
      journal = {\apjs},
         year = 2018,
        month = aug,
       volume = {237},
       number = {2},
          eid = {31},
        pages = {31},
          doi = {10.3847/1538-4365/aace58},
archivePrefix = {arXiv},
       eprint = {1710.02894},
 primaryClass = {astro-ph.CO},
       adsurl = {https://ui.adsabs.harvard.edu/abs/2018ApJS..237...31L}
}

@ARTICLE{Lapparent:1986,
       author = {{de Lapparent}, V. and {Geller}, M.~J. and {Huchra}, J.~P.},
        title = "{A Slice of the Universe}",
      journal = {\apjl},
         year = 1986,
        month = mar,
       volume = {302},
        pages = {L1},
          doi = {10.1086/184625},
       adsurl = {https://ui.adsabs.harvard.edu/abs/1986ApJ...302L...1D}
}

@ARTICLE{Bond:1996,
       author = {{Bond}, J. Richard and {Kofman}, Lev and {Pogosyan}, Dmitry},
        title = "{How filaments of galaxies are woven into the cosmic web}",
      journal = {\nat},
         year = 1996,
        month = apr,
       volume = {380},
       number = {6575},
        pages = {603-606},
          doi = {10.1038/380603a0},
archivePrefix = {arXiv},
       eprint = {astro-ph/9512141},
 primaryClass = {astro-ph},
       adsurl = {https://ui.adsabs.harvard.edu/abs/1996Natur.380..603B}
}

@ARTICLE{Joeveer:1978,
       author = {{Joeveer}, Mikhel and {Einasto}, Jaan and {Tago}, Erik},
        title = "{Spatial distribution of galaxies and of clusters of galaxies in the southern galactic hemisphere}",
      journal = {\mnras},
         year = 1978,
        month = nov,
       volume = {185},
        pages = {357-370},
          doi = {10.1093/mnras/185.2.357},
       adsurl = {https://ui.adsabs.harvard.edu/abs/1978MNRAS.185..357J}
}

@ARTICLE{Gregory:1978,
       author = {{Gregory}, S.~A. and {Thompson}, L.~A.},
        title = "{The Coma/A1367 supercluster and its environs.}",
      journal = {\apj},
         year = 1978,
        month = jun,
       volume = {222},
        pages = {784-799},
          doi = {10.1086/156198},
       adsurl = {https://ui.adsabs.harvard.edu/abs/1978ApJ...222..784G}
}

@ARTICLE{York:2000,
       author = {{York}, Donald G. and {Adelman}, J. and {Anderson}, Jr., John E. and {Anderson}, Scott F. and {Annis}, James and {Bahcall}, Neta A. and {Bakken}, J.~A. and {Barkhouser}, Robert and {Bastian}, Steven and {Berman}, Eileen and {Boroski}, William N. and {Bracker}, Steve and {Briegel}, Charlie and {Briggs}, John W. and {Brinkmann}, J. and {Brunner}, Robert and {Burles}, Scott and {Carey}, Larry and {Carr}, Michael A. and {Castander}, Francisco J. and {Chen}, Bing and {Colestock}, Patrick L. and {Connolly}, A.~J. and {Crocker}, J.~H. and {Csabai}, Istv{\'a}n and {Czarapata}, Paul C. and {Davis}, John Eric and {Doi}, Mamoru and {Dombeck}, Tom and {Eisenstein}, Daniel and {Ellman}, Nancy and {Elms}, Brian R. and {Evans}, Michael L. and {Fan}, Xiaohui and {Federwitz}, Glenn R. and {Fiscelli}, Larry and {Friedman}, Scott and {Frieman}, Joshua A. and {Fukugita}, Masataka and {Gillespie}, Bruce and {Gunn}, James E. and {Gurbani}, Vijay K. and {de Haas}, Ernst and {Haldeman}, Merle and {Harris}, Frederick H. and {Hayes}, J. and {Heckman}, Timothy M. and {Hennessy}, G.~S. and {Hindsley}, Robert B. and {Holm}, Scott and {Holmgren}, Donald J. and {Huang}, Chi-hao and {Hull}, Charles and {Husby}, Don and {Ichikawa}, Shin-Ichi and {Ichikawa}, Takashi and {Ivezi{\'c}}, {\v{Z}}eljko and {Kent}, Stephen and {Kim}, Rita S.~J. and {Kinney}, E. and {Klaene}, Mark and {Kleinman}, A.~N. and {Kleinman}, S. and {Knapp}, G.~R. and {Korienek}, John and {Kron}, Richard G. and {Kunszt}, Peter Z. and {Lamb}, D.~Q. and {Lee}, B. and {Leger}, R. French and {Limmongkol}, Siriluk and {Lindenmeyer}, Carl and {Long}, Daniel C. and {Loomis}, Craig and {Loveday}, Jon and {Lucinio}, Rich and {Lupton}, Robert H. and {MacKinnon}, Bryan and {Mannery}, Edward J. and {Mantsch}, P.~M. and {Margon}, Bruce and {McGehee}, Peregrine and {McKay}, Timothy A. and {Meiksin}, Avery and {Merelli}, Aronne and {Monet}, David G. and {Munn}, Jeffrey A. and {Narayanan}, Vijay K. and {Nash}, Thomas and {Neilsen}, Eric and {Neswold}, Rich and {Newberg}, Heidi Jo and {Nichol}, R.~C. and {Nicinski}, Tom and {Nonino}, Mario and {Okada}, Norio and {Okamura}, Sadanori and {Ostriker}, Jeremiah P. and {Owen}, Russell and {Pauls}, A. George and {Peoples}, John and {Peterson}, R.~L. and {Petravick}, Donald and {Pier}, Jeffrey R. and {Pope}, Adrian and {Pordes}, Ruth and {Prosapio}, Angela and {Rechenmacher}, Ron and {Quinn}, Thomas R. and {Richards}, Gordon T. and {Richmond}, Michael W. and {Rivetta}, Claudio H. and {Rockosi}, Constance M. and {Ruthmansdorfer}, Kurt and {Sandford}, Dale and {Schlegel}, David J. and {Schneider}, Donald P. and {Sekiguchi}, Maki and {Sergey}, Gary and {Shimasaku}, Kazuhiro and {Siegmund}, Walter A. and {Smee}, Stephen and {Smith}, J. Allyn and {Snedden}, S. and {Stone}, R. and {Stoughton}, Chris and {Strauss}, Michael A. and {Stubbs}, Christopher and {SubbaRao}, Mark and {Szalay}, Alexander S. and {Szapudi}, Istvan and {Szokoly}, Gyula P. and {Thakar}, Anirudda R. and {Tremonti}, Christy and {Tucker}, Douglas L. and {Uomoto}, Alan and {Vanden Berk}, Dan and {Vogeley}, Michael S. and {Waddell}, Patrick and {Wang}, Shu-i. and {Watanabe}, Masaru and {Weinberg}, David H. and {Yanny}, Brian and {Yasuda}, Naoki and {SDSS Collaboration}},
        title = "{The Sloan Digital Sky Survey: Technical Summary}",
      journal = {\aj},
         year = 2000,
        month = sep,
       volume = {120},
       number = {3},
        pages = {1579-1587},
          doi = {10.1086/301513},
archivePrefix = {arXiv},
       eprint = {astro-ph/0006396},
 primaryClass = {astro-ph},
       adsurl = {https://ui.adsabs.harvard.edu/abs/2000AJ....120.1579Y}
}

@ARTICLE{Driver:2011,
       author = {{Driver}, S.~P. and {Hill}, D.~T. and {Kelvin}, L.~S. and {Robotham}, A.~S.~G. and {Liske}, J. and {Norberg}, P. and {Baldry}, I.~K. and {Bamford}, S.~P. and {Hopkins}, A.~M. and {Loveday}, J. and {Peacock}, J.~A. and {Andrae}, E. and {Bland-Hawthorn}, J. and {Brough}, S. and {Brown}, M.~J.~I. and {Cameron}, E. and {Ching}, J.~H.~Y. and {Colless}, M. and {Conselice}, C.~J. and {Croom}, S.~M. and {Cross}, N.~J.~G. and {de Propris}, R. and {Dye}, S. and {Drinkwater}, M.~J. and {Ellis}, S. and {Graham}, Alister W. and {Grootes}, M.~W. and {Gunawardhana}, M. and {Jones}, D.~H. and {van Kampen}, E. and {Maraston}, C. and {Nichol}, R.~C. and {Parkinson}, H.~R. and {Phillipps}, S. and {Pimbblet}, K. and {Popescu}, C.~C. and {Prescott}, M. and {Roseboom}, I.~G. and {Sadler}, E.~M. and {Sansom}, A.~E. and {Sharp}, R.~G. and {Smith}, D.~J.~B. and {Taylor}, E. and {Thomas}, D. and {Tuffs}, R.~J. and {Wijesinghe}, D. and {Dunne}, L. and {Frenk}, C.~S. and {Jarvis}, M.~J. and {Madore}, B.~F. and {Meyer}, M.~J. and {Seibert}, M. and {Staveley-Smith}, L. and {Sutherland}, W.~J. and {Warren}, S.~J.},
        title = "{Galaxy and Mass Assembly (GAMA): survey diagnostics and core data release}",
      journal = {\mnras},
         year = 2011,
        month = may,
       volume = {413},
       number = {2},
        pages = {971-995},
          doi = {10.1111/j.1365-2966.2010.18188.x},
archivePrefix = {arXiv},
       eprint = {1009.0614},
 primaryClass = {astro-ph.CO},
       adsurl = {https://ui.adsabs.harvard.edu/abs/2011MNRAS.413..971D}
}

@ARTICLE{Laigle:2016,
       author = {{Laigle}, C. and {McCracken}, H.~J. and {Ilbert}, O. and {Hsieh}, B.~C. and {Davidzon}, I. and {Capak}, P. and {Hasinger}, G. and {Silverman}, J.~D. and {Pichon}, C. and {Coupon}, J. and {Aussel}, H. and {Le Borgne}, D. and {Caputi}, K. and {Cassata}, P. and {Chang}, Y.-Y. and {Civano}, F. and {Dunlop}, J. and {Fynbo}, J. and {Kartaltepe}, J.~S. and {Koekemoer}, A. and {Le F{\`e}vre}, O. and {Le Floc'h}, E. and {Leauthaud}, A. and {Lilly}, S. and {Lin}, L. and {Marchesi}, S. and {Milvang-Jensen}, B. and {Salvato}, M. and {Sanders}, D.~B. and {Scoville}, N. and {Smolcic}, V. and {Stockmann}, M. and {Taniguchi}, Y. and {Tasca}, L. and {Toft}, S. and {Vaccari}, Mattia and {Zabl}, J.},
        title = "{The COSMOS2015 Catalog: Exploring the 1 < z < 6 Universe with Half a Million Galaxies}",
      journal = {\apjs},
         year = 2016,
        month = jun,
       volume = {224},
       number = {2},
          eid = {24},
        pages = {24},
          doi = {10.3847/0067-0049/224/2/24},
archivePrefix = {arXiv},
       eprint = {1604.02350},
 primaryClass = {astro-ph.GA},
       adsurl = {https://ui.adsabs.harvard.edu/abs/2016ApJS..224...24L}
}

@ARTICLE{Sousbie:2011,
       author = {{Sousbie}, T. and {Pichon}, C. and {Kawahara}, H.},
        title = "{The persistent cosmic web and its filamentary structure - II. Illustrations}",
      journal = {\mnras},
         year = 2011,
        month = jun,
       volume = {414},
       number = {1},
        pages = {384-403},
          doi = {10.1111/j.1365-2966.2011.18395.x},
archivePrefix = {arXiv},
       eprint = {1009.4014},
 primaryClass = {astro-ph.CO},
       adsurl = {https://ui.adsabs.harvard.edu/abs/2011MNRAS.414..384S}
}

@ARTICLE{Barsanti:2022,
       author = {{Barsanti}, Stefania and {Colless}, Matthew and {Welker}, Charlotte and {Oh}, Sree and {Casura}, Sarah and {Bryant}, Julia J. and {Croom}, Scott M. and {D'Eugenio}, Francesco and {Lawrence}, Jon S. and {Richards}, Samuel N. and {van de Sande}, Jesse},
        title = "{The SAMI Galaxy Survey: flipping of the spin-filament alignment correlates most strongly with growth of the bulge}",
      journal = {\mnras},
         year = 2022,
        month = nov,
       volume = {516},
       number = {3},
        pages = {3569-3591},
          doi = {10.1093/mnras/stac2405},
archivePrefix = {arXiv},
       eprint = {2208.10767},
 primaryClass = {astro-ph.GA},
       adsurl = {https://ui.adsabs.harvard.edu/abs/2022MNRAS.516.3569B}
}

@ARTICLE{Malavasi:2017,
       author = {{Malavasi}, N. and {Arnouts}, S. and {Vibert}, D. and {de la Torre}, S. and {Moutard}, T. and {Pichon}, C. and {Davidzon}, I. and {Kraljic}, K. and {Bolzonella}, M. and {Guzzo}, L. and {Garilli}, B. and {Scodeggio}, M. and {Granett}, B.~R. and {Abbas}, U. and {Adami}, C. and {Bottini}, D. and {Cappi}, A. and {Cucciati}, O. and {Franzetti}, P. and {Fritz}, A. and {Iovino}, A. and {Krywult}, J. and {Le Brun}, V. and {Le F{\`e}vre}, O. and {Maccagni}, D. and {Ma{\l}ek}, K. and {Marulli}, F. and {Polletta}, M. and {Pollo}, A. and {Tasca}, L. and {Tojeiro}, R. and {Vergani}, D. and {Zanichelli}, A. and {Bel}, J. and {Branchini}, E. and {Coupon}, J. and {De Lucia}, G. and {Dubois}, Y. and {Hawken}, A. and {Ilbert}, O. and {Laigle}, C. and {Moscardini}, L. and {Sousbie}, T. and {Treyer}, M. and {Zamorani}, G.},
        title = "{The VIMOS Public Extragalactic Redshift Survey (VIPERS): galaxy segregation inside filaments at z $\simeq$ 0.7}",
      journal = {\mnras},
         year = 2017,
        month = mar,
       volume = {465},
       number = {4},
        pages = {3817-3822},
          doi = {10.1093/mnras/stw2864},
archivePrefix = {arXiv},
       eprint = {1611.07045},
 primaryClass = {astro-ph.GA},
       adsurl = {https://ui.adsabs.harvard.edu/abs/2017MNRAS.465.3817M}
}

@article{Wright:1999,
  title={Numerical optimization},
  author={Wright, Stephen and Nocedal, Jorge and others},
  journal={Springer Science},
  volume={35},
  number={67-68},
  pages={7},
  year={1999}
}
